\documentclass[twocolumn,floats,pra]{revtex4}

\def \uqsl2 {$U_q$(sl$_2$) }
\def \cA {{\cal A}}
\def \hcA {\widehat{\cal A}}
\newcommand {\be}{\begin{equation}}
\newcommand {\ee} {\end{equation}}
\newcommand {\bea}{\begin{eqnarray}}
\newcommand {\eea} {\end{eqnarray}}

\begin{document}


\title{Enlarged symmetry algebras of spin chains, loop models, and $S$-matrices}
\author{N. Read$^1$ and H. Saleur$^{2,3}$}
\affiliation{$^1$ Department of Physics, Yale
University, P.O. Box 208120, New Haven, CT 06520-8120, USA\\
$^2$ Service de Physique Th\'eorique, CEA Saclay, 91191 Gif sur
Yvette, France\\ $^3$ Department of Physics and Astronomy,
University of Southern California, Los Angeles, CA 90089, USA\\}
\date{January 11, 2007}

\begin{abstract}
The symmetry algebras of certain families of quantum spin chains
are considered in detail. The simplest examples possess $m$ states
per site ($m\geq2$), with nearest-neighbor interactions with
U$(m)$ symmetry, under which the sites transform alternately along
the chain in the fundamental $m$ and its conjugate representation
$\overline{m}$. We find that these spin chains, even with {\em
arbitrary} coefficients of these interactions, have a symmetry
algebra $\cA_m$ much larger than U$(m)$, which implies that the
energy eigenstates fall into sectors that for open chains (i.e.,
free boundary conditions) can be labeled by $j=0$, $1$, \ldots,
$L$, for the $2L$-site chain, such that the degeneracies of all
eigenvalues in the $j$th sector are generically the same and
increase rapidly with $j$. For large $j$, these degeneracies are
much larger than those that would be expected from the U$(m)$
symmetry alone. The enlarged symmetry algebra $\cA_m(2L)$ consists
of operators that commute in this space of states with the
Temperley-Lieb algebra that is generated by the set of
nearest-neighbor interaction terms; $\cA_m(2L)$ is not a Yangian.
There are similar results for supersymmetric chains with
gl$(m+n|n)$ symmetry of nearest-neighbor interactions, and a
richer representation structure for closed chains (i.e., periodic
boundary conditions). The symmetries also apply to the loop models
that can be obtained from the spin chains in a spacetime or
transfer matrix picture. In the loop language, the symmetries
arise because the loops cannot cross. We further define tensor
products of representations (for the open chains) by joining
chains end to end. The fusion rules for decomposing the tensor
product of representations labeled $j_1$ and $j_2$ take the same
form as the Clebsch-Gordan series for SU(2). This and other
structures turn the symmetry algebra $\cA_m$ into a ribbon Hopf
algebra, and we show that this is ``Morita equivalent'' to the
quantum group $U_q($sl$_2)$ for $m=q+q^{-1}$. The open-chain
results are extended to the cases $|m|< 2$ for which the algebras
are no longer semisimple; these possess continuum limits that are
critical (conformal) field theories, or massive perturbations
thereof. Such models, for open and closed boundary conditions,
arise in connection with disordered fermions, percolation, and
polymers (self-avoiding walks), and certain non-linear sigma
models, all in two dimensions. A product operation is defined in a
related way for the Temperley-Lieb representations also, and the
fusion rules for this are related to those for $\cA_m$ or \uqsl2
representations; this is useful for the continuum limits also, as
we discuss in a companion paper.

\end{abstract}

\pacs{PACS numbers: } 

\maketitle

\section{Introduction}
\subsection{Motivation}

For many years there has been wide interest in lattice and
continuum quantum field theory systems in two space-time (or two
Euclidean space) dimensions in areas ranging from statistical
mechanics to high-energy theory (including string theory). The
lattice models include those like the six-vertex model
\cite{baxter}, and when a continuum limit is taken in one of the
two dimensions (call it time) in an appropriate manner, the
transfer matrix becomes a local quantum-mechanical Hamiltonian
acting on a space of states which for many lattice models is of
the form of a quantum spin chain, that is a tensor product of
vector spaces, one for each site in the chain. In the most
interesting cases, the continuum limit in the second dimension, or
in both dimensions simultaneously, can also be taken to obtain a
continuum quantum field theory. This requires that the original
lattice or spin chain model be at or close to a critical point. In
this limit, attention is focused on the low-energy and
long-wavelength states of the system. The quantum field theory
description, including correlation functions, then yields the
universal properties of the underlying lattice model. In cases of
interest here, the quantum field theory of the nearby critical
point possesses conformal symmetry, which in two dimensions is
potentially very powerful in its analysis. At the same time, other
``internal'' (non-spacetime) symmetries that may be present in the
lattice model play an important role as they survive in the
continuum theory. Then a knowledge of the theory at a critical
point is also a starting point for the analysis of massive quantum
field theories obtained by applying a relevant perturbation (the
integral over space-time of a local operator) to the conformal
one.

Many lattice systems and conformal field theories (and their
massive deformations) have been understood for some time. But
there remain others that are not understood, and which have been
of recent interest in various diverse contexts:

(i) Among spin chains, there are the $m$, $\overline{m}$ chains
with SU($m$) symmetry, in which the fundamental representation,
denoted $m$, of SU($m$) alternates with its dual (or conjugate)
representation $\overline{m}$ along the chain, and the Hamiltonian
is the unique nearest-neighbor SU($m$)-invariant (``Heisenberg
exchange'') coupling. While the $m=2$ case is the usual spin-1/2
chain, the cases $m>2$ have not been well understood, but are of
interest in connection with valence-bond theories of high
temperature superconductors, for example. In the $m>2$ cases, only
first-order phase transitions occur.

(ii) Very similar spin chains, but with SU($m$) replaced by the
Lie superalgebra sl($2|1$), appeared in connection with a lattice
model that exemplifies a variant of the quantum Hall transition,
termed the spin Hall transition \cite{glr}. In these cases, there
is a second-order phase transition which was shown to have some of
the same exponents as ordinary classical percolation \cite{glr}.
Understanding this example of a localization transition in two
dimensions is of great interest for localization physics
generally.

(iii) There are various ``loop models'' that occur in statistical
mechanics, in which each configuration is a collection of loops in
two dimensions, which are assigned Boltzmann weights that depend
on the number of loops, the lengths of the loops, whether the
loops intersect, and so on \cite{nien}. The loops may be a subset
of the edges of some lattice, or may be in the continuum. Problems
such as polymers (self-avoiding random walks) and percolation have
formulations as loop models (see e.g. Refs.\
\cite{dfsz,baxter,nienrev}). These problems are among a subclass
in which the loops are required never to cross themselves or each
other. They are related also to stochastic Loewner evolution
(SLE).

(iv) There is a family of similar spin chain models generalizing
those in (ii) with Lie superalgebras as symmetries, and crucially,
the partition functions of all these models can be expanded as
sums over sets of non-crossing loops as in point (iii) (this is
how the relation with percolation was found
\cite{bb,affleck3,glr}). For these theories the complete exact
spectrum of the conformal field theory (CFT) at the critical point
has been found \cite{rs}, and exhibited surprisingly large
multiplicities that increased exponentially with energy (or
scaling dimension). Also, the natural massive perturbation has
been described at the level of S-matrices \cite{fr}. These
theories may also be viewed as critical points in other continuum
quantum field theories, such as non-linear sigma models that
possess the same Lie superalgebras as global symmetries.

(v) The CFTs in points (iii), (iv) are examples of irrational,
logarithmic CFTs. The majority of the CFTs that are understood are
what are known as {\em rational} theories, in which the local
fields in the theory fall into a finite set of representations of
some so-called chiral algebra, such as the Virasoro algebra
\cite{cft}. In irrational theories, this fails (though numbers
such as the scaling dimensions and central charge may still be
rational numbers). Moreover, the semisimplicity property (full
decomposability of operator products into direct sums of
irreducible representations, present in the rational case) of the
representations of the chiral algebra that occur is generally
lost, and this can lead to logarithms in correlation functions
\cite{gurarie}. This class of CFTs is of general interest, in view
of applications such as those already mentioned. We note that
logarithmic CFTs are non-unitary theories---the inner product on
the space of states is necessarily indefinite.

(vi) Irrational CFTs also occur in string theory, and in
particular we mention strings moving on a curved space such as
anti-de Sitter (AdS) space. These spaces have Lie superalgebras as
symmetries and are very similar to those mentioned above. (In
string theory applications, non-unitarity of the worldsheet CFT is
not necessarily a problem, due to the imposition of the Virasoro
constraints.) In some recent work, some of the same non-linear
sigma models with Lie superalgebra symmetry as in (iv) have been
considered using S-matrices \cite{polchinski}, in the hope of
reaching a better understanding of the AdS/CFT correspondence. In
these the scattering events are naturally pictured in space-time
using loops, as in the loop models. In these models, parameter
values at which crossings of loops are absent are special points.

In this paper we return to our study of these problems, and
concentrate here on symmetry aspects. (Apart from a brief
discussion, extensions to CFT will be left for a separate paper
\cite{rs3}.) We exhibit enlarged symmetry algebras of a class of
spin chains, loop models, and S-matrices, for both open and closed
boundary conditions in the ``space'' direction. We aim to show how
this can contribute to understanding this class of models, at both
the lattice and continuum (CFT) levels.

\subsection{A simple example}

Before surveying the rest of the paper, we will give a simple
example to show the flavor of the results; this example is
presented in detail (in the form of a spin chain) in the following
section. For the example, we consider identical particles (of mass
$M$ say), moving on a line (with open boundary condition). Each
particle carries an internal ``spin'' space consisting of $m$
possible states, which thus form an $m$-dimensional vector space.
When the particles collide, they scatter with some amplitudes,
most details of which will not be needed (the symmetry is {\em
independent} of these other details). The important point is that
the in- and out-going spin states of the pair of scattering
particles are related in a simple way. To describe it, we will
view the particles as always being reflected in the collision, so
that the particles labelled by $i$ are identified by their order
along the line, with coordinates $x_i$, subject to $x_i\leq
x_{i'}$ for all $i<{i'}$. In a collision of $i$, $i+1$, the spin
states afterwards are either the same as before, or if the spins
are the same, $a_i=a_{i+1}$, then they can instead annihilate and
be replaced by another pair $a_i'=a_{i+1}'$ (with coefficients
independent of $a_i$, $a_i'$). That is, the scattering matrix is
some amplitude (depending generally on the momentum of $i$, $i+1$)
times the identity matrix in spin space, plus another amplitude
times the projection operator $P_i$ onto the O($m$) singlet state
for the pair of spin states on $i$, $i+1$; here $P_i^2=P_i$. [In
the following sections we use notation $e_i=mP_i$ instead.] The
processes can be represented in a spacetime diagram with time
plotted vertically by continuous lines that represent particles,
with the spin state flowing along the line. In a collision the
particle worldlines reflect off each other (touch) without
crossing; the first amplitude can be represented by the lines that
approach but have an avoided crossing, and continue upward, while
for the second the two incoming lines can be joined as if they
annihilate, and a new pair of spin states created for the outgoing
particles (that is, the same as the first amplitude but rotated by
$\pi/2$). Thus the continuity of the lines shows the global
conservation of the spin. A general spacetime multiparticle
scattering process, or a contribution to the partition function
(in imaginary time) is then represented by a collection of
non-crossing lines and closed loops, which may touch but cannot
cross.

The model possesses more symmetry than the global O($m$). It has
global U($m$) symmetry, if we view the leftmost particle ($i=0$)
as transforming in the fundamental representation of U($m$), and
the particles as alternating between the fundamental and its
conjugate (or dual) representation thereafter. That is because the
only process of annihilation occurs between neighbors, which are
now always of the opposite type. This U($m$) symmetry has been
noticed repeatedly in the literature in contexts similar to this
one. In fact, if the particles are viewed as fixed on lattice
sites and interact via the Hamiltonian $H=-\sum_iP_i$, which has
the spin structure of this form, then we obtain the $m$,
$\overline{m}$ models described under point (i) above. Let the
generators of the Lie algebra gl$_m$ of U($m$) acting on the $i$th
particle be denoted $J_{ia}^b$ (details are given in the next
section). Then we will show that the Hamiltonian $H$ commutes with
a set of symmetry operators, that are a certain linear subspace of
the operators%
\be%
\widetilde{J}^{a_1a_2\ldots a_k}_{b_1b_2\ldots b_k}=\sum_{0\leq
i_1<i_2<\cdots <i_k} J_{i_1 b_1}^{a_1}J_{i_2
b_2}^{a_2}\cdots J_{i_k b_k}^{a_k} \ee%
for $k=1$, $2$, \ldots. The subspace is defined by conditions
involving all pairs of neighbors $J_i$, $J_{i+1}$ in this
expression; the operators in the subspace have components
$J^{a_1a_2\ldots a_k}_{b_1b_2\ldots b_k}$, with the condition that
any contraction of an $a_l$ with $b_{l\pm1}$ vanishes. The
symmetry algebra spanned by the $J$'s is {\em larger} than the
U($m$) symmetry, which is fully described by the action of
arbitrary products of the generators $J_a^b=\sum_iJ_{ia}^b$. The
algebra generated by the latter (an image of what is called the
universal enveloping algebra of gl$_m$) is a proper subalgebra of
the one found here. The energy eigenstates of the system form
multiplets of the enlarged symmetry algebra. As the symmetry
operators commute with {\em any} Hamiltonian (or S-matrix) that
has the specified form in its spin dependence, for example
$H=-\sum_i \lambda_iP_i$ with arbitrary real parameters
$\lambda_i$, there are minimal multiplicities that must be
possessed by the eigenstates of any such system, which are the
dimensions of the irreducible representations of the symmetry
algebra. (For particular cases, there can be further
``accidental'' degeneracies.) These
dimensions are given explicitly by %
\be%
D_j=(q^{2j+1}-q^{-2j-1})/(q-q^{-1}),\ee%
for the $j$th representation ($j=0$, $1/2$, $1$, $3/2$, \ldots),
where $q$ is defined by $m=q+q^{-1}$, and are positive integers
whenever $m$ is an integer $\geq 2$. When the number of particles
is even (odd), $2j$ must be even (resp., odd). Each such
representation can be decomposed into irreducible representations
of SU($m$), which are of increasing complexity as $j$ increases.
For example, for the SU(3) case, the numbers are Fibonacci
numbers, $D_j=F_{4j+2}=1$, $3$, $8$, $21$, \ldots, and these
multiplicities were observed in the $3$, $\overline{3}$ spin chain
model some time ago \cite{ha}.

The enlarged symmetry algebras result directly from the fact that
the worldlines of the particles never cross. This allows some
``flexibility'' to the symmetry transformations, as if U($m$)
could act in a position-dependent way, and still commute with the
interaction. The algebra is not the Yangian of U($m$). Unlike
Yangian symmetries, this symmetry algebra commutes with arbitrary
Hamiltonians such as $H$ above. This means the symmetry applies to
{\em random} non--translationally-invariant spin chain models as
well as translationally-invariant ones. Thus integrability of the
models, though present in many of those we mentioned earlier, is
{\em not} required for the enlarged symmetry to be present.

\subsection{Loop models}

Before describing more of our results, it may be useful to clarify
the relations among some loop models (readers more interested in
algebra and symmetry can skip this Section on a first reading). In
the above example, we began with identical particles and O($m$)
symmetry, but then noticed the U($m$) symmetry that is a simple
consequence of non-crossing of lines. This involved the use of
alternating fundamental and dual representations. In terms of
worldlines, the distinction between the two can be encoded by
adding an arrow to a line, so that the line represents the
fundamental of U($m$) when the arrow is in the positive time
direction (when the line is projected to the time axis), and the
dual when it is in the negative time direction. However, if the
system has a periodic boundary condition in the space direction,
then the orientations cannot consistently alternate all along the
system if there is an odd number of particles. In that case we
would be forced to drop the arrows and view the loops as without
an orientation. When studying partition functions of loop models
with a periodic boundary condition in both the space and time
directions, so that the system is topologically a torus, this of
course applies to both the space and time directions (or
``fundamental cycles'' on the torus). There are then two versions
of the loop models, independent of any other parameters: the {\em
unoriented} version, in which loops are viewed in general as
carrying no arrow to orient them, and any number, whether even or
odd, of loops can wind around either cycle of the torus; and the
{\em oriented} version, in which all loops carry a fixed
orientation that alternates along the intersection points with any
cycle on the torus, and the number of lines crossing any such
cycle must clearly be even. We emphasize that unlike in some other
models of loops, such as the loop formulation of the six-vertex
model, in the oriented loops models the orientations of loops are
fixed, and not summed over or treated as dynamical variables. The
assignment of orientations to all loop configurations can be made
unambiguous in the continuum models by the use of a base point and
an orientation for the torus (viewed as an orientable manifold).
The orientations are then specified by shading the region (bounded
by the loops) that contains the base point, and shading other
regions alternately so that each loop separates a shaded from an
unshaded region, then adding arrows to the loops such that the
shaded regions is to the left as one goes in the direction of the
arrow---this step uses the orientation of the torus.

The unoriented loop models can be given a lattice formulation in
which the loops occupy a subset of the edges of the honeycomb
graph (lattice) \cite{nien}, and no edge is occupied more than
once. The partition function of the model is a sum over these
allowed loop configurations, weighted with a factor $K$ for each
edge occupied by a loop, and a factor $m$ for each loop
\cite{nien}. For each value of the parameter $m$ in the range
$-2<m\leq 2$, these models possess a second-order phase transition
between a high-temperature (massive) phase, and a low-temperature
phase that is itself critical (massless). For each $m$ in
$-2<m<2$, all points in the low-temperature phase flow to the same
fixed point (CFT). There are thus two critical theories of
interest: dilute (the critical point) and dense (the
low-temperature phase), so-named because the density of loops
increases with decreasing temperature. The oriented loops models
can be obtained similarly as models of loops on a subset of the
edges of the square lattice, respecting rules for shading regions
as described in the previous paragraph, such that shading occurs
only on plaquettes on one of the two sublattices (this corresponds
to a Potts model with dilution [random removal] of the Potts
spins, which lie on the shaded plaquettes). The phase diagram
structure of the oriented loops models is similar to that of the
unoriented loops models. Then as $m$ varies between $2$ and $-2$,
there are four series of models (or CFTs) to study, corresponding
to the choices oriented or unoriented, and dilute versus dense.
(In Ref.\ \cite{rs}, only three of these series were described.
The fourth is the oriented dilute series, which contains the
exponents of the tricritical Potts models.) The oriented dense
loops phases can be obtained by working in the limit in which the
loops fill the edges of the square lattice (i.e.\ no dilution),
and that is the formulation with which we will begin here, in
which the transfer matrix corresponds to the Hamiltonian $H$ of
the $m$, $\overline{m}$ models. The more general models in which
not every edge is occupied correspond to models like the simple
example above, but with particles restricted to sites in a chain,
with the occupation number of each site taking the values 0 or 1
only. Further, the Hamiltonian contains terms that create or
destroy particles in O($m$)-singlet pairs on neighboring sites. It
should be clear that for the purposes of understanding the
symmetry, the presence of empty sites, and of single-particle
hopping (kinetic energy) and two-particle creation and destruction
terms in the Hamiltonian has no effect. Thus the distinction of
dilute versus dense can be ignored when analyzing the symmetry.
The oriented and unoriented cases turn out to be closely related
for open (i.e.\ free) boundary conditions in the space direction,
as we have already seen, but for closed (i.e.\ periodic) boundary
conditions there are some small differences between them that have
to be taken into account.

Again, the fact that the allowed configurations of loops in the
loop models do not include crossings (either between distinct
loops or parts of the same loop) is crucial for the symmetry
analysis. It is also a key feature of the physics. One can
consider modified models in which crossings are allowed, and are
weighted with some small weight. As crossing of two lines violates
the rules for orientation, this is allowed only within the
unoriented loops models (for the oriented loops models, the
simplest allowed crossing involves three lines intersecting at a
point). It has been shown \cite{jrs} that while this perturbation
is irrelevant in the dilute (critical) theories for $-2<m\leq2$,
in the dense theories with $-2<m<2$ it is a relevant perturbation,
and causes a flow to a different phase which was identified with a
Goldstone theory \cite{jrs}. The enlarged symmetry is broken by
this perturbation. On the other hand, in models or field theories
in which one can argue that there is a flow to the universality
class of one of the non-crossing loops models \cite{rs,fr}, the
enlarged symmetry is present in the infrared fixed point theory,
even though it was not present at shorter length scales. It is
also present if the infrared fixed point CFT is perturbed by a
relevant perturbation that does not involve crossings of loops or
otherwise break the symmetry, as in Ref.\ \cite{fr}.

\subsection{Summary of the paper}

Although the picture of loop configurations that we have just
described may be the best way to unify the diverse physical
applications mentioned earlier, for our purposes the most direct
way to analyze the symmetry is algebraic. Symmetry is after all an
algebraic concept, and symmetry generators are associated with a
cycle (or ``time-slice'') across the system. Symmetries are
defined as commuting with the Hamiltonian (or transfer matrix). In
fact, for $H=-\sum_i\lambda_iP_i$, commutation with $H$ for
arbitrary values of the coefficients $\lambda_i$ leads us to
search for operators that commute with all the operators $P_i$ (or
$e_i=mP_i$) considered earlier. The latter operators generate an
algebra, which for open boundary conditions is the Temperley-Lieb
(TL) algebra \cite{tl,baxter,bb,affleck3}. It is then a
well-defined algebraic problem to find the algebra of operators in
our Hilbert space that commutes with the given algebra. The
resulting algebras then are guaranteed to commute with {\em any}
Hamiltonian constructed from the operators $e_i$ by algebraic
operations. In the continuum limit (after the models have been
defined for $-2<m\leq 2$), the TL algebra becomes the Virasoro
algebra in the CFT, and we obtain an infinite-dimensional algebra
that commutes with the whole of the Virasoro algebra (or with two
copies of Virasoro in the closed case), and which is larger than
the global Lie superalgebra symmetry with which we began. The
symmetry also survives the effect of certain perturbations that
render the field theory massive.

For applications of symmetry in physics, one needs more than just
an associative algebra. One also needs to be able to analyze
tensor products of representations (defined as vector spaces) as
themselves representations of the same algebra, and for each
representation there should be a corresponding ``dual'', the
tensor product of which with the original representation contains
the singlet (trivial) representation (this is used for describing
antiparticles, for example). In terms of the algebra, these
requirements mean it must possess additional structures that make
it into a Hopf algebra. In our case, we find (for $m\geq2$ open
cases) that not only are the irreducible representations of our
algebra labelled by integers $j$, just as for SU($2$), but the
Clebsch-Gordan series for decomposing tensor products are also the
same. There are additional structures of ``braiding'' and
``twist'', and all of these agree with properties of the quantum
group deformation of the sl$_2$ Lie algebra, denoted
$U_q$(sl$_2$). More technically, we can say that the symmetry
algebra is Morita equivalent to the $U_q$(sl$_2$), and we also
obtain a version of this for the cases $|m|<2$ in which the
algebra is not semisimple.

The knowledge of these symmetry properties of generic Hamiltonians
will aid the numerical analysis of these chains, as the problem
reduces to diagonalizing a much smaller matrix (of size $d_j$ or
$\widehat{d}_j$ below) within each symmetry sector.

In the following, we begin with the symmetry algebras for the spin
chains, first in Section \ref{chains} for open boundary
conditions. We obtain the dimensions of the irreducible
representations of the symmetry algebras, and explicit formulas
for a basis for these algebras, for both oriented and unoriented
cases. In Section \ref{closch} we do the same for the closed
chains. In Section \ref{potts} we briefly discuss similar results
for the Potts model, and in Section \ref{susych} we address the
generalization of the spin chains so that they have supersymmetry
gl($m+n|n$) or osp($m+2n|2n$). Readers whose main interest is in
U($m$) spin chains can skip these two sections. Then in Section
\ref{tens} we introduce the structures that make the symmetry
algebras for the open cases, in the limit as the length
$\to\infty$, into Hopf algebras, for $|m|\geq 2$, and explain the
Morita equivalence with $U_q($sl$_2)$. In Section \ref{nonsemi} we
address the issues arising for the (supersymmetric) cases with
$|m|<2$ when the algebras involved are not semisimple, and show
the Morita equivalence with \uqsl2 in these cases also. This
section requires more mathematical sophistication than the rest of
the paper. In Section \ref{cont} we briefly discuss the continuum
limit, especially the case when it is a conformal field theory.
Finally, there is an Appendix which contains many technical
details. In this paper, we concentrate on stating the ideas and
results, and do not always give proofs of the statements; many are
elementary or can be found in the literature. The technical level
increases towards the end, but we hope that the early stages will
be widely accessible to physicists. We concentrate on the oriented
loop models, but also give some results for the unoriented loops
models.


\section{Spin chains: open boundary condition}
\label{chains}

In this section, we discuss the most basic examples, those of
U$(m)$ spin chains or oriented loops models, and also the
unoriented or O$(m)$ models. We explain the algebraic background
that is sufficient for analyzing these cases. We include explicit
results for dimensions and symmetry generators.


\subsection{Oriented loops models}

We consider an SU($m$) antiferromagnetic spin chain with open
(free) boundary conditions. It consists of $2L$ sites labelled
$i=0$, \ldots, $2L-1$, with an $m$-dimensional complex vector
space $V_i\cong {\bf C}^m$ at each site ($\bf C$ is the field of
complex numbers). The states can be represented using oscillator
operators $b_i^a$, $b_{ia}^\dagger$ for $i$ even,
$\overline{b}_{ia}$, $\overline{b}_i^{a\dagger}$ for $i$ odd, with
commutation relations $[b_i^a,b_{jb}^\dagger]=
\delta_{ij}\delta_b^a$ ($a$, $b=1$, \ldots, $m$), and similarly
for $i$ odd. The destruction operators $b_i^a$,
$\overline{b}_{ia}$ destroy the vacuum state, the daggers indicate
the adjoint, and the spaces $V_i$ are defined by the constraints
\begin{eqnarray}
b_{ia}^\dagger b_i^a&=& 1 \quad (i\hbox{ even}),\label{constr1}\\
\overline{b}_i^{a\dagger} \overline{b}_{ia} &=& 1 \quad (i\hbox{
odd})\label{constr2}
\end{eqnarray}
of one boson per site (we use the summation convention for
repeated indices of the same type as $a$). We define the
generators of U($m$) (or in fact of gl$_m$) acting in the spaces
$V_i$ by $J_{ia}^b=b_{ia}^\dagger b_i^b$ for $i$ even,
$J_{ia}^b=-\overline{b}_i^{b\dagger} \overline{b}_{ia}$ for $i$
odd, and the commutation relations among the $J_i$s (for each $i$)
are $i$-independent. Hence the global gl$_m$ algebra, defined by
its generators $J_a^b=\sum_i J_{ia}^b$, acts in the tensor product
$V=\otimes_{i=0}^{2L-1} V_i$ of copies of the fundamental
representation of gl$_m$ on even sites, alternating with its dual
on odd sites, as desired to construct an {\em antiferromagnetic}
spin chain. Though the U(1) subalgebra of gl$_m$ generated by
$J_a^a$ acts trivially on the chain (and by a scalar on each
site), it is often notationally convenient not to subtract this
trace from the generators $J_a^b$.

The SU($m$)-invariant nearest-neighbor coupling in the chain is
unique, up to additive and multiplicative constants. It is the
usual ``Heisenberg coupling'' of magnetism, and can be written in
terms of operators $e_i$, defined explicitly as %
\be e_i=\left\{\begin{array} {rl}
\overline{b}_{i+1}^{a\dagger}b_{ia}^\dagger
b_i^b \overline{b}_{i+1,b},&\hbox{ $i$ even,}\\
\overline{b}_i^{a\dagger}b_{i+1,a}^\dagger b_{i+1}^b
\overline{b}_{ib},&\hbox{ $i$ odd.}
\end{array}\right. \ee
The $e_i$'s are Hermitian, $e_i^\dagger=e_i$. Acting in the
constrained space $V$, they satisfy \cite{bb,affleck3} the
relations \cite{tl,baxter}
\bea e_i^2 &=& m e_i,\nonumber\\ e_i\, e_{i\pm 1}\, e_i &=&
e_i,\nonumber\\  e_i\,e_j&=&e_j\,e_i \qquad (j\neq i,\;i\pm 1).
\label{tlrel}
\eea %
We write the parameter $m$ as $m=q+q^{-1}$. The abstract
associative algebra over the complex numbers $\bf C$ generated by
unity and the $n-1$ generators $e_0$, \ldots, $e_{n-2}$ that
satisfy the relations (\ref{tlrel}) (and no other relations
algebraically independent of these) with parameter $q\in {\bf C}$
will be called the Temperley-Lieb (TL) algebra, TL$_{n}(q)$ (for
$n$ either even or odd). The representation we have constructed in
the space $V$ is faithful for $m\geq 2$. All algebras in this
paper are over $\bf C$ and are assumed to include unity.

In the space $V$, a much-studied Hamiltonian for a
nearest-neighbor antiferromagnetic spin chain is the $m$,
$\overline{m}$ model with open (free) boundary conditions,
\be H=-\epsilon\sum_{i\;{\rm even}}e_i -\epsilon^{-1}\sum_{i\;{\rm
odd}} e_i,\label{tlham} \ee%
where $\epsilon>0$ is a parameter; if $\epsilon \neq 1$, the model
is said to have staggered couplings. In the thermodynamic
($L\to\infty$) limit a phase transition occurs at $\epsilon=1$,
which is first order for $m>2$, second order for $m=2$. More
general Hamiltonians for the spin chain can be constructed, using
arbitrary elements of the TL algebra, that is any sum of products
of generators, perhaps with random coefficients.

There are also vertex models whose transfer matrices are TL
algebra elements. These models may be expanded as configurations
of loops that run along (and fill) the edges of the square
lattice, with avoided crossings at the vertices. This produces a
class of {\it loop model}. This uses a well-known graphical
representation of the TL algebra that has been described by many
authors (see e.g.\ Ref.\ \cite{baxter}). In our models, the loops
are viewed as oriented, with the fundamental of gl$_m$ running
along in the direction of the arrow. We emphasize again that these
orientations are fixed, not summed over (they are not dynamical
variables). The two ways of conserving directions of arrows on the
loops at a vertex represent the action either of $1\otimes 1$ or
of $e_i$ for the two sites $i$, $i+1$ in question. The symmetry
algebra we find in this paper determines the multiplicities in the
spectrum of {\em any} of these models.

The TL algebra arose in studies of the Potts model
\cite{tl,baxter}, to which Refs.\ \cite{bb,affleck3} thus found a
relation of the SU($m$) antiferromagnetic spin chains. In terms of
the TL generators $e_i$, the $Q$-state Potts model has the same
Hamiltonian, and the $e_i$ obey the TL algebra with $Q=m^2$, but
here for all non-negative integers $Q$ \cite{baxter}. In the Potts
model partition function, the loops are the boundaries of Potts
clusters \cite{baxter}. The possibility of using different
representation spaces for a given algebra was a main point of TL
\cite{tl}, and we will have much more to say about this below.

Because of the Hermiticity of the generators $e_i$ with respect to
the positive-definite inner product on the vector space $V$, it
follows that for any element $a$ of the algebra, $a^\dagger$ is
also an element. If such an algebra is finite dimensional, then it
is automatically {\em semisimple}, which implies that any of its
finite-dimensional modules (representations \cite{reps}) is also
semisimple, that is fully reducible into a finite direct sum of
irreducible representations (simple modules) \cite{pierce,anf}.
Because the representation of the TL algebra in $V$ is faithful
for $m\geq 2$, this shows that TL$_{2L}(q)$ is semisimple for
$m\geq2$ (in fact, this holds for all $q\geq 1$)
\cite{jones1,ghj,martin}.

The general algebraic results that will be used extensively in
this paper will now be summarized. In an irreducible
representation of dimension $N<\infty$ of an algebra $A$, $A$ acts
as the full matrix algebra $M_N({\bf C})$ of all complex $N\times
N$ matrices. Any semisimple algebra $A$ is isomorphic to a finite
direct product of such algebras, one for each distinct irreducible
(we refer to two irreducible representations as distinct if they
are not isomorphic): $A=\prod_j M_{N_j}({\bf C}) $, where $j$ runs
over the set of isomorphism classes of representations, of
dimensions $N_j$. In matrix language, this means that the algebra
$A$ is isomorphic to the algebra of all block-diagonal complex
matrices, where the blocks are $N_j\times N_j$. It follows that
the dimension of $A$ is ${\rm dim\,} A= \sum_j N_j^2$. The
commutant $B$ of such an algebra $A$ in a finite-dimensional
representation $V'$ (the commutant is the algebra of all linear
transformations of $V'$ that commute with all elements of $A$)
must also be semisimple, by Schur's lemma. If the $j$th distinct
irreducible representation of $A$ occurs in $V'$ with multiplicity
$M_j$, then the full matrix algebra $M_{M_j}({\bf C})$ commutes
with $A$ acting in the subspace of $V'$ spanned by the copies of
the $j$th irreducible. Assuming that $M_j>0$ for all $j$, which
means that $A$ is represented faithfully in $V'$, then there is a
one-one correspondence between the isomorphism classes of
irreducible representations of $A$ and $B$, and $B$ is the direct
product of algebras $M_{M_j}({\bf C})$ (so ${\rm
dim\,}B=\sum_jM_j^2$). In particular, (i) both algebras have the
same number of distinct irreducible representations; (ii) the
commutant of $B$ in $V'$ is $A$ (that is, the double commutant of
$A$, which necessarily contains $A$, is in fact equal to $A$);
(iii) the space $V'$ can be decomposed as
\be %
V'=\bigoplus_j {\bf C}^{N_j}\otimes {\bf C}^{M_j}\ee %
where ${\bf C}^{N_j}$ (${\bf C}^{M_j}$) stands for the irreducible
representation of $A$ (resp., $B$) of dimension $N_j$ (resp.,
$M_j$). Hence, ${\rm dim\,} V'=\sum_j N_j M_j$. Also, (iv) the
center of $A$ (i.e.\ the subalgebra of elements in $A$ that
commute with all elements of $A$) is also the center of $B$, and
both are isomorphic to $\prod_j M_1({\bf C})$ (i.e.\ a direct
product of one dimensional algebras, one for each $j$, each
isomorphic to the complex numbers). This correspondence between
representations of $A$ and $B$ is also a simple form of the more
general {\em Morita equivalence} of algebras (which applies to
algebras that are not necessarily semisimple).

We now apply these general results to TL$_n(q)$ acting in $V$, and
construct the commutant algebra explicitly. First, we require
information about the representations of the TL algebra TL$_n(q)$
(in this paragraph, we allow $n$ to be odd or even). The $m=2$
($q=1$) case of $V$ is instructive. This is just the su($2$)
spin-1/2 chain. The su($2$) symmetry commutes with permutations of
the sites, and the TL generators are essentially the
transpositions of neighbors, which generate the symmetric group
$S_n$ on $n$ sites. TL$_n(1)$ is isomorphic to the group algebra
of $S_n$, projected to the space of $S_n$ representations that
actually occur in $V$. The latter correspond to Ferrers-Young
diagrams with $n$ boxes and at most two rows. It follows that the
dimension of the $j$th irreducible representation of TL$_n(1)$,
which is the
multiplicity of the spin $j$ representation of su($2$) in the chain, is%
\be d_j=\left(\begin{array}{c}n\\ n/2+j
\end{array}\right)-\left(\begin{array}{c}n\\ n/2+j+1
\end{array}\right),\label{tldims}\ee
where $n/2+j$ must be an integer. For $n$ even, $j=0$, $1$,
\ldots, $n/2$. The sum of the squares of these dimensions is
$\sum_j d_j^2= (n+1)^{-1}{2n\choose n}$, the dimension of the TL
algebra \cite{ghj}. The same Ferrers-Young diagrams correspond to
the representations of su($2$) that occur in $V$; $2j$ equals the
difference in the number of boxes in the two rows in the diagram.
This well-known decomposition is called [the su($2$) case of]
Frobenius-Schur-Weyl duality, and is an example of the
correspondence discussed in the previous paragraph. For $q>1$, the
irreducible representations of TL$_n(q)$ retain the same
dimensions, because the algebra varies continuously with $q$
\cite{jones1,ghj,martin}, and hence the TL algebra also retains
the same dimension ${\rm dim}\,{\rm TL}_n(q)=(n+1)^{-1}{2n\choose
n}$. This formula is valid for all $m$, including $m=0$, as can be
readily seen from the diagrammatic definition of TL, in which each
element of a linear basis corresponds to a diagram
\cite{jones1,ghj,martin}.

We now describe the decomposition of our chain $V$ into
irreducibles of TL$_n(q)$, for $n=2L$. We use the following
non-orthogonal, but linearly-independent basis states. Each basis
state corresponds to a pattern of nested parentheses and dots,
such as $()\bullet(())\bullet$, with one symbol for each site of
the chain ($2L=8$ in the example). The parentheses must obey the
usual typographical rules for nesting, so that each ``$($''
corresponds to exactly one ``$)$''. Also, the dots must not be
inside of any parentheses. These rules imply that the $()$ pairs
consist of one even and one odd site, and that dots are
alternately on even and odd sites, starting with an even site at
the left. The states in the chain represented by such a diagram
are constructed by contracting the sites that correspond to each
$()$ pair into an SU($m$) singlet (``valence bond''). For the
dots, the state in the tensor product of spaces ${\bf C}^m$ (each
of which corresponds to a dot) must be chosen so that application
of the projection operator to the SU($m$) singlet for any two dots
that are adjacent (when parentheses are ignored) annihilates the
state. Thus, those sites are ``non-contractible''.

It is easily seen that the TL algebra applied to these basis
states does not mix states with different numbers of
non-contractible sites. Application of an $e_i$ always produces a
valence bond at $i$, $i+1$, together with a rearrangement of some
other contractions for sites that were contracted with $i$ or
$i+1$ before (if one of $i$, $i+1$ was a non-contractible dot, it
is moved to another position). Thus, the TL generators $e_i$
change a pattern to another valid pattern. However, when $i$,
$i+1$ are both non-contractible, $e_i$ annihilates the state. The
TL algebra never changes the state on the sequence of dots. The
number of valid patterns is independent of $m$, and one can use
the $m=2$ case to count them; in this case there is an invertible
mapping of the space of states, commuting with the action of
su($2$), that maps basis states corresponding to valid patterns
with $2j$ dots to those for ``standard'' Young tableaus with at
most two rows, such that the difference in length of the two rows
is $2j$ (a standard tableau is a Ferrers-Young diagram with one of
the numbers $1$, $2$, \ldots, $n$ inserted in each box, such that
the numbers are increasing both to the right along the rows and
down the columns). Hence for each number $2j=0$, $2$, \ldots, $2L$
of dots, the number of valid patterns coincides with the
dimensions $d_j$ of $S_{2L}$ representations \cite{sagan}. The
basis states are linearly independent and span the $j$th
irreducible representation of TL$_{2L}(q)$.

The number of states for each valid pattern with $2j$ dots
determines the dimension $D_j$ of the $j$th representation of the
commutant of TL$_{2L}(q)$ in $V$. These numbers can be found
inductively, by adding another pair of non-contractible dots to
the end of a sequence, and are independent of $L$. This leads
easily to the recurrence relation \cite{rs}
\be%
D_1 D_j=D_{j+1}+D_j+D_{j-1}.\label{recrel}\ee%
Also, it is clear that $D_0=1$, $D_1=m^2-1$ [$D_1$ is the
dimension of the adjoint representation of SU($m$)].
Using $m=q+q^{-1}$, the solution is%
\be%
D_j=[2j+1]_q,\label{dj}\ee where
$[n]_q=q^{n-1}+q^{n-3}+\ldots+q^{-n+1}=(q^n-q^{-n})/(q-q^{-1})$ is
the $q$-deformation of any integer $n$. As a check, the total
number of linearly-independent states we constructed is%
\be%
\sum_{j=0}^{L} D_j d_j=(q+q^{-1})^{2L}=m^{2L},\ee %
which is exactly ${\rm dim\,}V$. Note that these dimensions are
the multiplicities of energy eigenvalues for the generic
Hamiltonians in the TL algebra, mentioned earlier. For $m>2$, the
dimensions $D_j$ asymptotically increase exponentially with $j$.
For example, for $m=3$, the first few are $1$, $8$, $55$, $377$,
\ldots, and are the Fibonacci numbers $D_j=F_{4j+2}$. The $m=3$
cases were found for Hamiltonian (\ref{tlham}) previously
\cite{ha}. For $j>1$, the decomposition of these multiplets into
irreducible representations of su($m$) become increasingly
complicated.

To construct the commutant algebra explicitly, we introduce the
operators (for $k\leq 2L$)%
\be%
\widetilde{J}^{a_1a_2\ldots a_k}_{b_1b_2\ldots b_k}=\sum_{0\leq
i_1<i_2<\cdots <i_k\leq 2L-1} J_{i_1 b_1}^{a_1}J_{i_2
b_2}^{a_2}\cdots J_{i_k b_k}^{a_k} \ee%
(for $k=0$, we define $\widetilde{J}=1$, and for $k=1$,
$\widetilde{J}^a_b=J^a_b$ as defined earlier). For each $k=0$,
$1$, \ldots, these span a space of dimension $m^{2k}$. In this
space of operators we can impose linear conditions, that the
contraction of one of the indices $a$ with a {\em neighboring}
index $b$ [i.e.\ of $a_l$ with $b_{l+ 1}$ (resp., $b_{l-1}$), for
$l=1$, $2$, \ldots, $k- 1$ (resp., $l=2$, \ldots, $k$)] is zero.
This gives us a basis set $J^{a_1\ldots a_k}_{b_1\ldots b_k}$,
that are ``traceless'' in this sense.
For example, for $k=2$, we have %
\be%
J^{a_1 a_2}_{b_1 b_2}=\widetilde{J}^{a_1 a_2}_{b_1 b_2}%
-\frac{1}{m}\widetilde{J}^{a a_2}_{b_1 a}\delta^{a_1}_{b_2}%
-\frac{1}{m}\widetilde{J}^{a_1b}_{b b_2} \delta^{a_2}_{b_1}%
+\frac{1}{m^2}\widetilde{J}^{a b}_{b a}\delta^{a_1}_{b_2}\delta^{a_2}_{b_1}%
\ee%
and these span a space of dimension $(m^2-1)^2$. In general, the
dimension is $(D_{k/2})^2$. The exact forms are %
\be%
J^{a_1a_2\ldots a_k}_{b_1b_2\ldots b_k}=(P^\bullet P_\bullet
\widetilde{J})^{a_1a_2\ldots a_k}_{b_1b_2\ldots b_k},\label{commdef}\ee%
where $P^\bullet$ ($P_\bullet$) is the (Jones-Wenzl) projection
operator to the ``traceless'' sector on the vector space indexed
by $(a_1,b_2,\ldots)$ [resp., $(b_1,a_2,\ldots,)$], which can be
constructed recursively using the TL$_k(q)$ algebra in these
spaces \cite{fnsww}.

One can readily show that: (i) all $J^{a_1a_2\ldots
a_k}_{b_1b_2\ldots b_k}$ commute with all the $e_i$, hence with
all of TL$_{2L}(q)$ (they leave the patterns unchanged); (ii) all
$J^{a_1a_2\ldots a_k}_{b_1b_2\ldots b_k}$ with $k>2j$ annihilate
the $j$th irreducible representation of the commutant algebra;
(iii) the space of $J^{a_1a_2\ldots a_k}_{b_1b_2\ldots b_k}$s with
$k=2j$ acts as the matrix algebra $M_{D_j}({\bf C})$ on the $j$th
irreducible representation; (iv) $J^{a_1a_2\ldots
a_k}_{b_1b_2\ldots b_k}$ with $k<2j$ map the $j$th irreducible
representation into itself, and hence in that subspace can be
written as linear combinations of those with $k=2j$. In
particular, in our chain of $2L$ sites, the operators with $k$ odd
are linear combinations of those with $k$ even. Hence only even
$k$ are needed. These results show that {\em the algebra spanned
by $J^{a_1a_2\ldots a_k}_{b_1b_2\ldots b_k}$ ($k=0$, $2$, \ldots)
is the commutant algebra $\cA_m(2L)$ of TL$_{2L}(q)$ in $V$}, with
dimension ${\rm dim\, }\cA_m(2L)=\sum_j (D_j)^2$. Because the
dimensions $D_j$ are independent of $L$, the limit $L\to\infty$
exists, and we write $\cA_m=\lim_{L\to\infty}\cA_m(2L)$.

The ``obvious'' global symmetry algebra is gl$_m$, or more
accurately the universal enveloping algebra (UEA) $U({\rm gl}_m)$
of gl$_m$, which is the associative algebra generated by the
generators $J_a^b$ of gl$_m$, subject to the commutation relations
of gl$_m$ [or similarly for $U$(sl$_m$)] \cite{cp,kass}. For
$m>2$, our algebra $\cA_m$ is strictly larger than $U$(sl$_m$);
$U$(sl$_m$) is a proper subalgebra of $\cA_m$ \cite{quotalg}, and
hence the representations of $\cA$ can be decomposed into
representations of sl$_m$. The dimension of $\cA_m(2L)$ can be
found in closed form, and grows exponentially with $L$:%
\bea%
{\rm dim}\,\cA_m(2L)&=&
\frac{q^{4L+4}-q^{-4L-4}}{(q-q^{-1})^2(q^2-q^{-2})}-
\frac{2(L+1)}{(q-q^{-1})^2}\\
&=&\frac{[2L+2]_{q^2}-(2L+2)}{(q-q^{-1})^2}\\
&\sim&\frac{q^{4L}}{(1-q^{-2})(1-q^{-4})}\eea%
as $L\to\infty$; here we used $q>1$. By contrast, the dimension of
the quotient of sl$_m$ that acts faithfully in the chain is the
sum of squares of the dimensions of irreducibles that occur, and
the latter dimensions are known polynomials in the highest weight
of the representation, of degree at most $m(m-1)/2$ (the
dimensions of irreducibles of sl$_m$ are found by
Frobenius-Schur-Weyl duality, and given by the Weyl dimension
formula). The highest weights that occur are bounded by something
of order the length $L$ of the chain. We have not made a precise
estimate of the dimension of the resulting associative algebra,
but it is clear that it is bounded by a polynomial in $L$, and
thus much smaller than $\cA_m(2L)$ for large $L$.

We do not know of a ``small'' or ``simple'' set of generators for
$\cA_m$ (that would be analogous to the set of $J^a_b$ for $U({\rm
gl}_m)$). $\cA_m$ is not the Yangian of sl$_m$. However, the
properties above imply that $\cA_m(2L)$ is a {\em cellular}
algebra for all $L$, for which we have given a {\em cellular
basis} $J^{a_1a_2\ldots a_k}_{b_1b_2\ldots b_k}$ ($k=0$, $2$,
\ldots), in the sense defined in Ref.\ \cite{gl} (for an
exposition, see e.g.\ Ref.\ \cite{mathas}, or Sec.\ \ref{nonsemi}
below). This fact also generalizes to the unoriented and
supersymmetric versions below.


\subsection{Unoriented loops models}

The unoriented loops models \cite{nien} were discussed in the
Introduction. We consider an open spin chain in which each site is
in the vector representation of O($m$) [or of the Lie algebra
so$_m$ of O($m$)]. Such a chain can be represented by using
oscillator operators $b_i^a$, $b_{ia}^\dagger$, with commutation
relations $[b_i^a,b_{jb}^\dagger]= \delta_{ij}\delta_b^a$ ($a$,
$b=1$, \ldots, $m$)  for {\em all} $i$, $j=0$, \ldots, $n-1$. (In
this subsection, we again allow $n$ to be odd or even.) The
destruction operators $b_i^a$ destroy the vacuum state, and the
spaces $V_i\cong {\bf C}^m$ are defined by the constraints%
\be%
b_{ia}^\dagger b_i^a = 1 \ee%
of one boson per site for all $i$. TL generators can be written,
for $i=0$, $1$, \ldots, $n-2$, as%
\be%
e_i=\eta^{ab}\eta_{cd}b_{ia}^\dagger b_{i+1,b}^\dagger b_i^c
b_{i+1}^d.\ee%
Here $\eta_{ab}$ and its inverse $\eta^{ab}$ represent the
non-degenerate symmetric bilinear form, which can be taken to be
$\eta_{ab}=\delta_{ab}$ (the Kronecker delta); the invariance of
this form defines the symmetry group O($m$). The generators of the
so$_m$ Lie algebra on each site are %
\be%
G_{iab}=\eta_{bc}b_{ia}^\dagger b_i^c-\eta_{ac}b_{ib}^\dagger
b_i^c.\ee%
The interaction of sites $i$, $i+1$ given by the TL generator
$e_i$ is not the most general one (up to additive and
multiplicative constants) allowed by O($m$) symmetry. It generates
only loop configurations in which loops never cross, and it is
this feature that admits an enlarged symmetry. If we define
$\overline{b}_{ia}=\eta_{ab}b_i^b$,
$\overline{b}_i^{a\dagger}=\eta^{ab}b_{ib}^\dagger$ for $i$ odd,
then as noted in the Introduction, the operators $e_i$, which obey
the TL relations (\ref{tlrel}), are actually invariant under
U($m$) (with the odd sites transforming in the dual fundamental).
Generators $J_{ia}^b$ of gl$_m$ can be defined as in the oriented
loops models.

The results for the oriented case now generalize easily to the
unoriented case. The label $j$ for the representations now takes
values $0$, $1/2$, $1$, \ldots, $n/2$, with $2j$ odd (even) if and
only if $n$ is odd (resp., even). A linear basis for the commutant
algebra of TL$_n(q)$ is given by the same operators
$J^{a_1a_2\ldots a_k}_{b_1b_2\ldots b_k}$, eq.\ (\ref{commdef})
(with $2L$ replaced by $n$ in the summations), but now the algebra
for $n$ even (odd) is spanned by these operators with $k\leq n$
and $k$ even (resp., odd) only; clearly, the even sector is
isomorphic to that for the oriented case with the same $m$. The
dimensions of the irreducible representations are again given by
$D_j=[2j+1]_q$ for $q+q^{-1}=m$. As examples, $D_{1/2}=m$ is the
dimension of the vector representation of so$_m$, while
$D_1=m^2-1$ is that of the adjoint of sl$_m$, and decomposes into
the antisymmetric and traceless symmetric tensor irreducible
representations of so$_m$. We call the $n\to\infty$ limit of the
commutant algebras ${\cal B}_m$; it includes operators
$J^{a_1a_2\ldots a_k}_{b_1b_2\ldots b_k}$ for all $k$, and has
irreducible representations for all $j=0$, $1/2$, $1$, $3/2$,
\ldots.

One could also consider U($m$) chains with an odd number of sites,
on which the odd sector of ${\cal B}_m$ would act. However, this
would not fit naturally with the later developments below, unlike
the O($m$) models.


\section{Closed boundary condition}
\label{closch}

In this section, we perform an analysis similar to that of the
previous section for the case of the closed boundary condition.
This section can be skipped by readers mainly interested in Hopf
algebra structures.


\subsection{Oriented loops models}

We generalize the results to the closed (periodic) version of the
SU($m$) spin chain models. The space of states is the same (with
an even number $2L$ of sites), and the TL generators are defined
as there, but now there are $2L$ generators $e_i$, which obey the
relations (\ref{tlrel}) with $i\pm 1$ interpreted cyclically, with
$i=2L\equiv0$ (mod $2L$). In addition, there is now an obvious
cyclic symmetry of the system. We can introduce an operator $u^2$
(with inverse $u^{-2}$) which translates any state to the right by
$2$ sites (so as to be consistent with the distinction of two
types of sites carrying dual representations), so $u^{2L}=1$
(there are no odd powers of $u$, though there are in the
unoriented or O($m$) cases). We have $u^2 e_i u^{-2}=e_{i+2}$.
These operators generate an algebra.

The precise algebra can be defined abstractly as an algebra of
diagrams as for TL, but this time on an annulus (or finite
cylinder), in which a general basis element corresponds to a
diagram of $2L$ sites on the inner, and $2L$ on the outer
boundary; the sites are connected in pairs, but only
configurations that can be represented using lines inside the
annulus that do not cross are allowed \cite{jones2}. Further, for
the oriented loops models, the lines must be orientable, such that
the arrows emanate from the even sites and enter the odd sites on
the inner boundary, and the reverse for the outer boundary.
Multiplication is defined in a natural way on these diagrams, by
joining an inner to an outer annulus, and removing the interior
sites \cite{jones2}. We emphasize that whenever a closed loop is
produced when diagrams are multiplied together, this loop must be
replaced by a numerical factor $m$ (as for the TL algebra), even
for loops that wind around the annulus, as well as for those that
are homotopic to a point. The algebra is generated by the elements
$e_i$ and $u^2$, and they obey the above relations, which however
are not a complete set. (The numerical factor $m$ for winding
loops is not a consequence of the stated relations, but a separate
assumption.) We call this finite-dimensional ``annular'' algebra
\cite{jones2} the Jones-TL, or JTL algebra, JTL$_{2L}(q)$
\cite{jones2,eg} (the latter terminology is not standard). It is
easily seen that our definitions produce a representation of
JTL$_{2L}(q)$ in $V$, however it turns out that it is faithful
only when $m>2$. For $m=2$, the TL algebra already contains all
permutations of the sites, and the extra generators $e_{-1}$ and
$u^{\pm 2}$ acting in $V$ can be expressed in terms of the others.
Also, for real $q>0$, the JTL algebra is semisimple only for
$q\neq 1$ \cite{jones2}, unlike the TL algebra. We will see that
the JTL algebra is much richer than the TL algebra. For other
periodic generalizations of the TL algebra, which are infinite
dimensional, see e.g.\ Refs.\ \cite{ms1,ms2}.

On passing from TL$_{2L}(q)$ to JTL$_{2L}(q)$, some irreducible
representations of TL$_{2L}(q)$ will combine to form irreducibles
of JTL$_{2L}(q)$. On the other hand, since we work in the same
space $V$, when the algebra becomes larger, its commutant must
become smaller, and some irreducible representations of the
commutant will break into irreducibles of the commutant of
JTL$_{2L}(q)$. (These remarks assume the algebras involved are
semisimple.)

The dimensions of the irreducible representations of JTL$_{2L}(q)$
for $q>1$ are known \cite{jones2}. We construct representations of
the JTL algebra using parentheses and dots again, but now
parentheses can be paired cyclically, so $)\bullet(())\bullet($ is
a valid pattern (valid patterns may also be defined by drawing
them on a disk with the sites on the boundary, and lines within
the disk connect contracted sites without crossing, while
noncontractible sites can be reached within the disk from one
another without crossing a contraction line). Contractions that
cross the end of the chain, like one in the preceding example,
become pairs of dots if one reverts to the open TL point of view,
and so one finds for the number of valid patterns with $2j$ dots
\be%
\widehat{d}_j=\sum_{j'=j,j+1,\ldots} d_{j'}=\left(\begin{array}{c}2L\\
L+j
\end{array}\right). \label{hatdj}\ee
This is valid for $j>0$. For the $j=0$ case, all contraction lines
can be drawn without crossing the $0$, $2L-1$ link, so
$\widehat{d}_0=d_0$. [These formulas, which as we will see give
the dimensions $\widehat{d}_j$ of the irreducible representations
of JTL$_{2L}(q)$, also show how the representations decompose when
considered as representations of the subalgebra TL$_{2L}(q)$.]

For the set of valid patterns for each value $j=0$, $1$, \ldots,
$L$, one has a set of non-orthogonal but linearly independent
basis states, by again associating a singlet valence bond to each
pair $()$ of corresponding parentheses, and for the
non-contractible sites (now defined cyclically), states that
vanish if one such site is contracted with its neighbor on either
side (cyclically). The subspace spanned by these elements is a
representation of JTL$_{2L}(q)$ and of its commutant $\hcA_m(2L)$,
and its dimension is $\widehat{d}_j\widehat{D}_j$, where the
dimensions $\widehat{D}_j$ for each pattern will now be found. By
comparing with the definitions for the open case, we see that the
dimensions $\widehat{D}_j$ obey $D_j=\widehat{D}_j+D_{j-1}$
($j\geq2$),
$D_1=\widehat{D}_1$, $D_0=\widehat{D}_0$. That is,%
\be%
\widehat{D}_j=\left\{\begin{array}{l}q^{2j}+q^{-2j}\quad(j>1),\\
q^2+1+q^{-2}\quad(j=1),\\1\quad(j=0)  \end{array}\right. \ee Note
that again $\sum_j\widehat{d}_j\widehat{D}_j=m^{2L}={\rm dim}\,
V$.

Unlike the open chains, for the closed chains the representations
of JTL$_{2L}(q)\otimes\hcA_m(2L)$ of dimension
$\widehat{d}_j\widehat{D}_j$ that we have now constructed for each
$j$ are not irreducible when $j\geq2$. There is a non-trivial
center of the restriction of JTL$_{2L}(q)$ and of $\hcA_m(2L)$ to
the $j$th subspace. This may be seen most easily in terms of the
commutant $\hcA_m(2L)$. For any basis state in the $j$th subspace,
the states on the non-contractible sites can be cyclically
permuted by moving them two steps to the right, without affecting
the pattern. This operation clearly commutes with the JTL algebra,
so when viewed as acting on all the basis states simultaneously it
gives an operator which lies in the commutant. Further, it
commutes with all elements of the commutant (restricted to this
subspace), because as we will see in the explicit expressions
below, these elements involve sums over position which ensure that
they are invariant under these operations on the basis states
(ultimately this is because of the isomorphism of the JTL algebra
$e_i\to e_{i+1}$ for all $i$). Hence this operator lies in the
center of the commutant (acting in the $j$th subspace), and so
must also lie in the center of the JTL algebra. By Schur's lemma,
it acts as a root of unity $e^{2iK}$ in any irreducible
representation of either algebra. We call $K$ (defined modulo
$\pi$) the {\em pseudomomentum}. As translation of the
non-contractible sites by $2j$ steps brings the state back to
itself, we have $jK\equiv0$ (mod $\pi$). We may conclude that, for
each $j\geq 2$, though all irreducible representations of
JTL$_{2L}(q)$ have the same dimension $\widehat{d}_j$, they are
not all isomorphic, and there is a distinct irreducible
representation for each distinct allowed $K$, and thus $j$
distinct isomorphism classes of irreducibles in all \cite{jones2}.

The representations of the commutant of dimension $\widehat{D}_j$
can be decomposed into eigenspaces of $K$, with $K=\pi P/N$ where
$P\geq 0$ and $N$ are coprime ($N$ is a divisor of $j$, written
$N|j$). We will denote the dimensions of these subspaces by
$\widehat{D}_{jK}$, with $\sum_K \widehat{D}_{jK}=\widehat{D}_j$,
from which again, $\sum_{j,K} \widehat{D}_{jK}
\widehat{d}_j=m^{2L}$. For $j=0$, $1$, $K\equiv0$ and
$\widehat{D}_{jK}=\widehat{D}_j$. When the state on the sequence
of $2j$ non-contractible sites is periodic with period $d$, $1\leq
d<j$ (with $d|j$), it contributes only to pseudomomenta such that
$N|d$. Using M\"{o}bius inversion \cite{hw} (similarly to Appendix
A of Ref.\ \cite{rs}), we obtain the dimensions $\widehat{D}_{jK}$
of the representations with $j\geq2$ and given $K$ of the
commutant $\hcA_m(2L)$ of
JTL$_{2L}(q)$ for $m>2$,%
\be
\widehat{D}_{jK}=\sum_{d,d':N|d}\frac{\mu(d/d')}{d}(q^{2d'}+q^{-2d'}
),\ee%
where the sum is over all positive divisors $d$, $d'$ of $j$, and
$\mu(x)$ is the M\"{o}bius function \cite{hw}. Alternatively, by
calculating the trace of the projection operator
onto pseudomomentum $K$ for a fixed pattern, we obtain%
\be%
\widehat{D}_{jK}
=\frac{1}{j}\sum_{r=0}^{j-1}e^{2iKr}\left[q^{2(j\wedge
r)}+q^{-2(j\wedge r)}\right],\label{Djkn=0}\ee %
where $j\wedge r$ denotes the highest common divisor of $j$ and
$r$ ($j\wedge0=j$ for all integers $j\geq0$). These two
expressions are equal, again by using Ref.\ \cite{hw}. These
multiplicities were given in the second form by Jones
\cite{jones2} (for these oriented cases, we have corrected a small
error at the end of Ref.\ \cite{jones2}). These representations of
$\hcA_m(2L)$ are irreducible, and the dimension of the algebra is
${\rm dim\,}\hcA_m(2L)=\sum_{j,K}(\widehat{D}_{jK})^2$. In the
$L\to\infty$ limit, we obtain an algebra
$\hcA_m=\lim_{L\to\infty}\hcA_m(2L)$.

If we put $m=2$ (even though this is a case in which the JTL
algebra does not act faithfully in $V$), the multiplicities
correctly vanish whenever $K\not\equiv 0$ (mod $\pi$), but the
formula for $\widehat{D}_{j0}$ for $j>1$ is not correct for this
case. Here, because the image of the JTL algebra that acts
faithfully in $V$ is the same as TL, its commutant is a quotient
of $U$(sl$_2$), with irreducible dimensions $D_j=2j+1$.

Some elements of $\hcA_m(2L)$ can be constructed as in the open
case. We use, for $k\geq 1$,
\be%
\widetilde{J}^{a_1a_2\ldots a_k}_{b_1b_2\ldots b_k}=\sum_{
i_1<i_2<\cdots <i_k<i_1} J_{i_1 b_1}^{a_1}J_{i_2
b_2}^{a_2}\cdots J_{i_k b_k}^{a_k}, \ee%
where the summations extend periodically on the chain; these
commute with $u^2$. A set of elements of the commutant
$\hcA_m(2L)$ can now be written, for $k$ even, as
\be%
\widehat{J}^{a_1a_2\ldots a_k}_{b_1b_2\ldots b_k}=(P^\bullet
P_{\bullet}^{\vphantom\bullet}
\widetilde{J})^{a_1a_2\ldots a_k}_{b_1b_2\ldots b_k},\ee%
where, similarly to the open case, the projector $P^\bullet$
($P_\bullet$) is the projection operator to the ``traceless''
sector (annihilated by all $e_i$) on the vector space indexed by
$(a_1,b_2,\ldots)$ [resp., $(b_1,a_2,\ldots,)$]. The projectors
certainly exist, as they project onto (non-irreducible for $k>2$)
representations of the semisimple algebra JTL$_{k}(q)$. For $k$
odd, there is another set,%
\be%
\widehat{J}^{a_1a_2\ldots a_k}_{b_1b_2\ldots
b_k}=(P^\bullet_{\;\bullet}
\widetilde{J})^{a_1a_2\ldots a_k}_{b_1b_2\ldots b_k},\ee%
in which $P^\bullet_\bullet$ is the projection operator to the
traceless sector (annihilated by $e_i$) on the single vector space
indexed by $(a_1,b_2,a_3,\ldots,a_k,b_1,a_2,\ldots,b_k)$. These
projectors exist in the algebra JTL$_{2k}(q)$. (For $k=1$,
$\widehat{J}^a_b$ is the traceless generator of sl$_m$.) The
$\widehat{J}$s have the cyclic invariance property,%
\be%
\widehat{J}^{a_1a_2\ldots a_k}_{b_1b_2\ldots
b_k}=\widehat{J}^{a_ka_1\ldots a_{k-1}}_{b_kb_1\ldots b_{k-1}}\ee%
(the cyclic property is clear for the $\widetilde{J}$s, and for
the $\widehat{J}$s with $k$ even uses the fact that the matrix
elements of the projectors $P_\bullet$, $P^\bullet$ are the same
real numbers when written out in the respective bases).

Unlike the open case, for the closed case the $\widehat{J}$
operators with $k$ even do not form a linear basis for
$\hcA_m(2L)$; however, they do generate it. Let us study how they
act on the $j$, $K$ irreducible representations for $k=2j$. A
natural decomposition of this space of $\widehat{J}$s is obtained
by using the projector $P^\bullet_{(K)}$ [and $P_\bullet^{(-K)}$]
onto the subspace of the space indexed by $(a_1,b_2,\ldots)$
[resp., $(b_1,a_2,\ldots,)$] that has pseudomomentum $K$ (resp.,
$-K$) as well as being annihilated by all $e_i$. This is possible
because the $\widehat{J}$s do preserve pseudomomentum. We note
that the cyclic invariance property implies that these operators
for $K$ and $-K$ are the same (up to some relabelling). We choose
an orthonormal basis for this subspace, indexed by $\alpha$,
$\beta$, $\gamma$, \ldots (resp., $\alpha^*$, \ldots; there is a
correspondence between these bases as indicated by the notation),
and write these projected $\widehat{J}$s as
$\widehat{J}_{\alpha\beta}$. Among the irreducibles with $j=k/2$,
these operators annihilate those with pseudomomentum $\not\equiv
K$ or $-K$ (mod $\pi$). Strictly speaking, the cases in which
$K\equiv -K$ (mod $\pi$) should be distinguished from the more
general cases $K\not \equiv -K$ (mod $\pi$); we will return to
this after dealing with the generic case. Then in general one
finds that $\widehat{J}_{\alpha\beta}$ maps $\beta$ (with
pseudomomentum $K$) onto $\alpha$, but also $\alpha^*$ (with
pseudomomentum $-K$) to $\beta^*$. We may re-normalize such that
the first of these non-zero matrix elements is equal to one. Now
define operators ${\cal J}_{\alpha\beta}=
\widehat{J}_{\alpha\gamma}\widehat{J}_{\gamma\beta}$ where
$\alpha\neq\beta$, and hence also $\alpha^*\neq\beta^*$ (the
summation convention is not in force for the Greek indices). This
is possible because all the spaces for $j>0$ have dimension $>1$.
Then we see that ${\cal J}_{\alpha\beta}$ acts as the elementary
matrix $E_{\alpha\beta}$ (whose only non-zero entry is $1$ in
position $\alpha$, $\beta$) in the pseudomomentum $K$
representation, and annihilates all others with $j=k/2$, including
that for $-K$. Finally the diagonal entries are defined as ${\cal
J}_{\alpha\alpha}={\cal J}_{\alpha\beta}{\cal J}_{\beta\alpha}$,
(for any $\beta\neq\alpha$) which acts as $E_{\alpha\alpha}$ in
the $j$, $K$ irreducible. For the cases in which $K\equiv -K$ (mod
$\pi$), one should note that the basis states $\alpha^*$ are in
the same space, and are essentially a permutation of the basis
labelled $\alpha$. ${\cal J}_{\alpha\beta}=
\widehat{J}_{\alpha\gamma}\widehat{J}_{\gamma\beta}$ acts as
$E_{\alpha\beta}$ only if $\alpha^*$, $\beta^*$ and $\gamma$ are
all distinct. This works provided the space has dimension $>2$,
which they all do (for $j>0$). As the elementary matrices form a
linear basis for the full matrix algebra
$M_{\widehat{D}_{jK}}({\bf C})$, we have shown that the operators
$\widehat{J}$ do generate the algebra $\hcA_m(2L)$.

Thus finally, we have defined a set of operators ${\cal
J}_{\alpha\beta}$ for all even $k$ and for all allowed $K$ (mod
$\pi$), which are a linear basis for $\hcA_m(2L)$, and which act
on the corresponding ($j=k/2$) irreducible as elementary symmetric
matrices, while annihilating those with the same $j$ but different
$K$, as well as those with $j<k/2$. Once again, these properties
imply that this basis is cellular \cite{gl,mathas}.


\subsection{Unoriented loops models}

For the unoriented loops models, the only differences in the
algebra (for which we will use the, again not standard, notation
uJTL) are that the number of sites $n$ can be odd, and that there
is now an element $u$ (and its inverse $u^{-1}$) that is a
translation by one site to the right. This annular algebra
uJTL$_n(q)$ was also analyzed by Jones \cite{jones2}. The
dimensions of its irreducible representations are again given by
the same formula (\ref{hatdj}) for $\widehat{d}_j$ (with $2L$
replaced by $n$), $j\leq n/2$, where $n/2+j$ must be an integer.

The description of the center of the algebra is the same as
before, except that now translation of the states on
non-contractible sites by one step to the right is possible, and
we define the eigenvalue of this operation to be $e^{iK}$, where
$K$ (now defined modulo $2\pi$) is the pseudomomentum. It obeys
$2jK\equiv0$ (mod $2\pi$). Then we can write $K=2\pi P/N$, where
$P\geq0$ and $N$ are coprime, and $N|(2j)$. The dimensions
$\widehat{D}^{({\rm u})}_{jK}$ of the irreducible representations
of the commutant $\widehat{\cal B}_m(n)$ of uJTL$_n(q)$ in $V$
again obey $\sum_K \widehat{D}^{({\rm u})}_{jK}=\widehat{D}_j$,
and we note that for all odd $2j\geq 1$,
$\widehat{D}_j=q^{2j}+q^{-2j}$. The dimensions $\widehat{D}^{({\rm
u})}_{jK}$ are now given for $j\neq0$, $1$ by
\cite{jones2}%
\be%
\widehat{D}^{({\rm u})}_{jK}
=\frac{1}{2j}\sum_{r=0}^{2j-1}e^{iKr}\left[q^{(2j)\wedge
r}+q^{-(2j)\wedge r}\right].\ee %
(Again, these in fact depend only on the denominator $N$ in $K$.)
It is interesting to interpret the small $j$ cases, while giving
the result for $j=1$. $j=0$ (with $K\equiv0$ by definition) is the
singlet. For $j=1/2$, in which case $K\equiv0$ only, we have
dimension $\widehat{D}^{({\rm u})}_{1/2,0}=\widehat{D}_{1/2}=m$,
which is clearly the (irreducible) vector representation of the
so$_m$ Lie algebra. For $j=1$, there are now two possibilities,
$K\equiv0$ and $K\equiv \pi$ (mod $2\pi$). The dimensions are
$\widehat{D}^{({\rm u})}_{10}=m(m+1)/2-1$ and $\widehat{D}^{({\rm
u})}_{1\pi}=m(m-1)/2$, which are the dimensions of the traceless
symmetric, and the antisymmetric (or adjoint), irreducible
representations of so$_m$, respectively. For $j>1$, the
irreducible representations of $\widehat{\cal B}_m(n)$ are
reducible as representations of so$_m$.

On passing to the JTL$_{2L}(q)$ subalgebra of uJTL$_n(q)$
($n=2L$), for which $j$ must be integer, $K$ and $K+\pi$ become
identified, because only translations $u^2$ are in the algebra.
The corresponding pairs of irreducibles of $\widehat{\cal
B}_m(2L)$ combine into a single irreducible of $\hcA_m(2L)$, and
addition of the dimensions, $\widehat{D}^{({\rm
u})}_{jK}+\widehat{D}^{({\rm u})}_{j,K+\pi}=\widehat{D}_{jK}$,
reproduces eq.\ (\ref{Djkn=0}) (by redefining $r\to r/2$). It
follows that $\widehat{\cal B}_m(2L)$ is a proper subalgebra of
$\hcA_m(2L)$.

Some elements of the commutant $\widehat{\cal B}_m(n)$, can be
found, using the expressions found in the oriented case, together
with the same notation as in the unoriented open chains. For $n$
even, the operators $\widehat{J}^{a_1a_2\ldots a_k}_{b_1b_2\ldots
b_k}$ (for all $k$) still commute with all $e_i$, but do not
commute with $u$. This can be rectified by considering the
combinations $\widehat{G}^{a_1a_2\ldots a_k}_{b_1b_2\ldots
b_k}=\widehat{J}^{a_1a_2\ldots a_k}_{b_1b_2\ldots
b_k}+u\widehat{J}^{a_1a_2\ldots a_k}_{b_1b_2\ldots b_k}u^{-1}$,
which commute with $u$ (using the fact that $u^2$ commutes with
all $\widehat{J}^{a_1a_2\ldots a_k}_{b_1b_2\ldots b_k}$). The
effect of this may be understood as follows. In writing the
operators $J_{ia}^b$ for the unoriented loops models, a choice was
made [which was not forced by the underlying O($m$) symmetry of
the models] that the even sites transform as the fundamental of
gl$_m$, while the odd sites transform as the dual. The
simultaneous translation of all $k$ operators using conjugation by
$u$ reverses this assignment. Thus for example, for $k=1$, it
reduces to $\widehat{G}^a_b=
\widehat{J}^a_b+u\widehat{J}^a_bu^{-1}=G^a_b$, where
$G^a_b=\eta^{ac}\sum_iG_{ibc}$, the generators of global so$_m$,
as one would expect. If we lower indices using $\eta_{ab}$, %
\be%
\widehat{G}_{a_1\ldots a_k,b_1\ldots b_k}=\eta_{b_1c_1}\cdots
\eta_{b_kc_k} \widehat{G}^{c_1\ldots c_k}_{a_1\ldots
a_k}\ee%
and use a multi-index notation $A$ (resp., $B$) to stand for the
list $(a_1,a_2,\ldots,a_k)$ (resp., $(b_1, \ldots, b_k)$), then we
have explicitly,%
\bea%
\widehat{G}_{A,B}&=&\eta_{b_1c_1}\cdots \eta_{b_kc_k}
\widehat{J}^{c_1\ldots c_k}_{a_1\ldots a_k}\nonumber\\&&{}+(-1)^k
\eta_{a_1c_1}\cdots \eta_{a_kc_k} \widehat{J}^{c_1\ldots
c_k}_{b_1\ldots b_k}.\eea%
The operators thus have an additional (anti-) symmetry property%
\be%
\widehat{G}_{B,A}=(-1)^k \widehat{G}_{A,B}.\label{ansym}\ee%
This generalizes the antisymmetry of $G_{ab}$ to all $k$, and
resembles a property of the UEA $U$(so$_m$) of so$_m$, that an
product of generators $G_{a_1b_1}\cdots G_{a_kb_k}$ has the same
(anti-)symmetry. Hence this space of operators is smaller than
that for the oriented loops models.

For $n$ odd, if we try to construct corresponding operators
$\widehat{G}^{a_1a_2\ldots a_k}_{b_1b_2\ldots b_k}$, we notice
that the assignment of sites as alternately even and odd cannot be
performed consistently on the odd chain, though it can be if we
use a double cover. Consequently, the summations over sites (which
are needed to ensure commutation with $e_i$) which must extend
around the chain cyclically, automatically reverse the assignment
of sites as fundamental and dual when an operator $J_{ia}^b$ is
carried around the system. This leads to a modified cyclic
invariance property
\be%
\widehat{G}_{b_ka_1\ldots a_{k-1},a_kb_1\ldots b_{k-1}}
=-\widehat{G}_{a_1\ldots a_k,b_1\ldots b_k}\ee%
for $n$ odd (for $n$ even it has the same form as for the
$\widehat{J}$s in the oriented case). When all $k$ operators are
carried once around the system (by $i\to i+n$), all assignments
are reversed, and this already ensures commutation with $u$. That
is, for each set of $i_1$, \ldots, $i_k$ taking values in $0$,
$1$, \ldots, $n-1$ that are allowed by the cyclic ordering
inequalities as in the expression for $\widetilde{J}$, there are
two terms, instead of one as in $\widetilde{J}$ for $n$ even. Thus
for example for $k=1$, this again produces just $G_{ab}$. For the
resulting operators $\widehat{G}_{A,B}$, the conditions are that
the contractions%
\bea%
\eta^{a_lb_{l+1}}\widehat{G}_{a_1\ldots a_l\ldots a_k,b_1\ldots
b_l\ldots b_k}&=&\eta^{a_{l+1}b_l}\widehat{G}_{a_1\ldots a_l\ldots
a_k,b_1\ldots b_l\ldots b_k}\nonumber\\
&=&0\eea%
for $l=1$, \ldots, $k-1$, but for $l=k$ we have%
\be%
\eta^{a_ka_1}\widehat{G}_{a_1\ldots a_k,b_1\ldots
b_k}=\eta^{b_1b_k}\widehat{G}_{a_1\ldots
a_k,b_1\ldots b_k}=0\ee%
for all $k$, unlike the $n$ even case (this follows from the
others using the modified cyclic invariance). These can be
implemented using projectors of the types $P_\bullet^\bullet$ for
$k$ even, $P_\bullet P^\bullet$ for $k$ odd---the reverse of the
$n$ even case. The (anti-)symmetry property (\ref{ansym}) for all
$k$ holds here also and is another consequence of the modified
cyclic invariance. Note that explicit expressions for all the
operators $\widehat{G}_{A,B}$ for all $n$ and $k$ can be obtained
by using $J$s for the unoriented {\em open} chains, and taking
linear combinations so as to impose the additional conditions
[(modified) cyclic invariance and contractions] listed here.

The odd sector includes the $j=1/2$ representation, the vector
representation of so$_m$, and the algebra $\widehat{\cal B}_m(n)$
acts in this representation as $M_m({\bf C})$, which is not
spanned by the generators $G_{ab}$ (the antisymmetric matrices),
though they do generate it algebraically. They also generate the
full matrix algebra for each of the two $j=1$ irreducible
representations. We expect that the $\widehat{G}_{A,B}$s (ranging
over all $k$) generate $\widehat{\cal B}_m(n)$, similarly to the
oriented case, but here with $k$ even for the even sector, $k$ odd
for the odd sector.


\section{Potts models}
\label{potts}

This short section addresses corresponding results for (the
transfer matrix of) the Potts models, or the Hamiltonian of the
so-called quantum Potts models. It is not needed for the remainder
of this paper.

The preceding results for the open, oriented loops models also
hold for the Potts representation of the TL algebra, for $Q=m^2$
any positive integer. In the Potts representation, each site in
the chain on which the Potts transfer matrix acts corresponds to
two sites of the spin chain, and there are $Q$ states for each
such Potts site. The obvious global symmetry is the permutation
(or symmetric) group $S_Q$, which has only a finite set of
isomorphism classes of irreducibles. The commutant of the (open)
TL algebra in these representations is generally larger and for
$Q\geq 4$ (when the open symmetry algebras are semisimple),
possesses arbitrarily many irreducibles as $L\to\infty$. The
dimensions of the irreducible representations are given by the
same expressions as $D_j$ in terms of $q$, where now
$q^2+q^{-2}+2=Q$. Note that these dimensions are integers whenever
$Q$ is one. When $Q$ is a square, $Q=m^2$, the symmetry algebras
are isomorphic to those in the spin chains above, for $Q\geq4$, as
follows from Refs.\ \cite{bb,affleck3}. (We note that the proof of
the isomorphism of the Potts chain to the $m$, $\overline{m}$
U($m$) spin chain in Refs.\ \cite{bb,affleck3} can be completed
easily if one uses the algebraic perspective as in the present
paper.) For $Q=4$, the symmetry algebra of the open Potts chain is
just $U$(sl$_2$) (projected to the integer spins). For $Q<4$, the
TL algebras are not represented faithfully in the Potts
representation; instead, some semisimple quotient algebra
\cite{jones1} is represented faithfully. For $Q<4$ the symmetry
algebras are smaller, more like the group algebra of $S_Q$.

The closed cases require further study, even in the cases $Q>4$.
There is a loop representation of the Potts partition function
(see e.g.\ Ref.\ \cite{baxter}) on the torus, but the weights have
to be modified according to certain topological properties of the
clusters (shaded regions enclosed by a loop) \cite{dfsz}. It
follows that the algebra generated by the $e_i$s and the
translation $u^2$ in the Potts representation is {\em not}
isomorphic to the JTL algebra \cite{ms1,ms2}. While the $e_i$s and
the translation $u^2$ still obey the periodic version of relations
(\ref{tlrel}) and $u^2e_iu^{-2}=e_{i+2}$, we do not have the
complete set of relations that defines whatever algebra they
generate. Hence we have not established an algebraic analog of the
modified weights in the partition function, but it will change the
boundary condition built into the JTL algebra (and only the
boundary condition, as we know that the subalgebra associated to
any open portion of the chain is a TL algebra). This most likely
changes the symmetry (commutant) algebra. For example, for $Q=4$,
one knows that the CFTs for the continuum limit of the chains with
uniform coefficients, are different in the two cases, the periodic
spin-1/2 chain and the periodic $Q=4$ Potts model, though they are
closely related. In particular, the $Q=4$ periodic Potts model
does not have SU(2) symmetry, even in the continuum limit.


\section{Supersymmetric spin chains}
\label{susych}

This section briefly describes the results for the supersymmetric
generalizations of the spin chains. The detailed constructions are
given in the Appendix, and are essential for Sec.\ \ref{nonsemi},
where the nonsemisimple cases $m<2$ of the open chains are
considered. It is not needed for the following section on the
ribbon Hopf algebra structure of the U$(m)$ chains.

The spin chains for the oriented loops models can be generalized
so that each site carries a ${\bf Z}_2$-graded vector space of
dimensions $m+n$ for the even (bosonic), $n$ for the odd
(fermionic), subspace ($n\geq 0$ is an integer). This space is the
fundamental of the Lie superalgebra gl($m+n|n$) for $i$ even, and
its dual for $i$ odd. The chain is the graded tensor product of
these $V_i$ (it may be constructed \cite{rs} using fermion
operators $f^\dagger_{ia}$, $\overline{f}^{a\dagger}_i$ for
$a=m+n+1$, \ldots, $m+2n$, while $a=1$, \ldots, $m+n$ corresponds
to boson operators as in the $n=0$ special case; details are given
in the Appendix). The (J)TL algebra is again generated by
operators $e_i$ (and $u^2$). These models exist for all integer
$m$, provided $m+n$, $n\geq 0$ \cite{m=0}, and are non-trivial
when $m+2n>1$. The phase transition properties, including scaling
dimensions, are the same independent of $n$, though some
multiplicities may vanish for small $n$. Even though the
finite-dimensional representations of gl($m+n|n$) are not always
semisimple, the representations of TL$_{2L}(q)$ and its commutant
$\cA_{m+n|n}(2L)$ are still semisimple for $|m|\geq 2$, and
similarly for JTL$_{2L}(q)$ and its commutant $\hcA_{m+n|n}(2L)$
for $|m|>2$. (The commutant algebras here are actually
superalgebras when $n>0$; details about graded tensor products and
superalgebras can be found in the Appendix.) The notation
involving $m$, $n$ for these chains will be used consistently from
here on.

For the semisimple cases, the preceding constructions can be
carried through for all $n\geq 0$, with only minor variations. The
dimensions of the irreducible representations of the commutants
can be generalized to the total (usual) dimension, and the
superdimension (sdim) which is the dimension of the even (bosonic)
subspace minus the dimension of the odd (fermionic) subspace. The
{\em super}-dimensions will now be denoted $D_j$ or
$\widehat{D}_{jK}$, as they are determined by $m$ alone (in fact,
by $m^2$), independent of $n$, and are given by the same formulas
as above, which were the $n=0$ special cases. The total dimensions
will be denoted $D_j'$ or $\widehat{D}_{jK}'$, and involve also
$q'$ determined by $m+2n=q'+q'^{-1}$. For the open case, the total
dimensions are given by the same form $D_j'=[2j+1]_{q'}$
\cite{rs}. For the closed case, total dimensions can be obtained
by calculating the trace of the projection operator onto
pseudomomentum $K$ for a fixed pattern. One must be careful of
minus signs that arise when an odd state of a segment of the
non-contractible sites is translated around the system. For
$j>1$, the total dimensions for $|m|>2$ are%
\be%
\widehat{D}_{jK}'=\frac{1}{j}\sum_{r=0}^{j-1}e^{2iKr}w(j,j\wedge
r),\ee where $j\wedge r$ denotes the highest common divisor of $j$
and $r$ ($j\wedge0=j$ for all integers $j\geq0$), and %
\be%
w(j,d)=(q^{2d}+q^{-2d})\delta_{j/d\equiv
0}+(q'^{2d}+q'^{-2d})\delta_{j/d\equiv1},\ee%
where again $d|j$, and the congruences are modulo $2$. For $n=0$
this clearly reduces to eq.\ (\ref{Djkn=0}) \cite{jones2}. The
general case can be simplified and (using formulas from Ref.\
\cite{hw}) shown to be equal to the numbers $\Lambda_{\rm
mod}(M=j,N)$ for these models obtained by a different method in
Ref.\ \cite{rs} (up to a continuation to the different range
$|m|\leq2$ studied there). For $j=0$, $1$, we have $K\equiv0$
only, and $\widehat{D}_{j0}'=1$, $q'^2+1+q'^{-2}$, respectively
[the dimensions of the singlet and the adjoint of sl($m+n|n$)].

The supersymmetric versions of the unoriented loops models use the
defining (vector) representation of osp$(m+2n|2n$) on all the
sites. As for the O($m$) models, these representations are
self-dual, so the models still make sense for arbitrary numbers
$L$ of sites, even with periodic boundary conditions. For the open
chains, the commutant algebra becomes ${\cal B}_{m+2n|2n}(L)$,
which for $L$ even is isomorphic to ${\cal A}_{m+2n|2n}(L)$. The
superdimensions and total dimensions of the irreducibles (for
$|m|\geq 2$) are given by the same formulas as for the oriented
loops models, but with $j=1/2$, $3/2$, \ldots, allowed in addition
to non-negative integers. For the closed chains, the commutant is
$\widehat{\cal B}_{m+2n|2n}(L)$. The total dimensions (for
$|m|>2$) are given by formulas similar to those for the oriented
case (c.f.\ formulas for the super-dimensions $\widehat{D}_{jK}$
above), and again are related to the numbers $\Lambda_{\rm
mod}(M=2j,N)$ for these models that were obtained in Ref.\
\cite{rs}.

These results go far towards explaining the large multiplicities
found in the spectrum of scaling dimensions in the conformal field
theories of the loop models with the closed boundary condition in
Ref.\ \cite{rs}. They are still not a complete explanation because
so far we have analyzed only the semisimple cases $|m|>2$, while
the conformal cases occur for $|m|\leq2$. We comment on this in
Sec.\ \ref{nonsemi} below, but a full analysis for the closed
cases is beyond the scope of this paper.

On the other hand, as S-matrix formulations are based on open
boundary conditions, the unoriented $m=2$ open case with
underlying osp($2+2n|2n$) symmetry does describe the enlarged
symmetry of the theories in Ref.\ \cite{polchinski} at the special
point ($\chi=0$ in the notation there) at which the loops do not
cross. This point is also the end-point of the construction in
Ref.\ \cite{polchinski}, and coincides with the
Kosterlitz-Thouless transition point \cite{rs}.


\section{Tensor products, duals, and braiding}
\label{tens}

In this section we introduce natural operations turning a tensor
product of representations of the commutant algebras into
representations, and likewise for the dual representations, and
also operations of braiding and twist. These turn the commutant
algebras into ``ribbon Hopf algebras'', and we explain the close
relation, termed ``Morita equivalence'', of these to the quantum
group $U_q($sl$_2)$.

At this stage we have obtained semisimple associative algebras
$\cA_{m+n|n}(2L)$ with $|m|\geq 2$ for open chains, and
$\widehat\cA_{m+n|n}(2L)$ with $|m|>2$ for closed chains, and
their $L\to\infty$ limits $\cA_{m+n|n}$ and $\widehat\cA_{m+n|n}$.
(We also have algebras $\cA_m(2L)$ for the open Potts models for
integral $Q=m^2\geq 4$, which behave like the $n=0$ cases; the
following applies to these also, but we will not explicitly refer
to them again.) For finite chains of $2L$ sites, these possess
finite numbers of non-isomorphic irreducible representations.
Specializing until further notice to the open chains, these
representations are labelled simply by $j=0$, $1$, \ldots $L$.
Thus there is a correspondence between these representations for
different values of $n$, and even for different values of $m$. In
particular, the special case $n=0$, $m=2$ (the su$(2)$ spin-1/2
chain) played a special role. Now we must explain what the role of
the spin-1/2 chain is in the general case $m>2$. We focus on
$\cA_m$, as the extension to the superalgebras, though similar and
relatively straightforward, is notationally cumbersome; some of
the details are given in the Appendix and in the following
section.

As semisimple associative algebras, the representation theory of
each $\cA_m(2L)$ is completely characterized by the set of
isomorphism classes of simple modules (labelled by $j$) and their
dimensions $D_j$ (it is only slightly more complicated for the
superalgebra cases $n>0$, in which the modules are graded vector
spaces with a superdimension $D_j$ as well as a total dimension
$D_j'$). [Similarly the TL algebras TL$_{2L}(q)$, for which the
irreducible dimensions $d_j$ are independent of $q$, are hence
isomorphic {\em as associative algebras} for all $q\geq 1$, for
each $2L$.] For $q\neq 1$, a representation of TL$_{2L}(q)$ in a
``spin-1/2'' chain that consists of a tensor factor of ${\bf C}^2$
for each site was discovered by TL \cite{tl}. The Hamiltonian
(\ref{tlham}) is the so-called XXZ spin chain, with certain
boundary terms (see e.g.\ Ref.\ \cite{ps}). For $q\neq1$, it is
not invariant under su($2$). Instead, the commutant of
TL$_{2L}(q)$ in this chain is a finite-dimensional image, call it
$U_q({\rm sl}_2)^{(2L)}$, of the quantum group $U_q({\rm sl}_2)$
\cite{cp,kass}, which for now we view simply as an associative
algebra; this algebra is also referred to as the $q$-Schur algebra
$S_q(2L,2)$ \cite{dj,mathas}. [For generic $q$, \uqsl2 can be
defined by relations among its three generators, which are
$q$-dependent deformations of the commutation relations that
define $U({\rm sl}_2)$, such that for $q\to1$, \uqsl2 becomes
$U({\rm sl}_2)$ \cite{cp,kass}.] Hence $U_q({\rm sl}_2)^{(2L)}$ is
semisimple for $q\geq 1$, and its simple modules, labeled by
$j=0$, $1$, \ldots, $L$, have the same dimensions as for $q=1$,
namely $2j+1$. Then there is a one-one correspondence between
simple modules of $\cA_m(2L)$ and of $U_q({\rm sl}_2)^{(2L)}$ with
the same $j$, induced through their mutual relation to the
corresponding modules of the TL algebras TL$_{2L}(q)$. For these
semisimple algebras, that is the full content of Morita
equivalence. (The equivalence is {\em not} an isomorphism of the
algebras, as for example corresponding simple modules can have
different dimensions.) Now we introduce additional structures that
turn the $L\to\infty$ limits of $\cA_m(2L)$ and of $U_q({\rm
sl}_2)^{(2L)}$ into ``ribbon Hopf algebras''. For these, the
equivalence holds only for algebras with the same value of $m$ or
the corresponding $q$.

Frequently when dealing with symmetry in physics, we find more
structure than just the representations of an associative algebra.
In particular, we may find that a tensor product of two
representations (where the tensor product is just the usual one of
complex vector spaces, or graded tensor product for graded vector
spaces) can also be viewed as a representation of the same
algebra. For an arbitrary associative algebra it is not obvious
how this would be done in a consistent way [i.e.\ so that iterated
products are associative, that is so that the canonical
isomorphism of $({\cal V}_{j_1}\otimes{\cal V}_{j_2})\otimes{\cal
V}_{j_3}$ and ${\cal V}_{j_1}\otimes({\cal V}_{j_2}\otimes{\cal
V}_{j_3})$ as vector spaces is also an isomorphism of
representations of the algebra]. It requires some more structure
in addition to the associative algebra.

The relevant additional structure is a so-called {\em
comultiplication}, which is a linear map from the ``symmetry''
algebra $\cA$ into a tensor product with itself, $\Delta:
\cA\to\cA\otimes\cA$ \cite{cp,kass}. As $\cA\otimes\cA$ can be
viewed naturally as an associative algebra, the map $\Delta$ is
required to be a homomorphism of associative algebras. Then if
${\cal V}_{j_1}$, ${\cal V}_{j_2}$ are two representations of
$\cA$, the vector space ${\cal V}_{j_1}\otimes{\cal V}_{j_2}$, is
in a natural way a representation of $\cA\otimes\cA$, with the two
factors $\cA\otimes\cA$ acting on the respective factors in ${\cal
V}_{j_1}\otimes{\cal V}_{j_2}$. Then using the comultiplication,
${\cal V}_{j_1}\otimes{\cal V}_{j_2}$ becomes a representation of
$\cA$, with the element $a\in \cA$ acting by
$\Delta(a)\in\cA\otimes\cA$. In order to guarantee associativity
of the tensor product, we require {\em co-associativity} of the
co-multiplication. If the image of an arbitrary element $a$ of
$\cA$ is written in Sweedler's notation
\cite{kass} as%
\be%
\Delta(a)=\sum_{(a)} a'\otimes a'',\ee %
that is as a linear combination of elements of $\cA\otimes\cA$,
then co-associativity is the requirement that, for all $a$,%
\be%
\sum_{(a)} a'\otimes \Delta(a'')=\sum_{(a)} \Delta(a')\otimes a''\ee%
(both sides are elements of the algebra $\cA\otimes\cA\otimes
\cA$).

Familiar examples in physics include groups and Lie algebras (or
more accurately, the corresponding associative algebras, the group
algebras and UEAs, respectively). For the case of the su($2$) Lie
algebra, the comultiplication is the familiar notion of addition
of angular momentum, which means that the generators $J_a$ ($a=x$,
$y$, $z$) of su$(2)$ acting on a tensor
product are to be obtained as %
\be%
\Delta(J_a)=J_a\otimes 1 + 1\otimes J_a,\ee%
which is easily seen to be co-associative. With the introduction
also of a counit map, which is used to make the one-dimensional
vector space $\bf C$ into a representation of $\cA$, and an
antipode, which turns the dual vector space ${\cal V}^*$ of ${\cal
V}$ into a representation of $\cal A$, together with some further
compatibility requirements, one arrives at the definition of a
Hopf algebra \cite{cp,kass}. The notion of a symmetry as described
by a Hopf algebra embodies most of the usual features of
symmetries that arise in physics. Quantum groups are examples of
Hopf algebras that are not equivalent to groups or Lie algebras.
The quantum group \uqsl2 is a deformation of the UEA $U$(sl$_2$)
of the Lie algebra sl$_2$ \cite{cp,kass}.

Turning to our algebras ${\cal A}_m$, for the semisimple cases
$m\geq 2$, a comultiplication can be defined, by first defining
for
$\widetilde{J}$s:%
\be%
\widetilde{\Delta}(\widetilde{J}^{a_1a_2\ldots a_k}_{b_1b_2\ldots
b_k})= \sum_{k_1=0,1,\ldots,k}\widetilde{J}^{a_1a_2\ldots
a_{k_1}}_{b_1b_2\ldots b_{k_1}}
\otimes\widetilde{J}^{a_{k_1+1}\ldots
a_k}_{b_{k_1+1}\ldots b_k}.\ee%
Then the comultiplication is%
\be%
\Delta(J^{a_1a_2\ldots a_k}_{b_1b_2\ldots b_k})=(P^\bullet
P_\bullet\widetilde{\Delta}(\widetilde{J}))^{a_1a_2\ldots
a_k}_{b_1b_2\ldots b_k},\ee%
where the projectors $P^\bullet$, $P_\bullet$ act in the space
spanned by the indices as before. This is exactly what happens to
a $J$ acting on a chain of length $2L=2L_1+2L_2$, if the chain is
broken into two pieces of lengths $2L_1$, $2L_2$, by dropping the
generator $e_{2L_1-1}$. It is clear that $\Delta(\cA)$ lies in
$\cA_m\otimes\cA_m$ as required. It can be interpreted as saying
that two chains of lengths $2L_1$, $2L_2$ can be joined end to
end, and their states interpreted as representations of $\cA_m$ in
a natural way. Indeed, this is the algebraic origin of the
recursion relation eq.\ (\ref{recrel}) for the dimensions of the
irreducibles. Coassociativity of $\Delta$ is easily checked, using
simple properties of the projection operators $P_\bullet$ and
$P^\bullet$ under inclusions of chains into chains of longer
length. Using arguments similar to those in the recursion relation
for the dimensions, it is then not difficult to see that if ${\cal
V}_{j_1}$ and ${\cal V}_{j_2}$ are irreducible representations,
the tensor product decomposes as %
\be%
{\cal V}_{j_1}\otimes{\cal
V}_{j_2}\cong\bigoplus_{j=|j_1-j_2|}^{j_1+j_2}{\cal V}_j,
\label{fusrul}\ee%
and ${\cal V}_{j_2}\otimes{\cal V}_{j_1}$ decomposes the same way.
Notice that these ``fusion rules'' (i.e.\ the multiplicities with
which each simple module appears in a tensor product of two given
simple modules) are the same as for su(2), and that the total
dimensions of the two sides are equal. The latter is also true for
the superdimensions in the superalgebra $\cA_{m+n|n}$ ($n\geq 0$)
case.

Similarly, the antipode (which is a linear map, but reverses the
order of products in $\cA_m$) is defined by %
\be S(J^{a_1a_2\ldots a_k}_{b_1b_2\ldots b_k})=(-1)^k J^{a_k\ldots
a_2 a_1}_{b_k\ldots b_2 b_1},\ee %
and is invertible, $S^{-1}=S$. We note that, on the vector space
$V$, which is a tensor product $V=V_0\otimes V_2\otimes\cdots
V_{2L-1}$ (where the order in the tensor product denotes order in
the chain), the dual is naturally order-reversing \cite{kass},
$V^\ast=V_{2L-1}^\ast\otimes\cdots V_0^\ast$, which (because the
fundamental and its dual alternate in $V$ along the chain) is
another chain of the same type, and the same applies to the states
on the non-contractible sites in the $j$th representation. The
dual pairing on each of $V\otimes V^\ast$ and $V^\ast\otimes V$
(and hence those on ${\cal V}\otimes {\cal V}^\ast$ and ${\cal
V}^\ast\otimes {\cal V}$, where $\cal V$ is any representation of
$\cA_m$), that is the map from the tensor product to $\bf C$ that
commutes with the action of $\cA_m$, is then given by the
canonical (vector space) dual pairing between the corresponding
tensor factors $V_i$ in $V$ and and $V_i^\ast$ in $V^\ast$. The
double dual ${\cal V}^{\ast\ast}$ can be identified with $\cal V$
via the canonical vector-space isomorphism (equivalently, the
``left'' and ``right'' duals of $\cal V$ are the same, because
$S^{-1}=S$). More generally, for the superalgebras $\cA_{m+n|n}$,
the antipode is given below in eq.\ (\ref{antip}). In these cases
($n>0$), ${\cal V}^{\ast\ast}$ is not canonically isomorphic to
${\cal V}$, even though $S^{-1}=S$ (see the Appendix), but is
still isomorphic.

The comultiplication and antipode on $\cA_m$ agree with those on
$U({\rm gl}_m)$, an image of which is a subalgebra of $\cA_m$, so
that our Hopf algebra is an extension (of an image) of the Hopf
algebra $U({\rm gl}_m)$. This is not true for the additional
structures to which we turn next.

A braiding structure can be defined on the representations of
$\cA_m$. First, one can obtain a representation of the braid group
on the spaces $V$. The braid group can be defined by the
generators $\sigma_i$ ($i=0$, $1$, \ldots, $2L-2$) and their
inverses $\sigma_i^{-1}$ subject to the relations
$\sigma_i\sigma_{i+1}\sigma_i=\sigma_{i+1}\sigma_i\sigma_{i+1}$
for $i=0$, $1$, $2L-3$, and
$\sigma_{i}\sigma_{i'}=\sigma_{i'}\sigma_i$ for $i'\neq i\pm 1$.
It can be represented in the TL algebra by writing
$\sigma_i=i(q^{1/2}-q^{-1/2} e_i)$,
$\sigma_i^{-1}=-i(q^{-1/2}-q^{1/2}e_i)$ \cite{jones3,cfs}, where
the factor of $i$ in this and similar formulas below is a square
root of $-1$ (note that since the braid-group relations are
homogeneous, other representations can be obtained by multiplying
all $\sigma_i$ by a constant; the reason for our particular choice
is explained below). Hence there is an action of the braid group
on $V$, which commutes with $\cA_m(2L)$ (for $q\to-1$, the
$\sigma_i$ reduce to generators of the symmetric group, which obey
also $\sigma_i^2=1$ for all $i$; we explain below why it is $q\to
-1$ here, not $+1$). For $m>2$, this representation of the braid
group is not unitary for any choice of the common constant factor
in the $\sigma_i$ \cite{jones3}.

Next we construct an element
$\sigma_{2L_1,2L_2}$ in the braid group on $2L=2(L_1+L_2)$ sites as%
\be
\sigma_{2L_1,2L_2}=\stackrel{\rightarrow}{\prod_{i=0,1,\ldots,2L_1-1}}
\left[\stackrel{\leftarrow}{\prod_{i'=0,1,\ldots,2L_2-1}}
\sigma_{i+i'}\right],\ee %
where the ordered product of operators $x_i$ is defined by
$\stackrel{\rightarrow}{\prod}_{i=0,1,\ldots,n}x_i=x_0x_1\cdots
x_n$, and the reverse order in
$\stackrel{\leftarrow}{\prod}_{i=0,1,\ldots,n}x_i$. Then
$\sigma_{2L_1,2L_2}$ obeys
$\sigma_{2L_1,2L_2}\Delta(a)=\Delta(a)\sigma_{2L_1,2L_2}$ for any
$a\in \cA_m$, when acting on this chain. $\sigma_{2L_1,2L_2}$ can
be viewed as exchanging the two parts of a chain of length
$2L=2(L_1+L_2)$. From this one can obtain isomorphisms (which
commute with $\cA_m$) $\sigma_{{\cal U},{\cal V}}:{\cal
U}\otimes{\cal V}\to{\cal V}\otimes{\cal U}$ for each pair of
representations ${\cal U}$, ${\cal V}$. With the normalization of
the $\sigma_i$ specified above, if $\cal U$ or $\cal V$ is the
one-dimensional representation, then $\sigma_{{\cal U},{\cal V}}$
is the identity, and it follows that all these isomorphisms
$\sigma_{{\cal U},{\cal V}}$ are independent of the length of the
chain in which $\cal U$ or $\cal V$ appear. These isomorphisms
satisfy compatibility requirements, including some related to the
braid group relations. Note that $\sigma_{{\cal V},{\cal
U}}\sigma_{{\cal U},{\cal V}}$ is not the identity map when $q>1$.
$\sigma_{2L_1,2L_2}$ maps TL$_{2L_1}(q)\otimes$TL$_{2L_2}(q)$ to
TL$_{2L_2}(q)\otimes$TL$_{2L_1}(q)$ (acting on these algebras,
which are understood to be subalgebras of TL$_{2L}(q)$ in two
obvious ways, by conjugation). If we define the flip map $\tau$ on
the space for the two chains of total length $2L$ by
$\tau(v_1\otimes v_2)=v_2\otimes v_1$, then we find that
$\widetilde{\cal R}_{2L_1,2L_2}= \tau\circ\sigma_{2L_1,2L_2}$
commutes with the action of TL$_{2L_1}(q)\otimes$TL$_{2L_2}(q)$,
and hence is an element of $\cA_m(2L_1)\otimes\cA_m(2L_2)$. As
$L_1$, $L_2\to\infty$, $\widetilde{\cal R}_{2L_1,2L_2}$ becomes a
``universal R-matrix'' ${\cal R}\in \cA_m\otimes\cA_m$, making the
Hopf algebra into a braided Hopf algebra \cite{cp,kass}.

Similarly, we can define a ``twist'' map by%
\be%
\widetilde{\theta}_{2L}=(iq^{3/2})^{2L}\left(\stackrel{\leftarrow}
{\prod_{i=0,1,\ldots,2L-2}}
\sigma_i\right)^{2L}.\ee%
This commutes with TL$_{2L}(q)$, and hence lies in $\cA_m(2L)$ as
well as in TL$_{2L}(q)$, so it is in the center of both algebras,
and its $L\to\infty$ limit $\theta$ lies in the center of $\cA_m$.
It gives rise to natural isomorphisms $\theta_{\cal V}$ on each
module $\cal V$ that obey $\theta_{{\cal U}\otimes{\cal
V}}=(\theta_{\cal U}\otimes\theta_{\cal V})\sigma_{{\cal V},{\cal
U}}\sigma_{{\cal U},{\cal V}}$. The prefactor $(iq^{3/2})^{2L}$ in
$\widetilde{\theta}_{2L}$ represents a factor $iq^{3/2}$ for the
twist on the fundamental (or its dual) representation $V_i$ for
all $i$ [which correspond to the spin-1/2 module for $U_q({\rm
sl}_2)$], and ensures that $\theta$ acts as unity on the
one-dimensional module. $\theta$ evaluated on the $j$th simple
module of $\cA$ is $\theta_j=q^{2j(j+1)}$ (for $j\geq0$ an
integer), which is related to the Casimir $j(j+1)$ for simple
modules of sl$_2$ ($\theta_j$ is the same as for $U_q({\rm sl}_2)$
\cite{bk}). The twist maps on modules can alternatively be
constructed using the braiding and other maps already mentioned,
including the canonical isomorphism between the left and right
duals of $\cal V$ that exists because $S^{-1}=S$. With the twist
map included, the braided Hopf algebra becomes a ``ribbon'' Hopf
algebra \cite{cp,kass,bk} (our definition of $\theta$ would be
called $\theta^{-1}$ in these references). This structure allows
the definition of a ``quantum trace'' of any linear map on a
finite-dimensional representation $\cal V$, and in particular a
``quantum dimension'' \cite{cp,kass,bk,turaev}. In the case of
$\cA_m$, the quantum trace and dimension turn out to be simply the
ordinary vector-space trace and dimension. For the superalgebras,
the quantum trace and dimension turn out to be the supertrace and
superdimension.

Now we describe Morita equivalence more formally. Morita
equivalence of associative algebras means that their ``categories
of modules'' are equivalent (as categories) \cite{anf}. The
category of modules of an algebra $\cA$ consists of all the
modules of $\cA$, together with all the morphisms (or
homomorphisms---linear maps commuting with the action of $\cA$)
between them. Morphisms can be composed to form other morphisms.
Equivalence of these categories for two algebras $\cA$, ${\cal
B}$, means that there is mapping (a covariant functor) ${\cal F}$
from the modules of $\cA$ to those of $\cal B$, and from morphisms
of $\cA$ to those of $\cal B$, which is compatible with the
composition of morphisms. There is also a functor $\cal G$ in the
reverse direction, and the composite of these functors, in either
order, is a functor from one of the categories into itself, which
is required to be ``naturally isomorphic'' to the identity
functor. ``Naturally isomorphic'' to the identity means that for
each module of $\cA$ there is an isomorphism between it and its
image under the composite functor ${\cal G}{\cal F}$, and these
are compatible with the morphisms and their images under ${\cal
G}{\cal F}$ also. Morita showed that such an equivalence could
always be described in a certain canonical form involving a
bimodule over $\cA$ and $\cal B$, that is a module over $\cA$ and
$\cal B$ simultaneously, such that the actions on it of $\cA$ and
$\cal B$ commute \cite{anf}. For us, the module $V$ is a suitable
choice for this bimodule which yields an equivalence between
$\cA_m(2L)$ and TL$_{2L}(q)$, and likewise the spin-1/2 chain of
$2L$ sites yields the equivalence between $U_q({\rm sl}_2)^{(2L)}$
and TL$_{2L}(q)$. Under this equivalence, modules we labeled with
the same value of $j$ correspond. The composite of these
equivalences yields the desired Morita equivalence. For semisimple
algebras, a Morita equivalence is completely determined by a
correspondence of simple modules over the respective algebras.
This means that in fact as associative algebras, $\cA_m(2L)$ and
$U_q({\rm sl}_2)^{(2L)}$ for {\em all} $m\geq 2$ and {\em all}
$q\geq1$ are all Morita equivalent; however, this might not hold
when the additional structures of tensor product, duality,
braiding, and twist have been introduced. Moreover, for these
semisimple associative algebras, there are $(L+1)!$ distinct
Morita equivalences between each pair; the one under which the
modules with the same value of $j$ correspond is the only natural
choice for our purposes below.

When the tensor product is introduced, the category of modules
over $\cA_m$ becomes a ``tensor'' (or ``monoidal'') category. With
the duality of modules determined by the antipode, the tensor
category becomes a ``rigid tensor category''. The Morita
equivalence (under which modules with the same $j$ correspond)
respects the fusion rules and duality, so that the algebras
$\cA_m$ and \uqsl2 still appear to correspond for all $m\geq2$ and
all $q\geq1$, and indeed are still Morita equivalent (equivalence
of tensor categories is discussed in Refs.\ \cite{cp,kass,bk}).
With the introduction also of the braiding related to the
R-matrix, the rigid tensor category becomes a ``rigid braided
tensor category''. With the introduction of the twist, the rigid
braided tensor category becomes a ``ribbon category''
\cite{kass,bk,turaev}. Our statement in its most refined form is
then that {\em for each value of $|m|$, $|m|\geq 2$, all the
ribbon Hopf algebras $\cA_{m+n|n}$ for $n\geq0$, and \uqsl2 for
$m=q+q^{-1}$ are Morita equivalent, in the sense that their
categories of finite-dimensional modules are equivalent ribbon
categories\/}. [Here we mean the \uqsl2 modules that decompose
into integer spins only.] The equivalence given by the use of a
suitable bimodule maps all the ribbon category structures to each
other functorially. This implies numerical relationships between
the structures. For example, the quantum dimensions defined for
corresponding modules in each category should be the same. This
explains why the dimensions $D_j=[2j+1]_q$ of the simple $\cA_m$
modules (and the superdimensions of the simple $\cA_{m+n|n}$
modules for $n>0$), which are the quantum dimensions for $\cA_m$
(resp., $\cA_{m+n|n}$), are equal to the well-known quantum
dimensions of the corresponding simple \uqsl2 modules. This also
shows that the categories for distinct values of $|m|$ are not
equivalent as ribbon categories, because the values of these
numbers differ.

In viewing the spin-1/2 chain as a representation of TL$_{2L}(q)$
and of $U_q({\rm sl}_2)$, it is best to think of it too as
alternating the fundamental (spin-1/2) with its dual, and the TL
generators $e_i$ as constructed using the $U_q({\rm
sl}_2)$-invariant dual pairing of nearest neighbors. However, for
the unoriented O($m$) and osp($m+2n|2n$) models, with $m+2n$,
$n\geq 0$, (and again $|m|\geq 2$ in this paragraph) one is forced
to use an isomorphism ($\phi$, say) of the fundamental to its
dual, for example when concatenating two chains when the one on
the left has an odd number of sites. Then we must do the same for
the spin-1/2 chain also. Now many representations of the UEAs of
certain Lie algebras such as sl$_2$, so$_m$ and sp$_{2m}$ (and we
note that sl$_2\cong$ so$_3\cong$ sp$_2$) possess isomorphisms
$\phi_{\cal V}:{\cal V}^*\to{\cal V}$ to the representation from
its dual. This map can be iterated, $\phi_{{\cal
V}^*}\circ\phi_{\cal V}:V^{**}\to{\cal V}$, and the latter map is
either $+1$ or $-1$ times the canonical vector space map ${\cal
V}^{**}\to{\cal V}$ on irreducible representations $\cal V$; these
are then called real or pseudoreal, respectively. For so$_m$, all
tensor (i.e.\ non-spinor) representations are real, while for
sl$_2$, half-odd-integer spins are pseudoreal. In our O($m$)
chains [and hence also in the osp($m+2n|2n$) chains], all
representations are likewise real in this sense. The definition of
real and pseudoreal can be turned into an intrinsic property,
called the Frobenius-Schur indicator $\nu_{\cal V}$, of a ribbon
category that possesses maps $\phi_{\cal V}$. It is equal to $\pm
1$ on simple modules (or zero when no isomorphism $\phi_{\cal V}$
exists). It turns out to be $-1$ for half-integer spins for the
standard ribbon category structure on $U_q$(sl$_2$), as one would
expect from the above discussion of Lie algebras, but is $+1$ for
all simple modules of ${\cal B}_m$. This is then a discrepancy
with which we must deal.

We note that the twist can be redefined without changing the
braiding, though the map between right and left duals also
changes. For $U_{q''}$(sl$_2$) (the relevant value of $q''$ will
be determined below) and ${\cal B}_m$, the category of modules is
generated by iterating fusion with the $j=1/2$ module. The
properties of $\theta_j$ show that we can multiply $\theta_{1/2}$
by a constant and that determines the change for all other
modules; then the fact that $\theta_0=1$, together with the fusion
rules, implies that the only possible change is by multiplication
by $(-1)^{2j}$. Thus there are just two possible distinct twists
for given duality, braiding maps, etc. As explained in Ref.\
\cite{bk35}, the essential use of the map $\phi_{{\cal V}_{1/2}}$
implies that the variant twist is obtained, not the standard one
for $U_{q''}$(sl$_2$). For this definition, the Frobenius-Schur
indicator is $+1$ for all modules (this difference in connection
with TL algebras was noted in Ref.\ \cite{lw}), and thus agrees
with that for our algebras ${\cal B}_{m+2n|2n}$. The twist for a
simple module of ${\cal B}_{m+2n|2n}$ ($\widetilde{\theta}_{2L}$
is defined as above, but now we allow $2L$ to be odd) works out as
$\theta_j=i^{4j^2}q^{2j(j+1)}$, which is $q^{2j(j+1)}$ for $2j$
even, $iq^{2j(j+1)}$ for $2j$ odd. The quantum dimension of the
$j=1/2$ module of $U_{q''}$(sl$_2$) works out as $-q''-q''^{-1}$
because of the variant twist, so we must put $q''=-q$. Hence we
should compare with $U_{-q}$(sl$_2$) with the variant twist (which
can be obtained by applying the above construction of
$\widetilde{\theta}_{2L}$ in the spin-1/2 chain), and for this the
values on the simple modules are
$\theta_j=(-1)^{2j}(-q)^{2j(j+1)}=i^{4j^2}q^{2j(j+1)}$ [using
$(-q)^{1/2}=iq^{1/2}$]. This explains the factor $i$ in the
braiding $\sigma_i$ of $j=1/2$ with itself also; the $\sigma_i$s
in the $U_{-q}$(sl$_2$) spin-1/2 chain, when defined as above,
define the standard braiding which reduces to the usual exchange
(the flip $\tau$) when $-q=1$. We conclude that for the unoriented
loops models, ${\cal B}_{m+2n|2n}$ is Morita equivalent as a
ribbon Hopf algebra with $U_{-q}$(sl$_2$) with the {\em variant}
ribbon structure, where as usual $m=q+q^{-1}$. (For the oriented
loops models which involve only integer $j$, this reduces to the
same as before.)

Returning to the algebras $\cA_m$ (and $\cA_{m+n|n}$), we now
consider a further numerical relationship that follows from these
results. The Clebsch-Gordan coefficients for $\cA_m$, which
describe the explicit decomposition of a tensor product of simple
modules into simple modules, as in eq.\ (\ref{fusrul}), are
determined by the comultiplication and not only by the fusion
rules. They cannot be expected to correspond for the different
algebras as they depend on a basis for each representation, and
corresponding representations of $\cA_m$ for different $m>2$ and
those of \uqsl2 do not even have the same dimensions. However,
$6j$ symbols [or recoupling coefficients; coefficients that relate
alternative ways of constructing a tensor product of three
representations, again analogous to those familiar for su($2$)]
can be constructed as suitable sums (over the bases) of products
of Clesch-Gordan coefficients, and are independent of a basis. An
abstract construction of $6j$ symbols, using only the structure of
a ribbon category, is given in Ref.\ \cite{turaev}. Using that
formulation, we can show that the $6j$ symbols for $\cA_m$ are the
same as the quantum $6j$ symbols \cite{cfs} that apply for \uqsl2
at the corresponding value of $q$.




\section{Non-semisimple cases}
\label{nonsemi}

In this section, we turn to the non-semisimple cases $|m|<2$
($|m|\leq 2$) of the open (resp., closed) chains in the spaces
$V$. The cases $-2<m\leq 2$ are of particular interest as in these
the Hamiltonian in eq.\ (\ref{tlham}) with $\epsilon=1$ is at a
critical (second-order phase transition) point, and a continuum
limit can be taken that produces a conformal field theory (CFT)
\cite{glr,rs} (the case $m=2$, $n=0$ is the well-known spin-1/2
chain; all cases are non-trivial only if $m+2n>1$, so the
remainder require the use of the supersymmetric models). We will
again concentrate on the open oriented-loops models. In the
non-semisimple cases, all earlier results and formulas must be
reconsidered. We want to know whether the Morita equivalences of
the commutants as ribbon Hopf algebras that hold in the semisimple
cases also hold in the non-semisimple ones. We begin by showing
that the commutant is a cellular algebra, which produces insight
into the structure of the algebra, including a formula for its
dimension. Next we establish the Morita equivalence for the
finite-dimensional associative algebras in the finite-length
chains. Then we discuss carefully the $L\to\infty$ limits of the
TL algebra, its commutant, and of the categories of modules over
these algebras. Then we introduce the structures that turn the
$L\to\infty$ commutant algebras $\cA$ into ribbon Hopf algebras
and establish their Morita equivalence as such with $U_q($sl$_2)$.
In particular, this involves the interpretation of a product of
modules as a module over the commutant, and there is a
corresponding product for modules of the TL algebras, and both of
these behave stably in the $L\to\infty$ limit. Finally, we discuss
the closed chains, though not in as much detail.

The mathematical background required in this section is larger
than in the earlier sections. Accordingly, it will be appropriate
to refer to (simple) modules rather than (irreducible)
representations.

\subsection{Preliminary remarks}

The TL algebra acts faithfully in the spin-1/2 chain
representation discovered by TL, for all $q$. The delicate cases
occur when $q$ is a root of unity (but $q\neq \pm 1$), so $q^r=\pm
1$ for some integer $r>1$. With $m=q+q^{-1}$ and $m$ integer, $q$
is a root of unity for $m=-1$, $0$, $1$. In these cases, the TL
algebra acts faithfully in our spaces $V$ provided $m+2n>1$; it is
then not obvious that the TL algebra or its commutant are
semisimple, as we are not aware of a faithful representation on a
positive-definite space of states. In fact, it is not semisimple,
and neither is \uqsl2 in these cases. This means that there are
reducible modules that are not fully decomposable. Such a module
contains a proper submodule, but is not isomorphic to a direct sum
of simple modules. Such modules can however be decomposed into a
direct sum of indecomposable components; an indecomposable module
is one that cannot be decomposed further as a direct sum of its
submodules. Frequently, the decomposition into indecomposables is
unique up to isomorphism.

Here we will address the structure of these chains only at an
abstract level, leaving the details of the structure of the
modules for the companion paper \cite{rs3} (for TL and
$U_q$(sl$_2$), these are available in the literature
\cite{martin,ps,cp}). These details are used in only a few places
in these arguments. We will need to be more aware of the
difference between right and left modules than we have been up to
now. That is, if $v$ is an element of a right module $V$ over a
ring $R$, then an element $a$ of $R$ acts on it as $va$, and this
action obeys $v1=v$, $v(ab)=(va)b$. We note that any right module
over an algebra $R$ can be viewed as a left module over the
opposite algebra $R^{\rm op}$ (the opposite algebra is defined in
Appendix A). For TL$_{2L}(q)$, the opposite algebra is isomorphic
to TL$_{2L}(q)$ itself, because the defining relations
(\ref{tlrel}) are invariant if the order in all products of
elements is reversed (equivalently, there is an anti-isomorphism
of TL with itself). Hence there is a (Morita) equivalence between
the categories of right and of left modules over TL. (Indeed there
is more than one, as the isomorphism with the opposite algebra can
be composed with the automorphism of the TL algebra given by
$e_i\to e_{2L-2-i}$ for all $i$.) In particular, this applies to
the important class of modules known as {\em projective modules}
(a module is projective if and only if there is a free module in
which it is a direct summand \cite{pierce,anf,project}). Further,
as the TL algebra is a finitely-generated algebra over the field
of complex numbers $\bf C$, there is a Morita duality between its
categories of finitely-generated right and left modules
\cite{dual}; this duality is similar to Morita equivalence, except
that the functors are contravariant instead of covariant---they
reverse the direction of morphisms. In the present case, the
duality is simply obtained: the dual of any left module, viewed as
a vector space, is the dual vector space, on which the TL algebra
naturally acts on the right, and similarly for the dual of a right
module. In view of the equivalence between the categories of right
and left modules, this becomes a duality of the category of
finitely-generated left modules into itself. (This is similar to
the use of the antipode in a Hopf algebra to view the dual of a
left module as another left module.) This duality will be used
occasionally in the following. We must note that the dual of a
projective left module is not necessarily a projective right
module, and vice versa.

\subsection{Cellular structure}

It will be useful here to derive the cellular structure of the
commutant algebras $\cA$, which will give important insight into
their structure [or that of $U_q$(sl$_2)^{(2L)}$]. The derivation
of Morita equivalence with $U_q$(sl$_2)^{(2L)}$ in the next
subsection then allows the structure of the commutants to be
obtained by combining our analysis with the known structure of
$U_q$(sl$_2)^{(2L)}$ (we go into further detail about this
structure in the companion paper \cite{rs3}).

We will explain the notion of a cellular algebra \cite{gl}, using
the alternative basis-free definition of K\"onig and Xi
\cite{kx,mathas}. We will use the TL algebra as an example. To
show that an algebra $T$ [such as $T ={\rm TL}_{2L}(q)$] is
cellular, it is sufficient \cite{kx} to show the following: there
is a nested chain of ideals
$T^{(2j)}$,%
\be%
0=T^{(-2)}\subset T^{(0)}\subset T^{(2)}\subset \cdots \subset
T^{(2L-2)}\subset
T^{(2L)}=T\ee%
and an anti-involution $^s$ of $T$ to itself (an anti-involution
is an anti-isomorphism that squares to the identity isomorphism),
such that the quotients $T^{(2j)}/T^{(2j-2)}$, which are ideals of
$T^{(2L)}/T^{(2j-2)}$, have the following properties (we call the
anti-involution $^s$ rather than $^*$ as in Ref.\ \cite{gl} to
avoid possible confusion with the use of $^*$ for duals): (i)
$(T^{(2j)}/T^{(2j-2)})^s=(T^{(2j)}/T^{(2j-2)})$ for $j=0$, \ldots,
$L$; (ii) for each $j$, there is a left ideal ${\cal R}_j$ of
$T/T^{(2j-2)}$ such that $T^{(2j)}/T^{(2j-2)}$ is isomorphic to
${\cal R}_j\otimes {\cal R}_j^s$ as a
$T/T^{(2j-2)}$-$T/T^{(2j-2)}$ left-right bimodule, compatibly with
$^s$ (such an ideal is called a cell ideal, while the modules
${\cal R}_j$ for each $j$ are called cell or standard modules). In
the last formula, ${\cal R}_j^s$ is the right module with action
of $t\in T^{(2L)}/T^{(2j-2)}$ defined by $vt=t^sv$, where $v$ is
an element of ${\cal R}_j$ viewed as a vector space. ${\cal R}_j$
(resp., ${\cal R}_j^s$) can be viewed as a left (resp., right)
module of $T$ also, which is annihilated by $T^{(2j-2)}$. For TL,
the ideals $T^{(2j)}$ ($j\geq0$) can be described in the usual
diagrammatic picture as spanned by the elements that correspond to
diagrams with at most $2j$ lines connecting the top to the bottom
rows of dots; these are easily seen to form ideals, as the number
of lines is non-increasing under multiplication of diagrams. The
anti-involution can be taken as {\em either} of the two described
above; we construct the more appropriate choice below. Before
continuing, we point out that the cell (or standard) modules
${\cal R}_j$ of TL are also called Specht modules. They have
dimension $d_j$, and are a generalization of those we constructed
in the semisimple case using ``valid patterns''. For $q$ a root of
unity they are indecomposable, but not in general simple modules.

Now we derive the cellular structure of the commutant. As this is
a superalgebra, and as the cellular structure refers to the
opposite (super-)algebra, we actually need a slight generalization
to {\em cellular superalgebras}. This uses some definitions given
in the Appendix. First, we will obtain the nested chain of ideals
in $\cA=\cA_{m+n|n}(2L)$:%
\be%
0\subset \cA^{(2L)}\subset \cA^{(2L-2)}\subset \cdots \subset
\cA^{(0)}=\cA,\ee%
with quotients $J^{(k)}=\cA^{(k)}/\cA^{(k+2)}$ which are ideals of
$\cA/\cA^{(k+2)}$ (for $k$ even; the notation is consistent with
earlier usage of $k$). These ideals $\cA^{(k+2)}$ for $k=0$,
\ldots, $2L-2$ are defined as consisting of endomorphisms of $V$
as a TL-module (i.e., linear maps to itself that commute with the
TL algebra) that annihilate all elements of $T^{(k)}$ (from either
side, as they commute). That is, $a\in \cA$ is a member of
$\cA^{(k+2)}$ if $at=0$ for all $t$ in $T^{(k)}$. These spaces are
easily seen to form ideals in $\cA$, and to be nested as required.
Explicit relations that define elements in $\cA^{(k)}$ are given
in the Appendix. The elements constructed there for each $k$ are
representatives of the cosets forming the quotient spaces
$J^{(k)}$, and by abuse of notation we tend to call the spaces of
these elements $J^{(k)}$ also. We will refer to basis elements in
these spaces loosely as $J^{a_1\cdots a_{2L}}_{b_1\cdots b_{2L}}$,
as they correspond to those operators in earlier sections, however
caveats about this form of notation are discussed in the Appendix.
The construction shows that they lie in $\cA^{(k)}$, but we have
yet to show that they span it; we turn to this next.

First we consider $\cA^{(2L)}=J^{(2L)}$, which annihilates the
ideal $T^{(2L-2)}$. Note that $T^{(2L)}/T^{(2L-2)}\cong {\bf C}$
is one dimensional, and consists of cosets of $1$ (the identity)
only, while $T^{(2L-2)}/T^{(2L-4)}$ is spanned by the cosets of
all the $e_i$s. Hence $\cA^{(2L)}$ consists of elements of $\cA$
that annihilate the $e_i$s. This is exactly our set $J^{a_1\cdots
a_{2L}}_{b_1\cdots b_{2L}}$, because as $k=2L$ here, these consist
of a $J^a_b$ for each site of the chain, and the definitions
coincide. This space $J^{(2L)}$ has dimension $(D_{L}')^2$, and
superdimension $(D_{L})^2$, which are given by the same
expressions as before, but now for arbitrary $m$.

Now we consider the next case, $k=2L-2$. We require elements of
$\cA$ that annihilate all elements of TL with $2L-4$
through-lines. A subspace of this space $T^{(2L-4)}$ is spanned by
diagrams consisting of $e_{2L-2}$ (involving the last two sites),
and any diagram from the ideal $T^{(2L-4)}$ for the algebra
TL$_{2L-2}(q)$ on the first $2L-2$ sites. The commutant of this
subspace can be analyzed in the tensor product space of $V(2L-2)$
(for the first $2L-2$ sites) and $V(2)$ (for the last two). The
commutant of $e_{2L-2}$ in the space $V(2)$ for the last two sites
consists only of the identity, modulo operators that annihilate
$e_{2L-2}$ (the proof of this works differently for the cases
$m\neq0$ and $m=0$). Then we can tensor this with elements of
$\cA^{(2L-2)}(2L-2)$ for the first $2L-2$ sites, and these are the
span of $J^{a_1\cdots a_{2L-2}}_{b_1\cdots b_{2L-2}}$ that we have
from the previous step, as applied to the length $2L-2$ chain. The
ideal $\cA^{(2L-2)}$ that we seek is a subspace of this space of
operators. But in fact, we have such a space of operators that
commute with the full TL$_{2L}(q)$, and which reduce to this space
modulo elements that annihilate $e_{2L-2}$ (and hence are linearly
independent); these are just the $J^{a_1\cdots
a_{2L-2}}_{b_1\cdots b_{2L-2}}$ for the length $2L$ chain. Hence,
using induction, we conclude that each $J^{(k)}$ is spanned by the
cosets of the elements $J^{a_1\cdots a_k}_{b_1\cdots b_k}$.

To complete the cellular structure, and make some basic statements
about modules, we need an anti-involution $^s$; although one can
consider the possibility that $^s$ is only an anti-isomorphism, we
do find that there is such an anti-involution. First we obtain
(details are given in the Appendix) a natural anti-involution of
the algebra of endomorphisms of $V$ into itself, $E={\rm End}\,
V$. $E$ is naturally isomorphic to $V\otimes V^*$. As we have
noted previously, for the chain we have $V=V_0\otimes V_1\otimes
\cdots V_{2L-1}$, and the dual naturally reverses order, so
$V^*=V_{2L-1}^*\otimes\cdots V_0^*$. $V^*$ is not quite identical
to $V$, despite the alternation of $V_0$ and its dual $V_0^*$
along the chain in $V$; $V_0^{**}$ is not precisely the same as
$V_0$, but there is an isomorphism between them. (In the special
case $n=0$ treated earlier, this distinction can be ignored.) The
anti-involution is named $^s$ because it is so closely related to
the supertranspose map which takes an endomorphism $f$ of $V$ to
$f^*$, an endomorphism of $V^*$ (acting on the left on $V^*$).

In our representation in the chain $V$, both the TL algebra and
its commutant $\cA$ are subalgebras of $E$, and we expect that
$^s$ restricted to these subalgebras is the requisite
anti-involution. First, it is easily checked from the definitions
(again, see the Appendix) that $^s$ maps TL to itself: it takes
$e_i\to e_i^s=e_{2L-2-i}$, so it is the chain-reversing
anti-involution of TL mentioned above. Then if $a\in \cA$, $t\in$
TL, $at=ta$ implies $a^st^s=t^sa^s$ and hence $a^st=ta^s$ for all
$t\in$ TL. That is, the image $\cA^s$ is contained in $\cA$, and
vice versa, so $\cA^s=\cA$, and so $^s$ is an anti-involution of
$\cA$.

Next we consider the ideals $\cA^{(k)}$, $k=0$, $2$, \ldots, $2L$.
First, we wish to show that $J^{(k)}=\cA^{(k)}/\cA^{(k+2)}$ are
mapped onto themselves by $^s$, for all $k$. In fact, the ideals
$\cA^{(k)}$ are defined as the annihilators of $T^{(k-2)}$ from
either side. As these conditions are again mapped to themselves by
$^s$, we have $\cA^{(k)s} =\cA^{(k)}$ for all $k$. Finally we want
to show that $J^{(k)}\cong {\cal V}_{k/2}\otimes {\cal V}_{k/2}^s$
as $\cA$-$\cA$--bimodules, where ${\cal V}_j$ is a left
$\cA$-module and ${\cal V}_j^s$ is a right $\cA$-module related to
${\cal V}_j$ by using $^s$, compatibly with $J^{(k)s}=J^{(k)}$. It
will be sufficient to consider simply $J^{(2L)}=\cA^{(2L)}$ as all
$J^{(k)}$ are isomorphic to one of these (for some $L$) as ideals
of $\cA/\cA^{(k+2)}$. This is the subspace of $E\cong V\otimes
V^*$ annihilated by the generators $e_i$ of TL on both sides. We
note that $E$ itself has the required form, under the
anti-involution $^s$. Further, we can define ${\cal V}_L$ to be
the subspace of $V(2L)$ annihilated by all $e_i$ (acting on the
left), and ${\cal V}_L^s$ to be the subspace of $V(2L)^*$ (viewed
as a right module over $E$) annihilated by all $e_i$ (acting on
the right). Hence $J^{(k)}\cong {\cal V}_{k/2}\otimes {\cal
V}_{k/2}^s$ as $\cA$-$\cA$--bimodules, and $^s$ maps it to itself.
The modules ${\cal V}_j$ (${\cal V}_j^s$) are the left (resp.,
right) standard or cell modules for $\cA$, and have dimension
$D_j'$, and superdimension $D_j$. They correspond under Morita
equivalence to the standard modules of $U_q$(sl$_2)^{(2L)}$, which
are also referred to as Weyl modules, and have dimension $2j+1$.
Again, these standard modules are indecomposable but generally not
simple in these non-semisimple cases. Hence the dimension of each
space $J^{(k)}$ is $(D_{k/2}')^2$, and the superdimension is
$(D_{k/2})^2$. We note that $\cA/\cA^{(2)}\cong {\bf C}$ (given by
cosets of $1$), and $D_0'=D_0=1$. Finally, the total dimension of
the commutant $\cA$ is ${\rm dim}\,\cA=\sum_{j=0}^L {D_j'}^2$,
with similar growth behavior as in the semisimple cases.

Some further properties of the algebra follow from the cellular
superalgebra structure \cite{gl,mathas}. It follows from our
construction that $\cA(2L)$ is a quotient of $\cA(2L+2)$ by the
ideal $\cA^{(2L+2)}$ of $\cA(2L+2)$. The dimensions of the
standard modules are $L$-independent, provided that $j\leq L$. The
properties of a standard module ${\cal V}_j$ can then be found by
using the case $L=j$ of $\cA(2L)$. One notices that the inner
product on this module can be determined from the multiplication
in this algebra, because of the isomorphism $\cA^{(2L)}\cong {\cal
V}_{L}\otimes {\cal V}_{L}^s$ as $\cA$-$\cA$--bimodules. This
inner product may be degenerate in the non-semisimple cases of
$\cA$. There is a submodule of vectors in ${\cal V}_j$ that are
orthogonal to all vectors; the quotient by this submodule is
simple, and all isomorphism classes of simple modules are obtained
in this way \cite{gl,mathas}. For our cases $m=0$, $\pm 1$ (as
well as for $|m|>2$ where all standard modules are simple), this
simple quotient is non-zero for all $j$. The standard modules of
TL can also be analyzed in the same way. Each standard module is
either simple (i.e., the submodule mentioned is zero), or contains
a simple submodule, such that the quotient module is also simple.
It turns out that the simple quotient module is nonzero for all
real $m$ except $m=0$, for which the $j=0$ quotient module alone
is zero; thus there are less than $L+1$ isomorphism classes of
simple modules of TL in this case. For this reason also, the
singlet standard module of $U_q($sl$_2)^{(2L)}$ or
$\cA_{m+n|n}(2L)$ does not appear as a summand in the chain for
$m=0$: its multiplicity is the dimension of the $j=0$ simple TL
module, which is zero.

To conclude this subsection, we set our results in a larger
context. For $m\neq0$, the relation between the TL algebra and its
commutant is an example of what was studied in the wider context
of quasi-hereditary algebras \cite{ringel}, and is now frequently
known as ``Ringel duality'' (see Refs.\
\cite{mathas,smartin,dps,martin4}, and Ref.\ \cite{martin4} for a
pedagogical discussion). (It is not a duality in the categorical
sense, as the functor is covariant, not contravariant.) Cellular
algebras are quasihereditary provided a ``non-degeneracy''
property is satisfied \cite{mathas}; this fails for the TL algebra
in the case $m=0$ because, as mentioned above, there are less than
$L+1$ non-zero simple modules. For our purposes, the full strength
of this theory is not needed, and for the most part our arguments
go through in the case $m=0$ as well as $m\neq0$.


\subsection{Morita equivalence as associative algebras}

Now we consider the commutant of TL in the module $V$ again, from
a general point of view. For commutants, we will follow the useful
convention that the commutant of an algebra that acts on a module
on the right is viewed as an algebra acting on the same module on
the left, and vice versa. Here we will view $V=V_{\rm TL}$ as a
right module over the TL algebra. Thus we can write $V=\,_\cA\!
V_{\rm TL}$ to record this fact; $V$ is thus a left $\cA$-, right
TL-, bimodule (we let TL stand for the TL$_{2L}(q)$ and $\cA$ for
$\cA_{m+n|n}(2L)$). We will need the following useful Theorem
\cite{lemma}: Let $R$ be an algebra, $M$ a finitely-generated
right module over $R$, and $S$ the commutant of $R$ in $M$. Then
there is an equivalence between the category of direct summands
(as right $R$-modules) in direct sums of copies of $M$, and the
category of projective right modules over $S$ (and similarly with
``right'' replaced by ``left'' everywhere). For finite-dimensional
$R$ and $M$, we may think of the objects in the former category
more simply as direct sums of the summands in $M$. The equivalence
is constructed using the module $M$ itself \cite{lemma}, as we
will describe momentarily.

By taking TL$_{2L}(q)$ as $R$ and using either our spaces $V$, or
the spin-1/2 chain, for $M$, and the respective commutants as $S$,
we obtain category equivalences between the direct summands as
right TL modules, and the projective right modules over $S$. We
again denote the commutant algebra of TL$_{2L}(q)$ in $V$ by
$\cA_{m+n|n}(2L)$, and that in the spin-1/2 chain by $U_q({\rm
sl}_2)^{(2L)}$. (We should mention that when $q$ is a root of
unity, $U_q({\rm sl}_2)^{(2L)}$ is an image of the version of
\uqsl2 called the ``restricted specialization'' in Ref.\
\cite{cp}, which includes the so-called renormalized powers of the
generators that can be defined by a limiting process as $q$ tends
to the root of unity through the complex numbers.) The TL algebra
acts faithfully in both cases, and provided that the direct
summands under TL present in one chain are isomorphic to direct
summands present in the other, then for each $m$ and $L$ we obtain
equivalences between the categories of projective right modules
over $\cA_{m+n|n}(2L)$ (for all $n\geq0$ such that $m+2n>1$), and
over $U_q({\rm sl}_2)^{(2L)}$ for $q$ corresponding to $m$.

For the spin-1/2 chain, the TL algebra and the decomposition of
the chain have been much studied
\cite{martinch7,gw,martin1,martin2}. For our chains with TL acting
in the space $V$, we can apply an argument similar to that in
Sec.\ 4.1 of Ref.\ \cite{martin2}, to show that the direct sum
decomposition is the same as in the spin-1/2 chain, with non-zero
multiplicities of all summands provided $m+2n>1$. A sketch of the
argument goes as follows; it uses some of the concepts of the
preceding subsection. First we construct a set of states for each
$j$ as follows. We take each valid pattern as in Sec.\
\ref{chains}, with $2j$ dots, and as before form a corresponding
state of $V$ by forming a singlet for each contracted pair of
sites, and placing states on the dots that are annihilated by
$e_i$ acting on these dots as if the contracted pairs were absent.
The cardinality of this set is then $d_jD_j'$, but we have not
shown that the vectors are linearly independent, so the dimension
of the subspace ${\cal K}_j$ they span could be less. However, for
each fixed state on the dots, which is an element of the $j$th
standard module over the commutant $\cA$, the set of valid
patterns forms the standard module of TL. That is, there exists an
injective homomorphism of the $j$th standard module of TL into
${\cal K}_j$. Similarly, for any fixed valid pattern, the states
on the dots form the $j$th standard module of the commutant $\cA$,
so there is an injective homomorphism of the $j$th standard module
of $\cA$ into ${\cal K}_j$. For some $q$ and some $j$, the $j$th
standard module of TL is simple, and then ${\cal K}_j$ has
dimension $d_jD_j'$. But when, as mentioned in the previous
subsection, the standard module of TL contains a simple submodule
(with simple quotient module), ${\cal K}_j$ may decompose as a TL
module into a non-zero number of full standard modules, and some
number of simple quotient modules. If one takes the quotient by
all the remaining submodules, the number of simple quotients is
$D_j'$. One further point is that the subspaces ${\cal K}_j$ for
different $j$ are not necessarily linearly independent of each
other, but may ``overlap'', where this is allowed by the
representation theory of TL. That is, some of the copies of simple
quotient modules may coincide with the submodules in standard
modules in one of ${\cal K}_j$ at a smaller $j$. It is also true
that as a TL module, $V$ is self-dual under the natural duality of
left TL modules to left TL modules mentioned above. It must
decompose as a sum of indecomposable TL modules that are either
self-dual, or come in dual pairs. Now our task is to find a
faithful TL module that contains submodules isomorphic to ${\cal
K}_j$ for all $j$, is self-dual, and that has the same dimension
as $V$, namely $(m+2n)^{2L}=\sum_{j=0}^L d_j D_j'$. These
requirements are highly restrictive. Using the known
representation theory of the TL algebra \cite{martin}, the
structure (decomposition) of such a module is uniquely fixed, and
is the same as that of the spin-1/2 chain (except that the
multiplicities are different, though still nonzero); this then
must be the structure of $V$. It turns out that all indecomposable
summands are self-dual, and that there is the maximum possible
overlap of the spaces constructed above. The direct sum
decomposition is described for moderate $L$ in Ref.\
\cite{martin2}, and we have checked in several cases that we find
the same in $V$. We give further details in the cases of physical
interest, $m=0$, $\pm1$, in another paper \cite{rs3}.

The resulting equivalences of categories can be described
explicitly as follows \cite{lemma} [the structure is the same if
our chain $V$ is replaced by the spin-1/2 chain, and $\cA$ by
$U_q($sl$_2)$]. The functor from right $\cA$ modules to right TL
modules maps any right $\cA$ module $\cal V$ to the tensor
product, ${\cal V}\otimes_\cA V$, which is a right TL-module.
(Here we use the general tensor product over a non-commutative
ring $R$: if $M_R$ is a right $R$-module, and $_R N$ is a left
$R$-module, then $M\otimes_R N$ is the tensor product, which is
the usual space spanned by bilinears modulo the relations
$m\otimes an = ma\otimes n$ for all $m\in M$, $n\in N$, $a\in R$.
If $R$ is the complex numbers, we write simply $\otimes$ as
before.) The functor in the reverse direction is given on any
right TL-module $\cal W$ by Hom$_{\rm TL}(_\cA\!V_{\rm TL},{\cal
W})$, which is a right $\cA$ module. (Here we use the
Hom$_R(M_R,N_R)$ space, the vector space of $R$-homomorphisms
between right $R$-modules $M_R$, $N_R$. If $M_R$ is also a left
$S$-module, $M=\,_S\!M_R$, then the Hom$_R$ space is naturally a
{\em right} $S$ module. When $R$ is the complex numbers, it will
be denoted simply Hom.) Both maps are functors which also define
maps of morphisms to morphisms in the respective categories of
modules. When these functors are restricted to the subcategories
of summands in direct sums of copies of $V$ (as TL modules), and
projective modules over $\cA$, the equivalence of these latter
categories is obtained \cite{lemma}.

As the algebras are finite-dimensional, the resulting equivalence
of the categories of projective right modules over $\cA$ and over
\uqsl2 can be extended to an equivalence of the categories of all
right, and of all left, modules, that is for each $m$, a Morita
equivalence between $\cA_{m+n|n}(2L)$ (for all $n\geq0$ such that
$m+2n>1$) and $U_q({\rm sl}_2)^{(2L)}$. [These Morita equivalences
are given by similar functors as those described above, but with a
suitable $\cA$-TL bimodule (a projective generator) in place of
$V$ \cite{anf}.] Put more simply, one can read off the structure
of the commutant algebra of TL in the module $V$, or in the
spin-1/2 chain, from the structure of the chain as a TL module.
One only has to find all endomorphisms of the TL algebra in the
given module; these form the commutant. The structures of these
commutants are the same (though the multiplicities differ), so the
algebras are Morita equivalent.

In the non-semisimple cases, this argument does not establish
Morita equivalence of the TL algebra and its commutant, which does
not always hold, or else is not necessarily of the same form as in
the semisimple cases: for example, when $m=0$ the number of simple
modules in the two algebras is different for all $L$. For $|m|<2$,
$m\neq0$, it appears that there is a Morita equivalence between TL
and its commutant, but the correspondence is obtained by {\em
reversing} the ordering (analogous to our labels $j$ in the
semisimple cases) of the modules within each indecomposable block
of the algebras \cite{martin2}. This equivalence does not appear
to be useful for our purposes.

We also find that the commutant of $\cA_{m+n|n}(2L)$ in $V$ is
just the TL algebra (not larger), so they form a ``dual pair'', as
is also the case for $U_q({\rm sl}_2)^{(2L)}$ in the spin-1/2
chain \cite{martin2}. In the \uqsl2 case, this assertion is known
as quantum Schur-Weyl reciprocity \cite{dps,martin4}. Then we can
apply the Theorem with $\cA_{m+n|n}(2L)$ as $R$ and TL$_{2L}(q)$
as $S$, with $V$ as a left $\cA$-module. In the case of the
spin-1/2 chain, the decomposition into a direct sum under \uqsl2
was studied in Ref.\ \cite{ps}. The direct summands are a certain
type of \uqsl2 module, which (together with analogs for other
quantized UEAs) are now known to be examples of ``tilting
modules'' \cite{cp}. More precisely, there is another, more
general definition of tilting modules \cite{cp}, and the spin-1/2
chain is a ``full tilting module'', which means it is tilting, and
that its direct sum decomposition into indecomposable tilting
modules contains at least one copy of each indecomposable tilting
module over $U_q({\rm sl}_2)^{(2L)}$
--- except for $m=0$, when the $j=0$ singlet tilting module is not
a summand in the chain, [and the same is true for its Morita
equivalent module over $\cA_{m+n|n}(2L)$]. As the same (complete)
set of indecomposable projective left modules for the TL algebra
arises in connection with both $V$ and the spin-1/2 chain, it
follows that the category of left tilting modules for $U_q({\rm
sl}_2)^{(2L)}$ and its analog for $\cA_{m+n|n}(2L)$ are equivalent
(and this equivalence is the same as the restriction of the
previous one to the tilting modules). We will call these summand
modules for $\cA_{m+n|n}$ ``tilting modules'' also. We have also
checked explicitly for some cases that the modules of the
commutant in $V$ have the same submodule structure as those of
\uqsl2 in the spin-1/2 chain \cite{ps,martin2}. We will term the
direct summands of the TL algebra in $V$ ``tilting modules'' (of
TL) also.

The tilting modules, viewed as direct summands in $V$ as an
$\cA$-module, are related to the standard modules. Tilting modules
that are simple are standard, while tilting modules that are not
simple can be decomposed into a sub-module and a quotient module
that are both standard \cite{ps,cp,martin4}. Tilting modules are
generally defined as having such a series decomposition, and also
one for their dual. In the present cases, all tilting modules are
self-dual (when duality is viewed as a map from left modules to
left modules).

\subsection{$L\to\infty$ limit}

Next we discuss the $L\to\infty$ limits of the algebras, their
modules, and of the equivalence between them. Again, this requires
somewhat more care in the non-semisimple cases. There are two
different limits that can be taken for the TL algebra and for its
commutant. Just one of these is a limit of TL as an algebra, while
the other is a limit of the commutant algebras $\cA$ or
$U_q($sl$_2)$, but both also give rise to limits of the categories
of modules on {\em both} sides, that is TL and its commutant. The
more useful one of these two limits seems to be the one that
arises from the natural quotient (projection) maps of cellular
algebras on the commutant side. This emphasizes the usefulness of
the symmetry analysis, and we will learn more about this in the
following subsection when we consider the product operations.

For the TL algebras, TL$_{2L}(q)$ is a subalgebra of
TL$_{2L+2}(q)$. It can be defined as an injection of the former
into the latter, in an obvious way that takes $e_i$ to $e_i$ for
$i=0$, \ldots, $2L-2$, and the unit $1$ to $1$. This map from
$L\to L+1$ can be iterated, and one obtains a compatible system of
algebra homomorphisms from TL$_{2L_1}(q)$ to TL$_{2L_2}(q)$ for
all $L_2\geq L_1$. By a standard construction (see, e.g., Ref.\
\cite{wo}), one obtains from this the purely algebraic ``direct''
(or ``inductive'') limit of the algebra (the $C^*$-algebra version
of this was used extensively for semisimple quotients of TL in
Refs.\ \cite{jones1,ghj}). The direct limit algebra is generated
by $1$ and $e_i$, $i=0$, $1$, $2$, \ldots, subject to the
relations (\ref{tlrel}), and elements of the limit algebra are
finite linear combinations of products of a finite number of
$e_i$s, and thus every element lies in TL$_{2L}(q)$ for some $L$.
Notice that the left end of the chain is held fixed in the limit,
while the right end goes to infinity. One could also define the
direct limit in other ways, for example holding the right end of
the chain fixed, with generators $e_i$, with $i=\ldots$, $-1$,
$0$, by subtracting $2L-2$ from $i$ [for TL$_{2L}(q)$] before
defining the injection TL$_{2L}(q)\rightarrow$ TL$_{2L+2}(q)$ by
$e_i \rightarrow e_i$. Other ways include letting both ends go to
infinity. Thus there is more than one way to define such a limit,
because it depends on the choice of the system of injections. We
will nonetheless continue the discussion a little, using the
former definition of the direct limit.

Given a choice of injections TL$_{2L}(q)\rightarrow$
TL$_{2L+2}(q)$ for all $L$, we can use the induction functor to
map modules of the former to modules of the latter (this functor
may be familiar from the operation of inducing a representation of
a group from a representation of a subgroup). The induction
functor is defined as follows. If an algebra $R$ is a subalgebra
of $T$, then $T$ itself can be naturally viewed as left module
over itself, and as a right module over $R$; this is written as
$_{T}T_{R}$. If $M$ is a left $R$-module, then the induced left
$T$-module can be obtained as $_{T}T_{R}\otimes_R M$.  This map
from $L\to L+1$ can be iterated, and one obtains a compatible
system of functors from modules over TL$_{2L_1}(q)$ to modules
over TL$_{2L_2}(q)$ for all $L_2\geq L_1$. The direct (or
inductive) limit of this family can then be taken. It can be shown
that the induction functor maps projective modules of
TL$_{2L_1}(q)$ to projectives of TL$_{2L_2}(q)$. Then by composing
with the equivalence with the categories of direct summands
(tilting modules) of the commutants, we also obtain a functor from
the category of tilting modules of $\cA_{m+n|n}(2L_1)$ to that of
$\cA_{m+n|n}(2L_2)$. This functor can be described directly.
{}From $\cA_{m+n|n}(2L)$ one can form the Morita equivalent
algebra $\cA_{m+n|n}(2L_1)\otimes M_{(m+2n)^2}({\bf C})$, as
follows. All modules of the latter are obtained by tensoring a
module of the former with ${\bf C}^{(m+2n)^2}$, which describes
two additional sites of the chain, and this defines the Morita
equivalence. $\cA_{m+n|n}(2L+2)$ is a subalgebra of this larger
algebra. Then one can apply the functor of restriction of modules
to obtain a module over $\cA_{m+n|n}(2L+2)$ from a module over
$\cA_{m+n|n}(2L_1)\otimes M_{(m+2n)^2}({\bf C})$, and we have seen
that the latter are generated by tensor product modules. As
restriction maps direct summands to direct summands, we are done.

We have found a system of functors mapping modules over the
commutant for smaller to those for larger $L$, and we might expect
that there is a corresponding injection of the smaller algebra
into the larger. For the semisimple cases, there are injection
maps, but not for the non-semisimple cases. In no cases are there
injection maps of algebras that agree with the functors defined
above on the modules.

On the hand, as we have already seen, there is a compatible family
of quotient (or projection) homomorphisms from the commutant
$\cA_{m+n|n}(2L_2)$ to $\cA_{m+n|n}(2L_1)$ for $L_2>L_1$, which
can be obtained by iterating the maps $p_{2L+2}:\cA_{m+n|n}(2L+2)
\to \cA_{m+n|n}(2L)$ that we have already described; the kernel of
the map $p_{2L+2}$ is the cellular ideal $\cA^{(2L+2)}$ in
$\cA_{m+n|n}(2L+2)$. From this we can then define the limit
algebra $\cA_{m+n|n}$ as the ``inverse'' (or ``projective'') limit
of the system of projection (quotient) maps (see, e.g., Ref.
\cite{kass2}). In effect, this means that the limit algebra
$\cA_{m+n|n}=\cA$ is cellular with an infinite descending chain of
ideals
\be%
 \cdots\subset\cA^{(4)}\subset\cA^{(2)} \subset\cA^{(0)}=\cA.\ee%
(Thus the limit algebra is not artinian, though it is still
noetherian.) As the finite-dimensional quotient algebras
$\cA(2L)/\cA^{(k)}(2L)$ for each $k$ ($k\leq 2L$) are isomorphic
for all $L$, this limit seems extremely natural and corresponds
under Morita equivalence to the inverse limit \uqsl2 of the
finite-dimensional algebras. That is, we will define \uqsl2 itself
as the inverse limit of the $q$-Schur algebras
$U_q($sl$_2)^{(2L)}$, and this limit is Morita equivalent to our
algebras $\cA_{m+n|n}$. As usual, for the oriented-loops models we
mean here the quotient algebra of \uqsl2 with integer spin ($j$)
representations only.

Again, there are closely related functors on the categories of
modules over $\cA_{m+n|n}(2L)$. Given the projection maps
$p_{2L+2}$ for the algebras, there are natural ``pullback''
functors which ``lift'' any module over $\cA_{m+n|n}(2L)$ to a
module over $\cA_{m+n|n}(2L+2)$. The lifted module is the same as
the original when viewed as a (graded) vector space, so this
functor preserves dimensions, unlike the induction functor which
was applied on the TL side above. The system of compatible
pullback functors itself has a {\em direct} limit, and the limit
object is a category of modules over the limit algebra
$\cA_{m+n|n}$, such that any module over any of the
$\cA_{m+n|n}(2L)$'s is mapped to a module over the limit algebra
$\cA_{m+n|n}$, functorially. A natural category of such modules
over the limit algebra to consider is that in which every module
is the lift of a module over $\cA_{m+n|n}(2L)$ for some $L$; this
includes the lifts of all the finite-dimensional modules over
$\cA_{m+n|n}(2L)$ for all $L$, and is the smallest possible direct
limit category. Again, one can define a similar functor on, and
category of, \uqsl2 modules, and Morita equivalence is preserved
in the limit. The pullback functors map tilting modules to tilting
modules, and also standard modules to standard modules, in
particular mapping the $j$th indecomposable of either type to the
corresponding one for $L\to L+1$. It is then natural to speak of
tilting and standard modules for the limit algebra, as well as for
finite $L$.

In addition, the functor that yields an equivalence of the
categories of tilting modules for $\cA(2L)$ [or for
$U_q($sl$_2)^{(2L)}$] to that of projective modules over TL can be
composed with those above to produce functors from the category of
modules over TL$_{2L}(q)$ to that for TL$_{2L+2}(q)$, which are
{\em not} the same as the induction functors constructed above.
The present functors map the $j$th indecomposable projective
module for TL$_{2L}(q)$ to the corresponding (i.e., $j$th one) for
TL$_{2L+2}(q)$, unlike the induction functors above which act in a
more complicated way. We can take the smallest direct limit of
this compatible family of functors also. The result is a category
all of whose objects are infinite dimensional vector spaces; every
object is obtained by applying the composite of the infinite
sequence of functors to a module in the category of modules for
TL$_{2L}(q)$ for some size $L$, and similarly for the morphisms.
By using the alternative definition of projective modules that is
expressed entirely in terms of morphisms and objects \cite{anf},
it follows that the limit of a projective module taken in this way
is a ``projective object'' in the limit category. Because the
projection maps of the commutant algebras, and the functors that
follow from them, are defined in an essentially unique way, this
construction is not plagued by non-uniqueness issues as the
induction on the TL side was. These reasons of simplicity and
uniqueness are what make the present way of taking the limit seem
the most natural one for our purposes, and we will see more on
this in the next two subsections. Note that there are no
projection homomorphisms for the TL algebras taking $L+1$ to $L$,
even in the semisimple cases $|m|\geq2$, as can be seen by
considering the effect they would have on dimensions of modules as
vector spaces. While the direct limit of the functors exists as a
category, it is not clear to us whether there is a corresponding
(purely algebraic) limit of the TL algebras as well, for which
this would be a category of modules (but see also Sec.\ \ref{cont}
below).

\subsection{Tensor products and ribbon Hopf structure}

In this subsection we consider the additional structures that turn
the commutant algebras $\cA_{m+n|n}$ into ribbon Hopf algebras,
and show that these are Morita equivalent as ribbon Hopf algebras
to $U_q($sl$_2)$. The most important of these structures is the
comultiplication (or tensor product of modules).

A comultiplication for the commutant can be defined naturally.
First, a chain of $2L=2(L_1+L_2)$ sites can be broken into two of
lengths $2L_1$, $2L_2$ by removing the generator $e_{2L_1-1}$ that
connects them. This shows that TL$_{2L_1}\otimes$TL$_{2L_2}$ is
(isomorphic to) a subalgebra of TL$_{2L}$ for $L=L_1+L_2$ (and the
same value of $q$ throughout). Then, taking the commutants of
these algebras in the chain $V$ we see that $\cA_{m+n|n}(2L)$ is a
subalgebra of $\cA_{m+n|n}(2L_1)\otimes \cA_{m+n|n}(2L_2)$. We
define the natural inclusion (or injection) map
$\cA_{m+n|n}(2L)\to \cA_{m+n|n}(2L_1)\otimes \cA_{m+n|n}(2L_2)$ to
be $\Delta_{2L_1,2L_2}$. In order to take the inverse limits of
these algebras and so obtain an inclusion  $\Delta:\cA_{m+n|n}\to
\cA_{m+n|n}\otimes \cA_{m+n|n}$, we should show that the
inclusions for finite length are compatible with the projection
maps $p_{2L}$ (etc.).  That is, we should prove that for all
$L_1$, $L_2$ ($L_1+L_2=L$), $\Delta_{2L_1,2L_2}\circ
p_{2L+2}=({\rm id}\otimes p_{2L_2+2})\circ \Delta_{2L_1,2L_2+2}$
(and similarly for increasing $L_1\to L_1+1$), so that the double
inverse limit $\Delta$ of $\Delta_{2L_1,2L_2}$, as $L_1$,
$L_2\to\infty$, exists. This is fairly straightforward, and we
note only that ${\rm ker}\, \Delta_{2L_1,2L_2}\circ p_{2L+2}={\rm
ker}\,({\rm id}\otimes p_{2L_2+2})\circ
\Delta_{2L_1,2L_2+2}=\cA^{(2L+2)}_{m+n|n}(2L+2)$ for all $L_1$,
$L_2$. Then $\Delta$ is also co-associative, and is the required
comultiplication $\Delta:\cA_{m+n|n}\to \cA_{m+n|n}\otimes
\cA_{m+n|n}$. In terms of the elements $J^{a_1\cdots
a_k}_{b_1\cdots b_k}$, the comultiplication acts in exactly the
same way as in the semisimple case (except again that the use of
the Jones-Wenzl projectors should be avoided).

The main use of the comultiplication is its application to
modules. The pullback of the injection map $\Delta_{2L_1,2L_2}$
maps a module over $\cA_{m+n|n}(2L_1)\otimes \cA_{m+n|n}(2L_2)$ to
one over $\cA_{m+n|n}(2L)$. In particular, a tensor product of a
module over $\cA_{m+n|n}(2L_1)$ and one over $\cA_{m+n|n}(2L_2)$
is naturally a module over $\cA_{m+n|n}(2L_1)\otimes
\cA_{m+n|n}(2L_2)$. The pullback functor in the case when the
algebra homomorphism is an injection is more often called
``restriction'' (it is the ``right adjoint'' to the induction
functor for the same map \cite{anf}). It can be written explicitly
as the tensor product with%
\be%
_{\cA_{m+n|n}(2L)}\!\cA_{m+n|n}(2L)_{\cA_{m+n|n}(2L_1)\otimes
\cA_{m+n|n}(2L_2)},\ee%
which is the algebra $\cA_{m+n|n}(2L)$ viewed as a left
$\cA_{m+n|n}(2L)$-, right $\cA_{m+n|n}(2L_1)\otimes
\cA_{m+n|n}(2L_2)$-, bimodule. As the tensor product (over $\cA$)
is distributive over direct sums, the restriction functor, and
thus the product operation automatically maps a product of tilting
modules to a tilting module (this argument requires some
modification to handle the case of the $j=0$ tilting module for
$m=0$, which is not a direct summand in the chain; however, this
is easily done directly).

To obtain a similar restriction map for the limit, we may note
that the equal composite maps $\Delta_{2L_1,2L_2}\circ
p_{2L+2}=({\rm id}\otimes p_{2L_2+2})\circ \Delta_{2L_1,2L_2+2}$
have equal pullback functors, which are the composition of that
taking modules for $L$ to one for $L+1$, as discussed in the
preceding section, with the pullback of $\Delta_{2L_1,2L_2}$ or
$\Delta_{2L_1,2L_2+2}$. Thus the result of restricting using
$\Delta_{2L_1,2L_2}$ (or $\Delta_{2L_1,2L_2+2}$) is the same
either before or after applying the appropriate lift maps for
$L\to L+1$ (or $L_1\to L_1$, $L_2\to L_2+1$). Hence the limit of
the restriction maps exists, and we obtain a tensor product
operation from the category of modules over $\cA_{m+n|n}$ to
itself, that is it makes the product module over
$\cA_{m+n|n}\otimes \cA_{m+n|n}$ into one over $\cA_{m+n|n}$, and
further this tensor product is also associative. Thus the category
of modules over $\cA_{m+n|n}$ is now a tensor category, as is the
subcategory of tilting modules (as for \uqsl2 \cite{ps}). The
fusion rules for this product in non-semisimple examples will be
discussed in Ref.\ \cite{rs3}, but we note that the
one-dimensional ``singlet'' module, which is tilting, is the unit
for the tensor product. The unit and co-unit maps for the algebra
$\cA_{m+n|n}$ exist, the co-unit $\varepsilon$ being the map to
the complex numbers, defined as the limit of the projection maps
$\cA_{m+n|n}(2L)\to\cA_{m+n|n}(2L)/\cA_{m+n|n}^{(2)}(2L)\cong {\bf
C}$. The co-unit map is used to show that the singlet module is
the unit for the tensor product of modules \cite{cp,kass}. The
unit map $\eta$ simply maps a complex number $c$ to $c.1$ in
$\cA$.

To understand the antipode, we will begin by considering natural
operations on the space $V$ for $2L$ sites, as for the other
structures. The antipode is used to turn the left dual vector
space ${\cal V}^*$ of any module $\cal V$ into a left module. In
particular, we can consider the module $V$, and infer the result.
If we consider the birth maps discussed in the Appendix, and apply
to the case of $V$, we will obtain a map $b_V:{\bf C}\to V\otimes
V^*$. We can identify $V^*$ with $V$ as before, and then we
require that the image of $b_V$ in $V\otimes V$ be invariant under
$\cA$, that is the action of any $a\in\cA$ on this image is simply
given by multiplication by $\varepsilon(a)$ (this means that it is
annihilated by all ideals $J^{(k)}$ for $k>0$; it is isomorphic to
the singlet or $j=0$ standard module). The image of this map is
simply described as the state corresponding to the pattern (of
singlet contractions of $V_0$ with $V_0^*$ in one or the other
order) $((\cdots))$ of $2L$ left and $2L$ right parentheses. Now
we consider the action of $\cA$ on this module; this requires the
use of the comultiplication, while the action of $\cA$ on each
tensor factor is known because $\cA$ was defined as an algebra of
endomorphisms on such modules. From the definition of the
comultiplication $\Delta$, the action of $\cA$ on $V(2L)\otimes
V(2L)$ is identical to that on $V(4L)$. The image of $b_V$ is
clearly in the singlet subspace, but it may be useful to prove
this here as the corresponding proof was not given elsewhere in
this paper. We consider one of the operators $\widehat{j}$ defined
in the Appendix, that is a linear combination of the operators
$\widetilde{J}$ containing $k$ factors of $J_{ia}^b$. Consider
also the inner-most parentheses, the pair $()$. In the sum over
positions $i_l$, there are terms with zero, one or two operators
$j_i$ on these two sites [which are $2L-1$, $2L$ in the labeling
of $V(4L)$.] The terms with one or two operators at these sites
annihilate this singlet, as in the construction of $\widehat{j}$.
That leaves terms in which no operators are on those sites, and
that pair can effectively be deleted from the problem. Then
arguing by induction completes the proof. There are similar proofs
that operators with $k>2j$ annihilate the states described by
valid patterns with $2j$ non-contractible dots, which transform as
the $j$th standard module. Similarly, one can see directly that
the ideal $T^{(2j)}$ is annihilated by the operators $\widehat{j}$
with $k>2j$.

This calculation shows that the left dual $V^*$ of $V$ is
isomorphic to $V$ as a left $\cA$-module. The action of $a\in\cA$
on an element $w$ in the dual module $V^*$ is supposed to be given
by $S(a)^*w$, where $S$ is the antipode. On using the isomorphism
from $V^*$ to $V$, this becomes $S(a)^sv$ for the image $v\in V$
of $w$, where $^s$ is the anti-involution already discussed. But
$V$ is the defining representation for $\cA$, and so the $\cA$
action is also given by $av$. Taking matrix elements, $S(a)^s=a$
for all $a\in\cA$, and so (as $^s$ is an anti-involution),
$S(a)=a^s$. The explicit action of $^s$ is discussed further in
the Appendix, and then to the extent that we may use the notation
$J^{a_1\ldots a_k}_{b_1\ldots b_k}$, $S$ is given by
\bea%
S(J^{a_1\ldots a_k}_{b_1\ldots b_k})&=&
(-1)^k(-1)^{\sum_{l<l'}({\rm deg}\,a_l+{\rm deg}\,b_l) ({\rm
deg}\,a_{l'}+{\rm deg}\,b_{l'})}\nonumber\\
&&\mbox{}\times J^{a_k\ldots a_1}_{b_k\ldots b_1}.\label{antip}\eea %
Clearly this agrees with the cases $n=0$ discussed earlier; it is
the general formula for the cases $|m|\geq 2$ as well. This
construction of $S$ on the algebras for finite length chains
passes immediately to the $L\to\infty$ limit. We note that the
algebra $\cA$ in these oriented-loops cases is spanned by the
elements in the $J^{(k)}$s for $k$ even, so for these the sign
$(-1)^k$ can be dropped.

These make $\cA_{m+n|n}$ into a Hopf algebra, which is Morita
equivalent to \uqsl2 as a Hopf algebra. The braiding and twist can
be introduced using the TL generators exactly as in the semisimple
cases, and the Morita equivalence now extends to the algebras as
ribbon Hopf algebras. As a Hopf algebra, $\cA$ is an extension of
$U($gl($m+n|n$), but the braiding and twist are not extensions of
those that make $U($gl($m+n|n$) a ribbon Hopf algebra. We note
further that the unoriented cases carry through in an exactly
parallel way to the oriented and semisimple unoriented ones.

We have concentrated here on the faithful cases $m+2n\geq 2$. It
is worth noting that for $m=0$, $n=1$, the supersymmetric chain
has a formulation as a free (unconstrained) fermion system
\cite{rs}; see the Appendix. For this case $D_j'=2j+1$. In fact,
the commutant algebras $\cA_{1|1}(2L)$ and $U_q$(sl$_2)^{(2L)}$
with $q=i$ are isomorphic (and not only Morita equivalent) as
associative algebras for each $L$, and the limits $\cA_{1|1}$ and
\uqsl2 are isomorphic \cite{hubold} as ribbon Hopf algebras (the
difference between the graded tensor products used for
$\cA_{1|1}$, and the trivially graded ones for $U_q$(sl$_2)$, is
absorbed into the maps of the comultiplication and antipode under
this isomorphism).

\subsection{Induction product for TL modules}

There is also a product operation for the TL modules, which is the
analog of the tensor product of modules over the commutant, and is
again defined using the picture of joining chains end to end. It
has to be defined using the induction functor, and does not
preserve dimensions of modules. Here we define it, and show that
the fusion rules for the product of two projective modules over TL
agree with those for the corresponding tilting (summand) modules
over the commutant $\cA$ or $U_q($sl$_2)$.

For TL one uses again the inclusion of
TL$_{2L_1}\otimes$TL$_{2L_2}$ (in an obvious way) as a subalgebra
of TL$_{2L}$ for $L=L_1+L_2$. Then given any left TL-module $\cal
W$ over TL$_{2L_1}\otimes$TL$_{2L_2}$, one can apply the induction
functor to obtain a module over TL$_{2L}$, given explicitly by%
\be%
{\rm TL}_{2L}\otimes_{{\rm TL}_{2L_1}\otimes{\rm TL}_{2L_2}}{\cal
W},\ee%
where the algebra TL$_{2L}$ is viewed as a left TL$_{2L}$, right
TL$_{2L_1}\otimes$TL$_{2L_2}$ module (it is a right
TL$_{2L_1}\otimes$TL$_{2L_2}$ by restriction). It is a general
fact that any induction functor maps projective modules to
projective modules, so in particular, when applied to a projective
module over TL$_{2L_1}\otimes$TL$_{2L_2}$, such as a tensor
product of a projective of TL$_{2L_1}$ with a projective of
TL$_{2L_2}$, the result of induction is a projective module over
TL$_{2L}$. Notice that, like induction in general, this product
operation does {\em not} conserve the dimensions of the modules.

Given this product operation (functor), which maps a tensor
product of projective modules for lengths $2L_1$, $2L_2$, to a
projective module for length $2L$, it is natural to ask how it
compares with the product operation that was defined for modules
over the commutant, and in particular for the product of tilting
modules, since these correspond to the projective modules over TL
under the equivalence of categories. We can utilize the functor
from left TL modules to left modules over the commutant either
before or after taking the product. The functor for the length
$2L$ chain is the tensor product functor%
\be%
_{\cA_{m+n|n}(2L)}V_{{\rm TL}_{2L}(q)}\otimes_{{\rm TL}_{2L}(q)}
-,\ee%
and similarly for TL$_{2L_1}\otimes$TL$_{2L_2}$. Thus all
the functors involved can be written as tensor products with
suitable modules, either $V$ or one of the algebras. Using
associativity of the tensor product over non-commutative algebras,
we find that
both orders of functor operations reduce to the tensor functor%
\be%
_{\cA_{m+n|n}(2L)}V_{{\rm TL}_{2L_1}(q)\otimes {\rm
TL}_{2L_2}(q)}\otimes_{{\rm TL}_{2L_1}(q)\otimes {\rm
TL}_{2L_2}(q)}-.\ee%
As the functor from Ref.\ \cite{lemma} is an equivalence for
projectives/tilting modules, it follows that {\em the fusion rules
for the induction product of projective modules over TL are the
same as those for the direct summands (tilting modules) of the
commutants} at the corresponding values of $q$ and $m$. Since the
induction product of projective TL modules closes on projective
modules, and the tensor product of $\cA$ or \uqsl2 tilting modules
closes on tilting modules, these fusion rules can be characterized
completely by specifying the multiplicities in the decomposition
of a product of the respective indecomposable modules into
indecomposables, just as in the semisimple case. So this result
says that these multiplicities agree. This will be used
extensively in the companion paper \cite{rs3}.

We can also ask if the product on the TL side is compatible with
the lifting functor we defined above via the relation with the
lift on modules of the commutant under $L\to L+1$. We need to show
that one further square of functors commutes, namely the products
agree, whether taken before or after the lift. Making use of the
functor mapping to modules over the commutant, this is one face of
a cube whose faces are commuting squares, for which we have
already know that all the other squares commute (some by
definition of the functors). It follows that this one does also,
at least when restricted to projective TL modules.

\subsection{Closed chains}

We will briefly consider the closed chains in the not necessarily
semisimple context also. The JTL algebra is cellular \cite{gl},
and is semisimple when $q$ is not $\pm 1$ or a root of unity.
JTL$_{2L}(q)$ is represented faithfully in our space $V$ provided
$m+2n>2$. For the commutant, we can apply similar methods as in
the open case to show that the commutant in $V$ is also cellular.
The standard modules have dimensions $\widehat{D}'_{jK}$ for all
$q$, (and superdimensions $\widehat{D}'_{jK}$), and so the
commutant has dimension ${\rm dim}\, \widehat{\cA}_{m+n|n}=
\sum_{j,K}(\widehat{D}'_{jK})^2$, where $j=0$, \ldots, $L$ as
usual.

For the closed chains, the product operations, taking states of
two closed chains to those of a single closed chain, can be
thought of using a ``pants'' diagram (or trinion). In order to
join two incoming legs into a single torso, it is necessary first
to break each incoming closed chain, obtaining two open chains,
then join these end to end, and finally close the other two ends.
For the commutant algebra $\widehat{\cA}_{m+n|n}(2L_1)$ (and the
others with $L_2$, $L=L_1+L_2$), this produces a series of
operations: first induction from the closed to open case, for
breaking each leg; then the tensor product in the open case;
finally, the restriction map from open to closed algebras. For the
modules over the JTL algebras, there is a product defined
similarly (with induction and restriction interchanged). At least
in the semisimple cases, the fusion rules for the JTL algebra and
its commutant agree. Notice that, because of the use of induction,
the product of $\widehat{\cA}_{m+n|n}$ modules does not conserve
dimensions (it is not a tensor product of vector spaces), and also
that as $L\to\infty$ there is no upper limit on the $j$ values of
the representations in the fusion rules, though there is
necessarily such a limit on the highest weights in their
Clebsch-Gordan decomposition into $U({\rm gl}_m)$ modules. (Higher
$j$ values are generated during the induction step from closed to
open chains, as a given $j$ in the closed chain might arise from
an arbitrarily higher (as $L_1\to\infty$) $j$ in an open chain, by
making contractions to a singlet across the end of the chain.)


\section{Continuum limit and CFTs}
\label{cont}

In this last section before the conclusion, we comment a little on
the continuum limits of the chains. This topic will be taken up in
greater depth in a forthcoming paper \cite{rs3}.

In previous sections, we have considered the $L\to\infty$ of the
algebraic structures in the chains, especially the commutant of
the TL algebra and its modules, and the modules over the TL
algebra, from a purely algebraic point of view. But from a
physical point of view, more is required. Physically, we want to
choose a Hamiltonian $H$ for the chain, and examine low-energy
(and long-wavelength) properties in the $L\to\infty$ limit. We
will view the limit as taken with a lattice spacing distance
tending to zero as $L\to\infty$, such that the length of the chain
remains constant in the limit, equal to $1$, say (hence the term
``continuum limit''), and also with all parameters in $H$
proportional to $L$. Then low energies and long wavelengths mean
excitation energies and wavevectors of order $1$ in these units.
We are especially interested to begin with in cases where this
continuum limit is a non-trivial conformal field theory, which in
our units implies that excited states at energies of order 1 above
the ground state do exist. For the supersymmetric chains $V$
considered here, or for the spin-$1/2$ chain, this occurs for the
Hamiltonian $H=-L\sum_i e_i$, when $m$ is in the range $-2<m\leq
2$. It follows immediately from our analysis that our commutant
algebra is a symmetry of the low-lying spectrum of this
Hamiltonian for any finite $L$. It is not entirely clear how the
limit can be taken in a mathematically rigorous way, but roughly
we want to take the eigenvectors of $H$ that have low-energy
eigenvalues, and we expect that the inner products among these
vectors can be made to tend to some limits. Further, if we focus
on long wavelength Fourier components of the set of $e_i$, then we
expect their limits to exist, and their commutation relations to
tend to those of the Virasoro generators $L_n$
($L_n+\overline{L}{_n}$ in the closed chain case) \cite{ks}, in
the sense of weak convergence of matrix elements in this basis of
low-energy eigenvectors. Then the modules over the TL algebra
become modules over the UEA of the Virasoro, or possibly even a
larger, algebra. (For the closed chains, two copies of the
Virasoro algebra with generators $L_n$, $\overline{L}_n$ should
eventually emerge.)

The symmetry (commutant) algebra in the continuum limit, which
commutes with the Virasoro algebra, must be at least as large as
that in the finite-$L$ chains. For the open chains, it appears
that our commutant algebra [or \uqsl2 in the spin-1/2 chain] does
not become even larger in the limit. (In certain cases, such as
the open or closed $m=2$ spin-$1/2$ chain, the symmetry algebra
combines with the Virasoro algebra to form the sl$_2$ level 1
current [affine Lie] algebra, but this is not the case in general
\cite{rs}.) In the continuum limit, the basis for the commutant
algebra takes a similar form as on the lattice,
and can now be written for the open cases using%
\bea%
\lefteqn{\widetilde{J}^{a_1a_2\cdots a_k }_{b_1b_2\cdots b_k}
=}\nonumber\\&&\int_{0<x_1<x_2<\cdots<1}\prod_{i=1}^k dx_i\,
J^{a_1}_{b_1}(x_1)J^{a_2}_{b_2}(x_2)\cdots J^{a_k}_{b_k}(x_k)\eea%
where the integration is over $0<x_1<x_2<\cdots<x_k<1$ (where $xL$
is position on the chain, and all operators are at the same time),
$J^a_b(x)$ stands for the density of the generators $J_{ib}^a$ at
$x$, and the contraction of any upper with a {\em neighboring}
lower index is required to vanish as before (more precisely, one
must use the same relations derived in the Appendix for the
lattice case). The integration domain in these expressions
resembles that for the generators of the Yangian in an integrable
system, but we emphasize again that our algebra is not the
Yangian. The definition can be generalized using an arbitrary
Jordan curve $C$ with ends on the boundary (even with both ends on
the same boundary), with $J^a_b(x)$ replaced by the component of
the divergenceless currents $J_{\mu b}^a$ ($\mu=1$, $2$) normal to
the curve, and the integrations are long the curve. These
definitions also apply {\it mutatis mutandis} to the symmetry
algebras of the closed chain, with an arbitrary closed Jordan
curve $C$ in the most general form. These definitions for general
curves ensure that the enlarged symmetry really commutes with
conformal mappings of spacetime (which map curves $C$ onto one
another), and generalize those for the global symmetry generators
of gl($m+n|n$), etc. [We note that while, unlike in theories with
an affine Lie algebra, the Noether currents $J_{\mu b}^a$ do not
possess a decomposition into purely holomorphic and
anti-holomorphic parts \cite{rs}, their flux across a curve is the
Noether charge and is conserved and conformally invariant, and
thus this is expected to hold for our operators also.]

Except for $m=2$, the theories are not semisimple, and we expect
that the decomposition of the states under TL$\otimes \cA_{m+n|n}$
also determines the Virasoro structure. Hence the structures
studied here in the finite chains should be very useful for the
CFTs. In particular, much of the structure, including the fusion
rules, is dictated by the symmetry \cite{rs3}.

We emphasize that in the limit, the commutant algebras commute
with the full Virasoro algebra. We recall that a
(``fully-extended'') chiral algebra in a CFT is a maximal algebra
of integer--conformal-spin holomorphic fields that have abelian
monodromy and fusion rules. It seems that in most of the present
cases, the chiral algebra is just the Virasoro algebra. It is
unusual to find a large algebra that commutes with the full chiral
algebra (the Yangians, when present, do not commute with the full
chiral algebra, but only with a commutative subalgebra). There may
be cases of rational CFTs in which a finite group commutes with
the chiral algebra and fixes aspects of the CFT. But in the
present cases, the symmetry algebra is infinite-dimensional. Thus
we have begun the study of what we will, provisionally, call
``CFTs with symmetry'' in which some, possibly large, ``global''
symmetry algebra commutes with the chiral algebra of the CFT. We
expect that a correct use of symmetry can be an important guiding
principle in understanding irrational CFTs.

Some consequences of this procedure are worth emphasizing. For the
open oriented-loops models, the cases $m$ and $-m$ ($m\neq0$) were
the same, because the TL algebras are isomorphic. The symmetry
algebras were consequently also Morita equivalent, and this is a
consequence of only $q^2$ (not $q$) entering expressions. But for
the continuum limit defined here, a choice of Hamiltonian is an
essential part of the construction of the limit. The Hamiltonian
$H = -L\sum_i e_i$ is in isomorphic algebras in the two cases, but
the isomorphism involves reversing the signs of all the generators
$e_i$, and thus reversing the sign of $H$. Hence focusing on low
energies in the two cases produces different continuum limits. For
example, for $m=1$, the limiting CFT has central charge $c=0$,
while for $m=-1$ it has $c=-7$. These two theories have the same
commutant or symmetry structure, and the purely algebraic
$L\to\infty$ limits of the modules, and the fusion rules, are the
same, so that their Virasoro properties are still similar. [For
$m=2$, the limit has $c=1$, and is related by Morita equivalence
of symmetries to the limit of the antiferromagnetic spin-1/2
chain, which is a $c=1$ CFT, while for $m=-2$, the continuum limit
is related to the SU(2) ferromagnet, and is not even conformal.]
There are further Hamiltonians that may be of physical interest in
the same chains, that are elements of the TL algebra, and act
within constrained subspaces as in Ref.\ \cite{ks2} and Sec.\ 6 of
Ref.\ \cite{fr}, and these can produce further distinct CFTs in
the continuum limit. Similarly, the dilute-loops models also have
continuum limits in a similar fashion, and have the same symmetry
algebras commuting with the Virasoro algebra as in the
corresponding dense-loops cases.


\section{Conclusion}

As this paper has covered a lot of ground, let us try to summarize
a few points here. We began with simple models of spin chains with
$m$ states per site ($m\geq2$), with nearest-neighbor interactions
with U$(m)$ symmetry (under which the sites transform alternately
in the fundamental and its conjugate representation), the
interaction being essentially the projection onto a U$(m)$
singlet. We showed that such spin chains, even with arbitrary
coefficients of these interactions, have a much larger symmetry
algebra than U$(m)$, with representations labeled by $j=0$, $1$,
\ldots, $L$, for the $2L$-site chain, and these are irreducible in
the case of open chains (i.e., free boundary conditions). This
means that the spectrum of the chain is a lot simpler than might
have been expected, and can be computed more easily if this
symmetry is exploited. There are similar results for
supersymmetric chains with gl$(m+n|n)$ symmetry of
nearest-neighbor interactions ($m+n$, $n\geq0$, and here again
$|m|\geq2$), (open chain) Potts models in the Potts
representation, and also for closed chains (i.e., periodic
boundary conditions), though for the latter the representation
structure is richer. The symmetries also apply to the loop models
that can be obtained from the spin chains in a spacetime or
transfer matrix picture, and there are similar ones for dilute
loop models. In the loop language, the symmetries arise because
the loops cannot cross.

Full use of symmetry in physics requires more structure than just
an associative algebra. One wants to tensor representations and
decompose the product into representations, and there should be
dual representations for use as anti-particles. These structures
were obtained here (for the open chains) by considering joining
chains end to end. The ``fusion'' rules for decomposing the tensor
product of representations labeled $j_1$ and $j_2$ took the same
form as the Clebsch-Gordan series for SU(2). These structures turn
the symmetry algebra into a Hopf algebra, and this is said to be
Morita equivalent to a Hopf algebra \uqsl2 that is the quantum
group deformation of the familiar SU(2) algebra. We also
introduced structures of braiding and a twist; the former allows
us to exchange representations, and obeys Yang-Baxter--type braid
group relations. With the complete structure, one can compute, for
example, $6j$ symbols for the algebras.

Finally, we successfully extended all the open-chain results to
the cases $|m|< 2$, in the supersymmetric versions, for which the
algebras are no longer semisimple (that is, representations are
not fully decomposable into direct sums). This is of interest
because such models arise in connection with, for example,
disordered fermions, percolation, and polymers (self-avoiding
walks), all in two dimensions. The cases $|m|<2$, for suitable
Hamiltonians, possess continuum limits that are critical
(conformal) field theories, or massive perturbations of the same.
The structure of these open-chain conformal field theories is
discussed in a companion paper \cite{rs3}.

The present paper culminated in the following result, which is
worth stating again here: we find ribbon Hopf algebras
$\cA_{m+n|n}$ for all $m+n$, $n\geq0$ (resp., ${\cal B}_{m+2n|2n}$
for all $m+2n$, $n\geq0$) that are Morita equivalent as ribbon
Hopf algebras to \uqsl2 restricted to integer spin
finite-dimensional representations [resp., $U_{-q}($sl$_2)$
restricted to finite-dimensional representations, and with the
variant twist map] for $m^2=(q+q^{-1})^2$ (resp., $m=q+q^{-1}$).
(The two cases arise in correspondence with loop models that
respectively either have or do not have a fixed orientation on the
loops.) These algebras are thus ``quantum'', even though the
construction was very ``classical''. The algebras were also
analyzed as cellular algebras \cite{gl}.

A further result worth stating again concerns the Temperley-Lieb
algebra, which is generated by the nearest-neighbor interaction
terms in the spin chains. There is a notion of fusion for its
representations also, induced by joining chains end to end. The
fusion of projective modules closes on projective modules, and we
proved that the fusion rules are the same as for the direct
summand (or ``tilting'') modules for the corresponding symmetry
algebras $\cA$ or $\cal B$, whose fusion likewise closes on
themselves, as for \uqsl2 \cite{ps}. For the continuum limit, when
it is a conformal field theory, this gives the fusion rules for
the corresponding conformal fields. This is considered further in
the companion paper \cite{rs3}.

As open problems for future study, we have introduced a notion of
``conformal field theories with symmetry'', in which a large
symmetry algebra such as our $\cA_{m+n|n}$ commutes with the whole
Virasoro algebra (or with some larger chiral algebra). This may be
a key idea for understanding irrational conformal field theories
such as those for the critical points of disordered fermions in
two dimensions, as in the quantum Hall effect. More mathematical
open problems would include searching for Hopf
algebras---especially ones that are extensions of classical Hopf
(super-) algebras, such as $U($gl$(m|n))$ or
$U($osp$(m|2n))$---that are Morita equivalent to other quantum
groups. The corresponding notions starting from closed chains are
also a subject in urgent need of study.

\acknowledgments We are grateful to I. Frenkel and Z. Wang for
helpful remarks, discussions, or correspondence. Work by NR was
supported by NSF grant no.\ DMR-02-42949.


\appendix

\section{Superalgebra constructions}
\label{app}

\subsection{Graded tensor products}

A ${\bf Z}_2$-grading on an associative algebra $\cA$ can be
introduced if $\cA$ is a direct sum of two subspaces, $\cA=
\cA_{(0)}\oplus\cA_{(1)}$, with $1\in \cA_{(0)}$, and we can
associate with each element of these subspaces its {\em degree},
${\rm deg}\, a=0$, $1$, for $a\in\cA_{(0)}$, $\cA_{(1)}$
respectively, such that ${\rm deg}\,(a_1a_2)={\rm deg}\,a_1+{\rm
deg}\,a_2$ (mod $2$), for all $a_1$, $a_2$ that lie in either
$\cA_{(0)}$ or $\cA_{(1)}$, so that the degree map is a
homomorphism of the multiplicative (monoid) structure of $\cA$
into ${\bf Z}_2$ (viewed as a monoid under addition). The
existence of such a grading implies that a grading can be defined
on a module ${\cal V}$ over $\cA$, by finding subspaces ${\cal
V}_{(0)}$ and ${\cal V}_{(1)}$ (of degrees $0$, $1$ respectively)
with ${\cal V}={\cal V}_{(0)}\oplus{\cal V}_{(1)}$, such that if
$v$ is an element of either ${\cal V}_{(0)}$ or ${\cal V}_{(1)}$
and $a$ is an element of either $\cA_{(0)}$ or $\cA_{(1)}$, then
${\rm deg}\,av={\rm deg}\,a+{\rm deg}\,v$ (mod 2).

The grading plays a role in tensor products (we follow Kassel
\cite{kass}, but generalized to include grading). The tensor
product of $\cA$ with itself is defined as a vector space in the
usual way, and similarly for the tensor product ${\cal
V}_1\otimes{\cal V}_2$ of two modules (representations) ${\cal
V}_1$ and ${\cal V}_2$ over $\cA$. The latter becomes a
representation of $\cA\otimes\cA$ by the action $(a_1\otimes
a_2)(v_1\otimes v_2)=(-1)^{{\rm deg}\,a_2{\rm deg}\,v_1}a_1
v_1\otimes a_2 v_2$ when $a_1$, $a_2$, $v_1$ and $v_2$ lie in the
subspaces on which the grading is defined. In particular, to make
$\cA\otimes\cA$ a left (and right) module over itself, we must
have $(a_1\otimes a_2)(a_3\otimes a_4)=(-1)^{{\rm deg}\,a_2{\rm
deg}\,a_3}a_1a_3\otimes a_2a_4$ when each of $a_1$, \ldots, $a_4$
lies in either $\cA_{(0)}$ or $\cA_{(1)}$. The grading on ${\cal
V}_1\otimes{\cal V}_2$ itself is defined by ${\rm
deg}\,(v_1\otimes v_1)={\rm deg}\,v_1+{\deg}\,v_2$ for $v_1\in
{{\cal V}_1}_{(0)}$ or ${{\cal V}_1}_{(1)}$, and similarly for
$v_2$. (From here on we omit further specifications of elements or
vectors as belonging to either one of the graded subspaces when it
is obvious from the context that such a condition is needed in
order that an expression containing the degree of an element or
vector be well-defined.)

Finally, the flip map $\tau$ is a homomorphism of graded modules
${\cal V}_1\otimes{\cal V}_2\to {\cal V}_2\otimes{\cal V}_1$ such
that $\tau^2={\rm id}$, and in particular an involutory
automorphism of $\cA\otimes\cA$. It is required to reduce to the
usual flip map on the tensor product of the even (i.e.\ degree
zero) subspaces. Hence it is given in general by $\tau(v_1\otimes
v_2)=(-1)^{{\rm deg}\, v_1{\rm deg}\,v_2}v_2\otimes v_1$, and
similarly on $\cA\otimes\cA$. The flip map is also used in
defining the ``opposite'' algebra $\cA^{\,\rm op}$ (in which the
order of multiplication is reversed), because multiplication can
be regarded as a bilinear map from the vector space
$\cA\otimes\cA$ to $\cA$. So in $\cA^{\rm op}$, we first apply
$\tau$ to $\cA\otimes\cA$, then multiply (in $\cA$), to obtain
$a\cdot b=(-1)^{{\rm deg}\, a\,{\rm deg}\, b}ba$. $\cA$ is
(graded) commutative if $a\cdot b=ab$ for all $a$, $b$ in $\cA$.

More generally, if $f$ is a linear map of graded vector spaces,
$f:{\cal V}_1\to{\cal V}_2$ that respects the grading, then we can
define ${\rm deg}\,f=0$ if it maps ${\cal V}_{1(0)}$ to ${\cal
V}_{2(0)}$ and ${\cal V}_{1(0)}$ to ${\cal V}_{2(0)}$, while ${\rm
deg} \, f=1$ if it does the reverse. The space of all linear maps
${\cal V}_1\to{\cal V}_2$, called ${\rm Hom}\,({\cal V}_1,{\cal
V}_2)$, then becomes a graded vector space. If $f$, $g$ are linear
maps of graded vector spaces, $f:{\cal V}_1\to{\cal V}_2$, and
$g:{\cal W}_1\to{\cal W}_2$, then on the tensor product we have
$f\otimes g:{\cal V}_1\otimes{\cal W}_1\to{\cal V}_2\otimes{\cal
W}_2$ defined on $v_1\in {\cal V}_1$ and $v_2\in {\cal W}_1$ by
$(f\otimes g)(v_1\otimes v_2)=(-1)^{{\rm deg}\, g {\rm
deg}\,v_1}fv_1\otimes gv_2$.

\subsection{Dual spaces and spin chain}

In this subsection we address the left and right duals of a graded
vector space, and the basic definition for our spin chain space
$V$.

Because we write maps on the left, it is natural to define the
(left) dual ${\cal V}^*$ of a graded vector space $\cal V$ to be
the space of all graded $\bf C$-linear maps to the complex
numbers, ${\cal V}^*={\rm Hom}\,({\cal V},{\bf C})$. Equivalently,
there is a dual pairing (or evaluation map) $d_{\cal V}:{\cal
V}^*\otimes {\cal V}\to{\bf C}$. Note that this is similar to
conventional Dirac notation in quantum mechanics, except that no
complex conjugation is involved in the definition of the dual.
There is also a ``co-evaluation'' map $b_{\cal V}:{\bf C}\to{\cal
V}\otimes {\cal V}^*$ that is compatible with $d_{\cal V}$. It is
defined via the natural identification of the action of $\bf C$ on
$\cal V$ as elements in ${\rm End}\,{\cal V}={\rm Hom}\,({\cal
V},{\cal V})$, and ${\rm End}\,{\cal V}$ is naturally isomorphic
to ${\cal V}\otimes {\cal V}^*$ (all as graded objects, compatibly
with the grading). Then $\cal V$ becomes a left module over ${\rm
End}\,{\cal V}$.

Given a map of graded spaces $f:{\cal V}\to{\cal W}$, and the dual
spaces, we can define the super-transpose map $f^*:{\cal
W}^*\to{\cal V}^*$, such that $d_{\cal V}(f^*\otimes{\rm
id})=d_{\cal W}({\rm id}\otimes f)$ on ${\cal W}^*\otimes {\cal
V}$, which leads to $d_{\cal V}(f^*(\alpha)\otimes v)=(-1)^{{\rm
deg}\,\alpha{\rm deg}\, f} d_{\cal W}(\alpha\otimes f(v))$, where
$v\in{\cal V}$, $\alpha\in {\cal W}^*$. This defines $f^*$ through
its matrix elements, in analogy with the usual definition of the
transpose of a matrix.

The right dual $^*{\cal V}$ of a graded vector space can also be
defined, via compatible maps $d_{\cal V}':{\cal V}\otimes{^*{\cal
V}}\to{\bf C}$ and $b_{\cal V}':{\bf C}\to{^*{\cal V}}\otimes{\cal
V}$. This may be identified with ${\cal V}^*$ through the flip
map, $d_{\cal V}'=d_{\cal V}\circ\tau$ and $b_{\cal V}'=\tau\circ
b_{\cal V}$. Now we can point out that if ${\cal V}^*$ is the left
dual of $\cal V$, then $\cal V$ is the {\em right} dual of ${\cal
V}^*$, $^*({\cal V}^*)\cong (^*{\cal V})^*\cong{\cal V}$
(canonical isomorphisms). Consequently, ${\cal V}^{**}$ cannot be
canonically identified with $\cal V$, as would be the case for
ordinary vector spaces, though they are isomorphic. On elements,
if we view a nonzero vector $v\in {\cal V}$ as part of a basis for
${\cal V}$, and define $v^*$ to be the dual basis vector, and
similarly for $v^{**}$, then the isomorphism maps $v^{**}\in{\cal
V}^{**}$ to $(-1)^{{\rm deg}\,v}v\in{\cal V}$, because of the use
of the flip map $\tau$ that relates the right and left duals.

Our supersymmetric ``spin chain''  can be constructed using these
ideas. The space $V=V_0\otimes V_1\otimes\cdots V_{2L-1}$, where
$V_i$ is isomorphic to $V_0$ for $i$ even [$V_0$ is referred to as
the fundamental or defining representation of gl$(m+n|n)$], while
$V_i$ for $i$ odd is viewed as the left dual of $V_0$. The
definitions above are compatible with, and sufficient to
establish, the constructions of the supersymmetric chains in Ref.\
\cite{rs}. The elements $e_i$ can be identified with
$e_i=b_{V_0}d_{V_0}'$ acting on the pair $i$, $i+1$ for $i$ even,
and similarly for $i$ odd, where $V_0$ is the graded vector space
of dimensions $m+n|n$ for the oriented loops models (for any
non-negative values of $m+n$ and $n$).

\subsection{Explicit construction of spin chain and endomorphisms}

We now describe in detail the explicit construction of the space
$V$ using boson and fermion oscillators with constraints. We also
include the space of endomorphisms of the chain, and an
anti-involution on these endomorphisms.

For $i$ even we have boson operators $b_i^a$, $b_{ia}^\dagger$,
$[b_i^a,b_{jb}^\dagger]=\delta_{ij}\delta_b^a$ ($a$, $b=1$,
\ldots, $n+m$), and fermion operators $f_i^\alpha$,
$f_{i\alpha}^\dagger$,
$\{f_i^\alpha,f_{j\beta}^\dagger\}=\delta_{ij}\delta_\beta^\alpha$
($\alpha$, $\beta=1$, \ldots, $n$); here labels like $\alpha$ on
the fermion operators stand for $\alpha=a-(m+n)$ for a
corresponding $a$ index. For $i$ odd, we have similarly boson
operators $\overline{b}_{ia}$, $\overline{b}_i^{a\dagger}$,
$[\overline{b}_{ia},\overline{b}_j^{b\dagger}]=\delta_{ij}\delta^b_a$
($a$, $b=1$, \ldots, $n+m$), and fermion operators
$\overline{f}_{i\alpha}$, $\overline{f}_i^{\alpha\dagger}$,
$\{\overline{f}_{i\alpha},\overline{f}_j^{\beta\dagger}\}=
-\delta_{ij}\delta^\beta_\alpha$ ($\alpha$, $\beta=1$, \ldots,
$n$). Notice the minus sign in the last anticommutator; since our
convention is that the $\dagger$ stands for the adjoint, this
minus sign implies that the norm-square of any two states that are
mapped onto each other by the action of a single
$\overline{f}_{i\alpha}$ or $\overline{f}_i^{\alpha\dagger}$ have
opposite signs, and the ``Hilbert'' space has an indefinite inner
product. The space $V$ is now defined as the subspace of states
that obey the constraints
\begin{eqnarray}
\sum_ab_{ia}^\dagger b_i^a+\sum_\alpha f_{i\alpha}^\dagger
f_i^\alpha &=& 1 \quad (i\hbox{ even}),\\
\sum_a\overline{b}_i^{a\dagger}
\overline{b}_{ia}-\sum_\alpha\overline{f}_i^{\alpha\dagger}
\overline{f}_{i\alpha} &=& 1 \quad (i\hbox{ odd}).
\end{eqnarray}
The sums here and below are over $a=1$, \ldots, $m+n$, and
$\alpha=1$, \ldots, $n$; for clarity, we are {\em not} using the
summation convention on indices $a$, $\alpha$ in this Appendix.

The generators of the Lie superalgebra gl$(m+n|n)$ acting on each
site of the chain are the bilinear forms $J_{ia}^b= b_{ia}^\dagger
b_i^b$, $f_{i\alpha}^\dagger f_i^\beta$, $b_{ia}^\dagger
f_i^\beta$, $f_{i\alpha}^\dagger b_i^b$ (depending on whether $a$,
$b$ on the left hand side are in $1$, \ldots, $m+n$ or $m+n+1$,
\ldots, $m+2n$) for $i$ even, and correspondingly
$J_{ia}^b=-\overline{b}_i^{b\dagger} \overline{b}_{ia}$,
$\overline{f}_i^{\beta\dagger} \overline{f}_{i\alpha}$,
$-\overline{f}_i^{\beta\dagger} \overline{b}_{ia}$,
$-\overline{b}_i^{b\dagger} \overline{f}_{i\alpha}$ for $i$ odd,
which for each $i$ have the same (anti-)commutators as those for
$i$ even. Under the transformations generated by these operators,
$b_{ia}^\dagger$, $f_{i\alpha}^\dagger$ ($i$ even) transform as
the fundamental (defining) representation $V_0$ of gl($n+m|n$),
$\overline{b}_i^{a\dagger}$, $\overline{f}_i^{\alpha\dagger}$ ($i$
odd) as the (left) dual fundamental $V_0^\ast$. Hence the space
$V$ is the graded tensor product of alternating irreducible
representations $V_0$, $V_0^\ast$ as desired; the signs in the
$J_{ia}^b$ for $i$ odd can be understood as these generators are
minus the supertranspose of the action on the fundamental. In the
spaces $V_0^\ast$ on the odd sites, the odd states (those with
fermion number $-\overline{f}_i^{\alpha\dagger}
\overline{f}_{i\alpha}$ equal to one) have negative norm-square.
The destruction operators $b_i^a$, $f_i^\alpha$ on the even sites
transform in the dual $V_0^*$. Those on the odd sites
$\overline{b}_{ia}$, $\overline{f}_{i\alpha}$ do not transform in
the double dual $V_0^{**}$, which is the dual of $V_0^*$ and would
be obtained if the anticommutator for $\overline{f}$ contained the
usual plus sign. Instead, because of the minus in the relevant
anticommutators, they transform in $V_0$. Thus upper (resp.,
lower) indices always transform in the same way. (We note that for
the osp($m+2n|2n$) spin chains or unoriented loops models, in
which all sites are supposed to be equivalent, the negative signs
can be assigned to $n$ of the $2n$ fermion components on every
site, in a translationally-invariant fashion, with similar
results.)

The TL generators are constructed as follows. First, we note that
for any two sites $i$ (even), $j$ (odd), the combinations
\begin{equation}
\sum_a  \overline{b}_{ja} b_i^a + \sum_\alpha
\overline{f}_{j\alpha}f_i^\alpha,\quad
\sum_ab_{ia}^\dagger\overline{b}_j^{a\dagger} + \sum_\alpha
f_{i\alpha}^\dagger\overline{f}_j^{\alpha\dagger}
\end{equation} %
are invariant under gl($n+m|n$). (The order of the operators in
these expressions differs from Refs.\ \cite{rs,fr}, but so do some
signs in the expressions for the generators of gl($n+m|n$) above,
so that invariance still holds.) Then the evaluation and
co-evaluation maps applied to each pair of neighbors $i$, $i+1$
can be written in terms of such combinations:%
\bea%
d_{V_{i+1}}&=&\sum_a \overline{b}_{i+1,a}b_i^a  + \sum_\alpha
\overline{f}_{i+1,\alpha}f_i^\alpha\quad (i\hbox{ even}),\\
d_{V_{i+1}}&=&\sum_a \overline{b}_{ia}b_{i+1}^a  + \sum_\alpha
\overline{f}_{i\alpha}f_{i+1}^\alpha\quad (i\hbox{ odd}),\\
b_{V_i}&=&\sum_ab_{ia}^\dagger\overline{b}_{i+1}^{a\dagger} +
\sum_\alpha
f_{i\alpha}^\dagger\overline{f}_{i+1}^{\alpha\dagger}\quad
(i\hbox{ even}),\\
b_{V_i}&=&\sum_ab_{i+1,a}^\dagger\overline{b}_i^{a\dagger} +
\sum_\alpha
f_{i+1,\alpha}^\dagger\overline{f}_i^{\alpha\dagger}\quad (i\hbox{
odd}).
 \eea
As $V_0^{**}$ does not appear in $V$, we have used the isomorphism
of $V_0^{**}$ with $V_0$ (or equivalently of $V_0^*$ with $^*V_0$)
in the cases of $d_{V_{i+1}}$ for $i$ even and $b_{V_i}$ for $i$
odd; we have abused notation a little and not recorded this in the
notation. (More accurately, we could write the operators for the
latter two cases as $d_{V_0}'$ and $b_{V_0}'$, with the
understanding that sites $i$, $i+1$ are meant with $V_0$, $V_0^*$
in their correct positions in the chain; this produces the correct
signs.) Then the TL generators can be written
as%
\be%
e_i=b_{V_i}d_{V_{i+1}}\ee for all $i$.

Next we describe the endomorphisms of $V$ in terms of explicit
expressions. We will revert to using indices $a=1$, \ldots,
$m+2n$, and it will be less confusing to drop the use of raised
indices for the time being. Then we define a basis for the space
of endomorphisms by the operators %
\bea%
\lefteqn{E_{a_0,a_1,\ldots,a_{2L-1},a_0',\ldots,a_{2L-1}'}=}&&\nonumber\\
\qquad&&\qquad b_{0a_0}^\dagger \overline{b}_1^{a_1\dagger}\cdots
\overline{b}_{2L-1}^{a_{2L-1}\dagger}\overline{b}_{2L-1,a_{2L-1}'}\cdots
b_0^{a_0'}\eea %
These are written for the case all $a_i$, $a_i'$ in the range $1$,
\ldots, $m+n$; for any $a_i$ ($i$ even) that lie in $m+n+1$,
\ldots, $m+2n$, $b_{ia}^\dagger$ must be replaced by
$f_{i\alpha}^\dagger$, and similarly for $i$ odd and for $a_i'$.
We will simplify notation and write ${\rm deg}\, a=0$ for the
degree of boson or fermion operators with $a$ in the range $1$,
$m+n$, ${\rm deg}\, a=0$ for $a$ in $m+n+1$, \ldots, $m+2n$. Then
the multiplication of the $E_{\ldots}$s is given by%
\bea%
\lefteqn{E_{a_0\ldots a_{2L-1},a_0'\ldots a_{2L-1}'}E_{b_0\ldots
b_{2L-1},b_0'\ldots b_{2L-1}'}=}&&\nonumber\\
&&\qquad E_{a_0\ldots a_{2L-1},b_0'\ldots
b_{2L-1}'}(-1)^{\sum_{i\,{\rm odd}}{\rm deg}\,a_i'} \prod_i
\delta_{a_i',b_i}
\eea%
The minus signs can be removed by defining
$\widetilde{E}_{a_0\ldots a_{2L-1}'}=E_{a_0\ldots
a_{2L-1}'}(-1)^{\sum_{i\,{\rm odd}}{\rm deg}\,a_i'}$. Then the
$\widetilde{E}_{\ldots}$s are a basis for the space of
endomorphisms ${\rm End}\, V$, which is naturally isomorphic to
$V\otimes V^*$, and as the dual $V^*$ of $V=V_0\otimes
V_0^*\otimes\cdots$ is naturally isomorphic to $V_0^{**}\otimes
V_0^*\otimes\cdots$ with the reverse ordering, the
$\widetilde{E}_{\ldots}$s refer to the natural basis for this
space. Again, the minus signs for odd elements on odd sites are
due to the map from $V_0^{**}$ to $V_0$, which are used to map
$V^*$ to $V$.

The anti-isomorphism on the endomorphisms is given by %
\bea%
\lefteqn{(E_{{a_0}\ldots a_{2L-1}'})^s =}\nonumber\\
&&\qquad (-1)^{(\sum_i{\rm deg}\,a_i)(\sum_{i'}{\rm
deg}\,a_i')}E_{a_{2L-1}'\ldots a_0}.\eea%
One can check that this is an algebra isomorphism to $({\rm End}\,
V)^{\rm op}$ as required. Further, it is clear that
$((E_{{a_0}\ldots a_{2L-1}'})^s)^s= E_{{a_0}\ldots a_{2L-1}'}$, so
$^s$ is an anti-involution. The sign in the definition of $^s$
resembles that in the supertranspose $(E_{{a_0}\ldots
a_{2L-1}'})^*$ of $E_{{a_0}\ldots a_{2L-1}'}$, however the
supertranspose of an endomorphism of $V$ would be an endomorphism
of $V^*$ (acting on the left). Our anti-isomorphism $^s$ is
obtained by combining the supertranspose with the map $V^*\to V$
given by multiplication by $(-1)^{\sum_{i\,\rm odd}{\rm deg}\,
a_i}$.

One can easily check that the identity in ${\rm End}\, V$,%
\be%
1=\sum_{a_0\ldots a_{2L-1}}(-1)^{\sum_{i\,\rm odd}{\rm deg}\,
a_i}E_{a_0\ldots a_{2L-1},a_0\ldots a_{2L-1}}\ee%
is invariant under $^s$, and that the TL generators map as $e_i\to
e_i^s=e_{2L-2-i}$. The anti-isomorphism $s$ is not determined
uniquely by these properties. One can obtain other
anti-isomorphisms by conjugating with, for example, elements of
the supergroup GL$(m+n|n)$, obtained by exponentiating the action
of the generators $\sum_i J_{ia}^b$ of gl$(m+n|n)$ on $V$. A
particular case is the involutive automorphism of ${\rm End}\, V$
given by multiplication of an element by $-1$ to its degree, which
is conjugation by the similarly-defined map on $V$. However, this
freedom may be reduced if we also insist that the anti-isomorphism
be an anti-involution.

\subsection{Symmetry generators}

Now we obtain an explicit description of a set of operators that
commute with the action of the TL algebra in $V$, which is shown
in the main text to be a basis for the commutant algebra
$\cA_{m+n|n}(2L)$. We again begin with a set of operators (for $k\leq 2L$)%
\be%
\widetilde{J}^{a_1a_2\ldots a_k}_{b_1b_2\ldots b_k}=\sum_{0\leq
i_1<i_2<\cdots <i_k\leq 2L-1} J_{i_1 b_1}^{a_1}J_{i_2
b_2}^{a_2}\cdots J_{i_k b_k}^{a_k}, \ee%
where $J_{ia}^b$ for $a$, $b=1$, \ldots, $m+2n$ have been defined
above. As the notation may suggest, each $J_{ia}^b$ transforms as
the representation $V_0\otimes V_0^*$ (with the graded tensor
product in that order), whether $i$ is odd or even. This is clear
for $i$ even as $J_{ia}^b=b_{ia}^\dagger b_i^b$ [again here, the
notation is that $b_{ia}$ (resp., $\overline{b}_i^a$) means
$f_{i\alpha}$ (resp., $\overline{f}_i^\alpha$) with
$\alpha=a-(m+n)$ when $\alpha>0$, and similarly for the adjoints
of these operators]. For $i$ odd there is an additional sign:
$J_{ia}^b=-(-1)^{({\rm deg}\,a)({\rm deg}\,
b)}\overline{b}_i^{b\dagger}\overline{b}_{ia}$. The overall minus
is irrelevant, while $(-1)^{({\rm deg}\,a)({\rm deg}\, b)}$ is
exactly the sign produced by flipping the order of the factors in
the tensor product. Hence the set of $\widetilde{J}$s for each $k$
transform as $V_0\otimes V_0^*\otimes\cdots\otimes V_0^*$, with
$2k$ factors, and in particular this is true for all terms in the
summation over the $i_l$s.

The $\widetilde{J}$s are to be used as a basis set for a space of
operators within which we construct the commutant of the TL
algebra. It will suffice to consider linear combinations from a
subset all having the same value of $k$. Then a linear combination
may be written %
\be%
\widehat{j}=\sum_{a_1,a_2,\ldots,b_k}j_{a_1\ldots a_k}^{b_1\ldots
b_k}\widetilde{J}^{a_1\ldots a_k}_{b_1\ldots b_k},\ee%
where $j_{a_1\ldots a_k}^{b_1\ldots b_k}$ are numerical
coefficients. For each $i$ in the range $i=0$, \ldots, $2L-2$, the
terms in the summation over $i_l$s in $\widehat{j}$ can be grouped
into those in which zero, one, or two of the sites $i$, $i+1$  are
occupied by a factor $J_i$ or $J_{i+1}$. Those in which none of
the $i_l$, ($l=1$, \ldots, $k$) are equal to $i$ or $i+1$ clearly
commute with $e_i$. Also, the terms in which just one $i_l=i$ or
$i+1$ may be grouped in pairs, such that they contain
$J_{ib_l}^{a_l}+J_{i+1,b_l}^{a_l}$ as a factor (all other operator
factors commuting with $e_i$), and these also commute with $e_i$;
indeed, they annihilate it from either side. Finally there are
terms in $\widehat{j}$ in which $i_l=i$, $i_{l+1}=i+1$ for some
$l$. For these it will clearly be sufficient if the linear
combination of expressions containing
$J_{ib_l}^{a_l}J_{i+1,b_{l+1}}^{a_{l+1}}$ annihilates $e_i$ from
either side. (In the main text, the use of this stronger condition
is motivated, and shown to produce the whole commutant.) This
condition reduces further to the conditions that $d_{V_{i+1}}$
annihilates these terms from the left, and $b_{V_i}$ annihilates
them from the right. This imposes a set of linear relations on the
coefficients $j$. (It is here that the above transformation
properties of the terms in $\widetilde{J}$ are important, as the
coefficients $j$ contain no $i_l$ dependence.) These relations
involve adjacent indices of the
coefficients $j$. By explicit calculation we find:%
\bea%
\sum_{a}j^{\ldots b_l a\ldots}_{\ldots a a_{l+1}\ldots}&=&0,\label{jrel1}\\
\sum_{a}j^{\ldots a b_{l+1}\ldots}_{\ldots a_l a\ldots}(-1)^{{\rm
deg}\,a +{\rm deg}\,a({\rm deg}\,a_l+{\rm deg}\,b_{l+1})}
&=&0,\label{jrel2} \eea for $l=1$, \ldots, $k-1$; indices not
displayed (those with subscript $<l$ or $>l+1$) are free. The
number of independent solutions to these relations for $k$ even is
$(D_{k/2}')^2$. This can be shown by using the independence of the
coefficients from the length $L$, so that we may consider a chain
with $k=2L$; then reorder the operators in the $\widetilde{J}$s
into the form of the basis elements $E_{\ldots}$ above (this is an
isomorphism), and consider the subspace annihilated by all $e_i$
from either side; as discussed in the main text, the dimension of
this space is clearly the square of the dimension $D_{k/2}'$ of
the standard module ${\cal V}_{k/2}$.

In each space $J^{(k)}$ of operators constructed using solutions
to the relations eqs.\ (\ref{jrel1}) and (\ref{jrel2}), we can
attempt to produce a complete set of operators $J^{a_1\ldots
a_k}_{b_1\ldots b_k}$ that are not linearly independent, but obey
relations on adjacent indices similar to those for $j$ (this is
how these spaces of operators are referred to in the main text).
These are supposed to be constructed (and normalized) in the form
$J^{a_1\ldots a_k}_{b_1\ldots b_k}=\widetilde{J}^{a_1\ldots
a_k}_{b_1\ldots b_k}$ $\pm$ corrections that involve contractions
of pairs of adjacent indices. From the gl$(m+n|n)$ symmetry
considerations above, the relations must take the form%
\bea%
\sum_a J^{\ldots a a_{l+1}\ldots}_{\ldots b_l a\ldots}(-1)^{{\rm
deg}\,a}&=&0, \label{Jrel1}\\
\sum_a J_{\ldots a b_{l+1}\ldots}^{\ldots a_l a\ldots}(-1)^{{\rm
deg}\,a({\rm deg}\,a_l+{\rm deg}\,b_{l+1})}&=&0,\label{Jrel2}
\eea%
for $l=1$, \ldots, $k-1$. For $n=0$ (and thus $m\geq 2$), these
reduce to the tracelessness conditions used in the earlier
sections of this paper. This works also for $n>0$, at least when
$|m|\geq2$, when the sets of $j$-coefficients involved are
essentially Jones-Wenzl projectors acting in the index spaces;
these spaces are isomorphic to $V(k)\otimes V(k)^*$. For
example, for $k=2$ the expression is, for $m\neq0$,%
\bea%
J^{a_1a_2}_{b_1b_2}&=&\widetilde{J}^{a_1a_2}_{b_1b_2}-\frac{1}{m}
\delta^{a_1}_{b_2}\sum_a\widetilde{J}^{aa_2}_{b_1a}(-1)^{{\rm
deg}\,a}\nonumber\\
&&\mbox{}-\frac{1}{m}\delta^{a_2}_{b_1}(-1)^{{\rm deg}\,a_2{\rm
deg}\,b_1+{\rm deg}\,a_1{\rm deg}\,b_1+{\rm deg}\,a_2{\rm
deg}\,b_2}\nonumber\\
&&\mbox{}\quad\times\sum_a\widetilde{J}^{a_1a}_{ab_2}(-1)^{{\rm
deg}\,a({\rm deg}\,a_1+{\rm deg}\,b_2)}\nonumber\\
&&\mbox{}+\frac{1}{m^2}\delta^{a_1}_{b_2}\delta^{a_2}_{b_1}(-1)^{{\rm
deg}\,a_2{\rm deg}\,b_1+{\rm deg}\,a_1{\rm deg}\,b_1+{\rm
deg}\,a_2{\rm deg}\,b_2}\nonumber\\
&&\mbox{}\quad\times\sum_{a,b}\widetilde{J}^{ab}_{ba}(-1)^{{\rm
deg}\,a};
\eea%
this satisfies relations (\ref{Jrel1}), (\ref{Jrel2}), and the
coefficients $j$ in its expansion in terms of $\widetilde{J}$s
obey relations (\ref{jrel1}), (\ref{jrel2}). Obviously this
expression fails for $m=0$. There are similar problems for some
other $k$ values for the cases $m=0$, $\pm1$, in which the
algebras, and some of the standard modules, are not semisimple;
the problem occurs whenever the $k/2$th standard TL module over
TL$_{k}(q)$ is not simple, in which case there is no corresponding
Jones-Wenzl projector (an idempotent element of the TL$_{k}(q)$
algebra) onto that module. When a complete set $J^{a_1\ldots
a_k}_{b_1\ldots b_k}$ obeying all the conditions stated above does
not exist for some value of $k$, the construction using
coefficients $j$ can always be used instead.

For the action of the anti-involution on the symmetry generators,
we first note that by writing $J_{ia}^b$ in terms of the basis
elements $E_{\ldots}$, we can show that
$(J_{ib}^a)^s=-J_{2L-1-i,b}^a$. Then from the anti-involution
property of $^s$, %
\bea%
(\widetilde{J}^{a_1\ldots a_k}_{b_1\ldots b_k})^s&=&
(-1)^k(-1)^{\sum_{l<l'}({\rm deg}\,a_l+{\rm deg}\,b_l) ({\rm
deg}\,a_{l'}+{\rm deg}\,b_{l'})}\nonumber\\
&&\mbox{}\times \widetilde{J}^{a_k\ldots a_1}_{b_k\ldots b_1}.\eea %
We can then see that the relations (\ref{jrel1}), (\ref{jrel2}) on
the coefficients $j$ in the operators $\widehat{j}$ are invariant
under $^s$, and so the transformation maps $J^{(k)}$ into itself.
One may obtain an idea of how the transformation looks from the
action on the ${J}^{a_1\ldots a_k}_{b_1\ldots b_k}$, when these
are available, which takes the same form as for
$\widetilde{J}^{a_1\ldots a_k}_{b_1\ldots b_k}$.


\subsection{Hopf superalgebras}

When the definitions for superalgebras and graded tensor products
of their modules are used to replace the usual ones \cite{cp,kass}
in the definitions for Hopf algebras, ribbon Hopf algebras, etc,
we obtain what may be called Hopf superalgebras, ribbon Hopf
superalgebras, etc. We note that Hopf superalgebras arise
naturally as the UEAs for Lie superalgebras, and the constructions
for the action of gl$(m+n|n)$ within the modules $V$ serve as
examples. Needless to say, a Hopf superalgebra $\cal A$ for which
${\cal A}_{(1)}=0$ is just an ordinary Hopf algebra. Here we will
mention the main variations in the definitions and properties for
Hopf superalgebras.

First, the comultiplication $\Delta:\cA\to\cA\otimes\cA$ is a
superalgebra homomorphism; we write %
\be%
\Delta(a)=\sum_{(a)}a'\otimes a''.\ee%
There are unit $\eta:{\bf C}\to\cA$ and counit
$\varepsilon:\cA\to{\bf C}$ maps, as usual; these maps are even.
The antipode $S$
obeys%
\be%
\sum_{(a)}a'S(a'')=\sum_{(a)}S(a')a''=\eta(\varepsilon(a)),\ee%
and is an anti-linear homomorphism to $\cA^{\rm op}$, thus
$S(ab)=(-1)^{{\rm deg}\,a\,{\rm deg}\, b}S(b)S(a)$. If it is
invertible, and if $S^{-1}=S$, as in our cases, then it is an
anti-involution. Dually, $S$ is also a graded co-algebra
homomorphism to the opposite co-algebra, in which the coproduct is
$\Delta^{\rm op}=\tau\circ\Delta$; this means that it obeys
$S\otimes S\circ\Delta^{\rm op}(a)=\Delta\circ S(a)$, which is %
\be%
\sum_{(a)}(-1)^{{\rm deg}\,a'{\rm
deg}\,a''}S(a'')\otimes S(a')=\sum_{(a)}S(a)'\otimes S(a)''.\ee%
The antipode is used in defining dual representations. For
example, if $\cA$ acts on $\cal V$ as $v\to av$ for $a\in \cA$,
$v\in {\cal V}$, then the left dual vector space ${\cal V}^*$ of
$\cal V$ becomes the left dual as an $\cA$-module, on which the
$\cA$ action is $w\to S(a)^*w$, where $w\in V^*$ and we used the
supertranspose $S(a)^*$ of $S(a)$ so that ${\cal V}^*$ becomes a
left $\cA$-module. Readers can check that these definitions are
satisfied by the UEA of gl$(m+n|n)$, with the comultiplication and
antipode defined on the generators $J_a^b$ by
$\Delta(J_a^b)=J_a^b\otimes 1+ 1\otimes J_a^b$ and
$S(J_a^b)=-J_a^b$, and then extended to the rest of the algebra by
the homomorphism properties, and also that this agrees with the
construction of our spin chain. In particular, the odd sites
indeed transform as the left dual of $V_0$.

The definitions of braiding and twist are unchanged when written
in terms of the graded $\Delta^{\rm op}$, and so on, and with
these additional structures we are led to definitions of ribbon
Hopf superalgebras. The definitions of tensor and ribbon
categories (as abstract categories) are unchanged, and are
satisfied by the categories of graded modules over ribbon Hopf
superalgebras, and hence equivalences of these categories still
make sense in this broader context. The ordinary supertrace and
superdimension on modules over the UEA of a Lie superalgebra can
then be viewed as examples of the more general concepts of quantum
trace and dimension.


\subsection{Enlarged symmetry algebra for open gl($1|1$) chain}

Here we give details of our construction for the open gl($1|1$)
spin chain, which is a free fermion system \cite{rs}. The free
fermion form of the model is defined using fermion operators $f_i$
and their adjoints $f_i^\dagger$, $i=0$, $1$, \ldots, $2L-1$,
which obey $\{f_i,f_{i'}\}=0$,
$\{f_i,f_{i'}^\dagger\}=(-1)^i\delta_{ii'}$. The TL generators can
be written as%
\be%
e_i=(f_i^\dagger+f_{i+1}^\dagger)(f_i+f_{i+1}),\ee%
for $i=0$, \ldots, $2L-2$. Some symmetry operators in $J^{(k)}$,
for $k=1$, $2$, which commute with all the $e_i$s, are%
\bea%
F&=&\sum_if_i,\\
F^\dagger&=&\sum_if_i^\dagger,\\
F_{(2)}&=&\sum_{i<i'}f_if_{i'},\\
F_{(2)}^\dagger&=&\sum_{i<i'}f_{i'}^\dagger f_{i}^\dagger,\\
N&=&\sum_i(-1)^if_i^\dagger f_i -L.\eea %
The remaining symmetry operators turn out to be sums of products
of these, so this set of five operators (together with $1$) is a
set of generators of the full algebra $\cA_{1|1}$. The graded
commutators of these five close on themselves, so they form a Lie
superalgebra. $F$, $F^\dagger$ generate a Lie sub-superalgebra
isomorphic to psl($1|1$). $F_{(2)}$, $F_{(2)}^\dagger$, and $N$
generate an sl$_2$ Lie subalgebra, with $N$ as $2S_z$, and $F$,
$F^\dagger$ transform as a doublet under this sl$_2$, and so form
a Lie ideal. (We note that $(-1)^if_i^\dagger f_i$ is the fermion
number $=0$, $1$ at site $i$.) The Lie superalgebra is thus not
semisimple, but is a semidirect product of these two, and can be
viewed as the superalgebra of translations and sl$_2$ rotations of
the superplane with anticommuting coordinates $f$, $f^\dagger$.
The associative (universal enveloping) algebra it generates,
$\cA_{1|1}$, is isomorphic to $U_q($sl$_2)$ (modulo the
restriction to modules with integer $j$). Representations of this
algebra acting in the spin chain can be easily constructed.

For the corresponding closed chain, generators like $F_{(2)}$ are
lost, as the summation must be extended around the chain, and then
anticommutation of $f_i$ and $f_{i'}$ makes it vanish.

We emphasize that for open gl($n|n$) chains with $n>1$, the
symmetry algebra $\cA_{n|n}$ is not an enveloping algebra of a
finite-dimensional Lie superalgebra. For the closed chains, some
symmetry operators of the open version are lost on closing the
chain, for reasons similar to the case of $F_{(2)}$ above, but for
general $n>1$ many operators remain.


\begin{references}
\bibitem{quotalg} Strictly, here we mean the image of $U({\rm
sl}_m)$ in the finite-dimensional representations that are trivial
under the center ${\bf Z}_m$ of SU($m$).
\bibitem{m=0} The TL algebras for $m$ and $-m$ are isomorphic. Our
definition of the algebra by relations (\ref{tlrel}) differs from
some authors by a rescaling by $m$, which makes a difference only
in the case $m=0$ (or $\infty$).
\bibitem{project} Projective modules \cite{pierce,anf} are {\em not}
the same as the ``projective representations'' which are more
familiar to most physicists, and which are not needed here.
\end{references}
\end{document}